\documentclass[12pt]{article}
\usepackage{amsmath, amsthm, amssymb}
\usepackage[dvips]{graphics}
\usepackage[usenames,dvipsnames]{xcolor}
\usepackage{subfigure}
\usepackage{multirow}
\usepackage{graphicx} 
\usepackage{bm}     
\usepackage{setspace} 
\usepackage{booktabs} 
\usepackage{verbatim}
\usepackage{appendix} 
\usepackage{natbib} 
\usepackage{changepage}
\usepackage{caption}

\usepackage{hyperref}

\newcommand{\beginsupplement}{%
        \setcounter{section}{0}
        \renewcommand{\thesection}{S\arabic{section}}
        \setcounter{table}{0}
        \renewcommand{\thetable}{S\arabic{table}}%
        \setcounter{figure}{0}
        \renewcommand{\thefigure}{S\arabic{figure}}%
        \setcounter{page}{1}
        \renewcommand{\thepage}{S\arabic{page}}
        \setcounter{equation}{0}
        \renewcommand{\theequation}{S\arabic{equation}}
     }

\usepackage[bb=boondox,scr=rsfso]{mathalfa} 

\usepackage{soul}

\theoremstyle{plain}

\theoremstyle{definition}

\theoremstyle{remark}

\input{commands.sty}

\bibpunct{(}{)}{;}{a}{}{,}

\makeatletter
\def\moverlay{\mathpalette\mov@rlay}
\def\mov@rlay#1#2{\leavevmode\vtop{%
   \baselineskip\z@skip \lineskiplimit-\maxdimen
   \ialign{\hfil$\m@th#1##$\hfil\cr#2\crcr}}}
\newcommand{\charfusion}[3][\mathord]{
    #1{\ifx#1\mathop\vphantom{#2}\fi
        \mathpalette\mov@rlay{#2\cr#3}
      }
    \ifx#1\mathop\expandafter\displaylimits\fi}
\makeatother

\begin{document}
\def\spacingset#1{\renewcommand{\baselinestretch}%
{#1}\small\normalsize} \spacingset{1}

  \title{\bf A Projection Approach to Local Regression \\
with Variable-Dimension Covariates}
  \author{Matthew J. Heiner\thanks{
    The authors gratefully acknowledge partial funding from grant FONDECYT 1220017.} \ and  Garritt L. Page \\
    Department of Statistics, Brigham Young University, Provo, Utah\\
    and \\
    Fernando Andr\'es Quintana \\ Departamento de Estad\'{i}stica, \\ Pontificia Universidad Cat\'{o}lica de Chile, Santiago
         \\ and Millennium Nucleus Center for the \\ Discovery of Structures in Complex Data
    }
  \maketitle

\begin{abstract}
Incomplete covariate vectors are known to be problematic for estimation and inferences on model parameters, but their impact on prediction performance is less understood. We develop an imputation-free method that builds on a random partition model admitting variable-dimension covariates. Cluster-specific response models further incorporate covariates via linear predictors, facilitating estimation of smooth prediction surfaces with relatively few clusters. We exploit marginalization techniques of Gaussian kernels to analytically project response   distributions according to any pattern of missing covariates, yielding a local regression with internally consistent uncertainty propagation that utilizes only one set of coefficients per cluster. Aggressive shrinkage of these coefficients regulates uncertainty due to missing covariates. The method allows in- and out-of-sample prediction for any missingness pattern, even if the pattern in a new subject's incomplete covariate vector was not seen in the training data. We develop an MCMC algorithm for posterior sampling that improves a computationally expensive update for latent cluster allocation. Finally, we demonstrate the model's effectiveness for nonlinear point and density prediction under various circumstances by comparing with other recent methods for regression of variable dimensions on synthetic and real data.
\end{abstract}


\noindent%
{\it Keywords:} dependent random partition models, clustering, indicator missing, pattern missing, Bayesian nonparametrics
\vfill

\newpage
\spacingset{1.5} 

\section{Introduction}

It is common in applied settings that one or more covariates are unsuccessfully measured on a subset of subjects.  As a result, incomplete covariate vectors are often encountered (\citealt{molenberghs2014handbook}).  In clustering and regression settings, variable dimension covariate vectors can be problematic, necessitating methods that appropriately accommodate them.  Recently, \cite{ppmxMissing} developed an elegant and uncomplicated variable dimension regression approach (VDReg) that is based on the covariate-dependent random partition models (PPMx) of \citet{PPMxMullerQuintanaRosner}.   Their approach avoids imputation, seamlessly accommodates mixed-type covariate vectors (i.e., vectors populated with continuous and categorical variables), and provides prediction for any missingness pattern regardless of whether it appears in the training data set. However, their method only incorporates covariates in the prior on partitions.  It seems reasonable that including covariates in the likelihood/sampling model (e.g., through a regression) could improve prediction rates and/or result in a more parsimonious partition estimate. In this paper, we detail an approach that includes a regression in the likelihood of a hierarchical model that continues to avoid imputation and accommodates all missingness patterns.  The response models are analytically marginalized according to the pattern of missing covariates, yielding a local regression with internally consistent uncertainty propagation.  This results in a coherent nonlinear modeling approach that is consistent and parsimonious across missingness patterns.  


\begin{figure}[b!]
    \centering
    \includegraphics[width=3.5in]{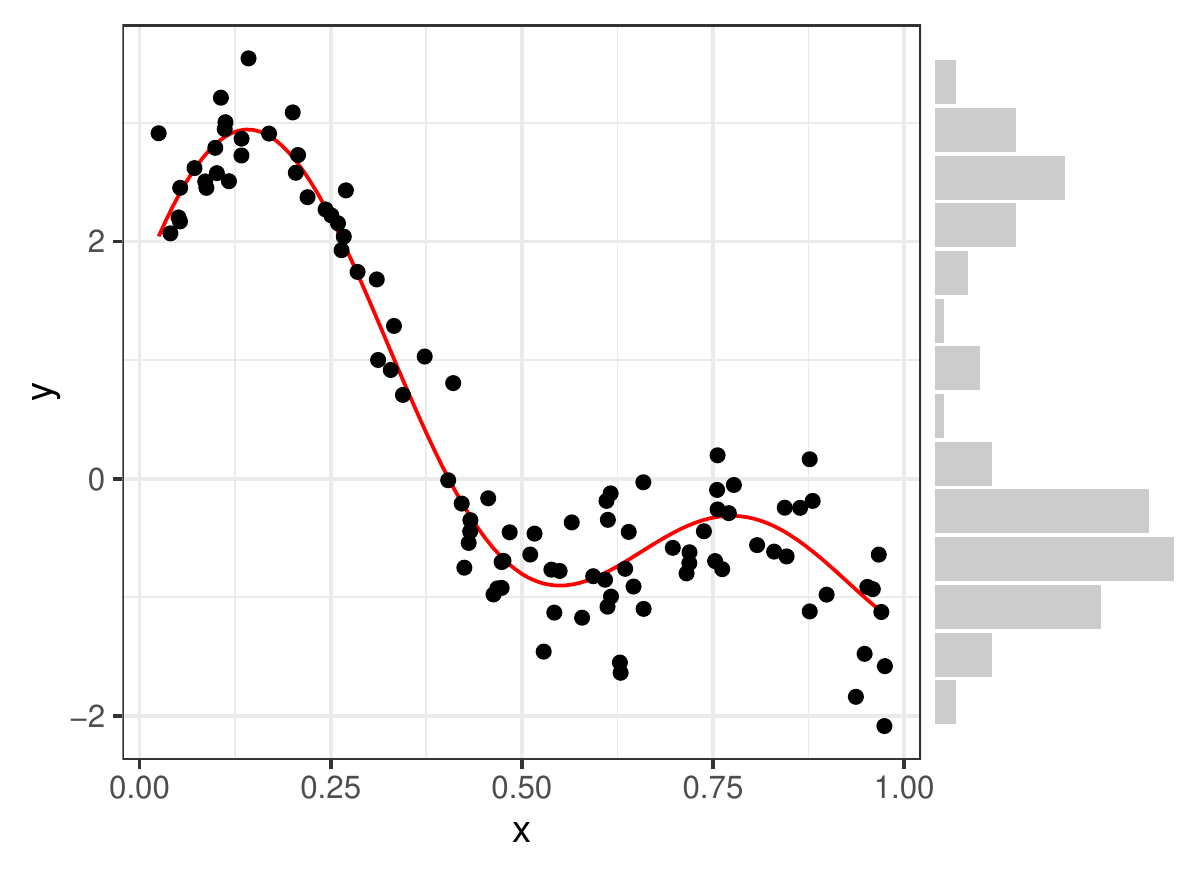}
    \caption{Scatter plot of data generated from a nonlinear function (red curve) and Gaussian noise. When $x$ is missing at random, regression methods that do not account for uncertainty in $x$ can fail to capture the distribution of the response $y$, represented by the marginal histogram on the right.}
    \label{fig:simpleMarg}
\end{figure}

Incomplete predictor vectors can have adverse effects on prediction rate, particularly if an influential predictor is missing (\citealt{Mercaldo2020patternSub}).  This is particularly true when considering local regressions.  For example, it is not obvious how to make predictions for units that exhibit a missingness pattern not present in the training data.  
In addition, omission of a relevant covariate can inflate uncertainty, or perhaps even worse, in some cases introduce multimodality to a regression response distribution, as illustrated in Figure \ref{fig:simpleMarg}.  
This is problematic because most ensemble prediction methods create point estimates from a weighted mean and unimodal error and while that error may be conditionally heteroscedastic, it would miss multimodality. In contrast, random partition models and imputation methods behave like mixture models (mixtures of densities/predictions) and are able to accommodate multimodal error  distributions that precipitate from marginal, or partially conditional, response distributions. 
However, point predictions with these methods can fall in areas of relatively low predictive density, say, a valley between two modes.  For these reasons it is important to carefully select the performance metric when studying the behavior of prediction methods developed to accommodate incomplete predictor vectors as the typically employed ones (i.e., mean squared prediction error) may not be appropriate.

Unlike the statistical literature dedicated to missing response values which has become quite rich (\citealt{daniels&hogan:2008}, \citealt{molenberghs2014handbook}), the literature dedicated to missing covariate values is less developed.  This may be due to thinking that methods developed for missing response can be applied in the case of missing covariates.   Although it is possible to apply missing response methods to the missing covaraite case, there are added complications when considering prediction like those mentioned above.   Due to this, methods that focus on incomplete predictor vectors have begun to appear in the literature.  \citet{AdaptiveBayesianSLOPE} consider variable selection  in high-dimensional settings with missing covariates. Specifically, they combine SLOPE (sorted $L^1$ regularization) with spike-and-slab LASSO, and a stochastic approximation of expected maximization algorithm to impute missing data. In the context of the analysis of electronic health records, \cite{https://doi.org/10.48550/arxiv.2201.00068} develop a Bayesian nonparametric common-atoms regression model for generating synthetic controls in clinical trials that incorporates variable dimension covariates. They do so by adopting the same approach to covariate-dependent priors that we follow here.  Although variable-dimension covariates are not considered, \citet{friedberg2020local} show an advantage of using local linear predictors in random forests which is similar to our desire to include local linear regression in a partition model.  

In this paper we also consider the value added of including a local regression in the data model.  Since adding covariates to the mean model often times improves model fit and out-of-sample prediction rates even for variables that are not deemed ``significant,'' intuition would dictate that this would hold in a local regression setting as well.  However, contrary to previously held intuition, including covariates in the data model of a covariate dependent clustering method does not guarantee improved prediction rates and in fact can result in degraded performance.  Because of this, and due to the added computational cost of including covariates, we develop a quick procedure that provides some insight into whether including covariates in the likelihood will provide benefit beyond employing a covariate-dependent clustering method. 

The rest of the paper is organized as follows.  In Section \ref{sec:background} we provide the necessary notation and background to make the paper self contained.  In Section \ref{sec:model} we detail our approach to regression with variable-dimension covariates and provide computational details in Section \ref{sec:computation}.  Section \ref{sec:simulations} details a simulation study while Section \ref{sec:applications} describes results from two applications commonly encountered in the literature.  We provide some concluding remarks in Section \ref{sec:discussion}.

\section{Background and Preliminaries}
\label{sec:background}

We define notation that will be used throughout and provide the relevant details of VDReg.  Let  $i = 1, \ldots, m$ index $m$ experimental units. Let $\rho_m = \{S_1,
\ldots, S_{k}\}$ denote a partition (or clustering) of the $m$ units
into $k$ nonempty and exhaustive subsets so that $\{1,\ldots,m\} =
\bigcup_j S_j$, for disjoint subsets $S_j$. In addition, we will use cluster membership indicators $c_i = j$ if $i \in S_j$ when describing the model.   Let $\bx_i = (x_{i1}, \ldots, x_{ip})$ denote a $1\times p$ complete covariate vector measured on unit $i$ and $\bx = \{\bx'_1, \ldots, \bx'_m\}$ the $m\times p$ covariate matrix.
Further, $\bx^{\star}_j = \{\bx_i:i \in S_j\}$ denotes the collection of covariate vectors belonging to units that are members of the $j$th cluster.  We introduce missing covariates by denoting as $\OO_i$ the collection of covariate indices that are observed for subject $i$.  Thus, the $i$th subject's observed covariate vector is denoted as $\bx_i^o = \{x_{i\ell} : \ell \in \OO_i\}$ and the collection of observed covariate vectors that belong to the $j$th cluster is $\bx_j^{\star o} = \{\bx^{o}_{i} : i \in S_j\} = \{x_{i\ell} : \ell \in \OO_i, i\in S_j\}$ and the collection that are missing is $\bx_j^{\star m} =\{x_{i\ell} : \ell \not\in \OO_i, i\in S_j\}$.  

VDReg is based on the PPMx prior on partitions, for generic partition $\rho$, given by 
\begin{align} \label{eq:ppmx}
  p(\rho) \propto \prod_{j=1}^k c(S_j \mid M)\prod_{\ell=1}^pg_{\ell}(\bm{x}^{\star}_j \mid \bm{\xi}_{\ell}).
\end{align}
Here $c(\cdot \mid M)$ is called a cohesion function and $g_{\ell}(\cdot \mid \bm{\xi}_{\ell})$ a similarity function.  The main contribution of VDReg is evaluating the similarity function using only subjects $i \in \Cjl =\{i:i\in S_j, \ell \in \OO_i\}$, i.e., those with observed covariate $\ell$. Letting $\xjlo=\{ x_{i \ell}: i \in \Cjl\}$, the modified similarity function in VDReg is 
\begin{align} \label{eq:VDRegSim}
\tg(\bx_j^{\star o}\mid\bxi)
\defeq  \prod_{\ell=1}^p \tg_\ell(\xjlo \mid \bxi_{ \ell}) \defeq
        \prod_{\ell=1}^p \int \prod_{i\in \Cjl} q(x_{i \ell}  \mid \bzeta_{j \ell}) \,
        q(\bzeta_{j \ell} \mid \bxi_{\ell})
        \diff\bzeta_{j \ell},
\end{align}
where $q(\cdot \mid \bzeta_{j \ell})$ and $q(\bzeta_{j \ell} \mid \bxi_{\ell})$ represent a conjugate pair of densities, and $\bxi_\ell$ is a vector of hyperparameters influencing how covariates inform clustering. 
In \eqref{eq:VDRegSim} we adopt the convention that $\prod_{i\in\varnothing} q(x_{i \ell}  \mid \bzeta_{j \ell}) = 1$. 
An appeal to VDReg is that algorithms commonly employed in Bayesian nonparametric methods (e.g., Algorithm 8 of \citealt{neal:2000}) readily apply, with the slight adjustment that a missing indicator matrix be carried along when evaluating \eqref{eq:VDRegSim}. 

\section{Modeling Approach}
\label{sec:model}

We first motivate our proposed model by observing that the VDReg similarity function can be derived from 
a marginalization of a product partition model (PPM) for $(y, \bx)$ jointly. 
We then propose a conditional PPMx model for variable-dimension local regression in Section \ref{sec:condmod}. 
The remainder of the section develops the modeling approach, including priors and prediction, and introduces a procedure to screen for local linearity.

\subsection{Alternate perspective on original variable-dimension PPMx}
\label{sec:alternateView}

Consider a PPM prior on $\rho$, i.e., $g_\ell(\bx^\star_j) = 1$ for all $\ell$, $j$ in \eqref{eq:ppmx}, 
and a joint sampling model for $(y, \bx)$ that employs the conjugate densities from the similarity functions in \eqref{eq:VDRegSim} to model each $x_\ell$ independently. Then analytically integrating over the missing $\bm{x}^{m}$ results in the same distribution of $(y, \rho \mid \bx)$ as that obtained by modeling $(\bm{y} \mid \rho)$ with $(\rho \mid \bm{x})$ based on $\tg(\bx^{\star o}_j\mid\bxi)$ in \eqref{eq:VDRegSim}. That is, dropping similarity contributions from missing covariates is equivalent to integrating them out of a PPM for random $\bx$, under certain conditions.

Assuming joint Gaussianity of $(y, \bx)$ and relaxing independence of $y$ with each $x_\ell$ yields a sampling model that can still be analytically marginalized over $\bm{x}^{m}$ and factored into the product over $q_{\ell}(x_\ell)$ for $\ell \in \OO$ and a density for $y \mid \bm{x}^o$ whose mean is linear in the observed covariates. Specifically, take $q_\ell(x_\ell) = \mathcal{N}\big(x_\ell ; \, \mu^{(x)}_\ell, \sigma^{(x)2}_\ell \big)$ for $\ell = 1, \ldots, p$, and $y \mid \bx \sim \mathcal{N}\big(\mu + \sum_{\ell=1}^p \beta_\ell z_\ell, \, \sigma^2\big)$, where $z_\ell = (x_\ell - \mu^{(x)}_\ell) / \sigma^{(x)}_\ell$. Integrating the joint density with respect to $\bx^m$ yields
\begin{align}
    \label{eq:normal_reg_marg}
    \int p(y \mid \bx ) \prod_{\ell=1}^p q_\ell(x_\ell) \diff \bx^m = \mathcal{N}\Big(y ; \, \mu + \sum_{\ell \in \OO} \beta_\ell\, z_\ell, \, \sigma^2 + \sum_{\ell \notin \OO} \beta^2_\ell \Big) \, \prod_{\ell \in \OO} q_\ell(x_\ell) \, ,
\end{align}
where $\ell \notin \OO$ is taken to mean $\ell \in \{1, \ldots, p \} \setminus \OO$. Note that centering and scaling each $x_\ell$ stabilizes the mean and simplifies the expression for inflated variance of the conditional distribution of $y$ for all possible missingness patterns in $\OO$. The product of $q_\ell(x_\ell)$ densities over the indices of observed covariates ($\OO$) resembles again the construction of the similarity function $\tg$ in \eqref{eq:VDRegSim}. 

Introducing parameters that are shared between the sampling model for $(y \mid \bx)$ and an auxiliary model for $\bx$ complicates the connection and idea of an ``equivalent" joint model. However, the mechanics of marginalizing in \eqref{eq:normal_reg_marg} offer a simple and coherent framework that \emph{motivate} the conditional model, in the spirit of PPMx, that we next propose.

\subsection{Variable-dimension local regression model}
\label{sec:condmod}

Rather than model $\bx$ as a random quantity, we adopt the VDReg similarity function in \eqref{eq:VDRegSim} and develop a purely conditional sampling model for $(y \mid \bm{x}^o, \rho)$ that mimics the marginalization behavior of the joint model based on \eqref{eq:normal_reg_marg}. Although this model admits a corresponding imputation scheme, our implementation proceeds without imputation.

The univariate normal densities $\{ q_\ell(\cdot) \}$ contain parameters that in the original PPMx model are either fixed or integrated out of the similarity function as part of $\bzeta$. The joint model in \eqref{eq:normal_reg_marg} also employs these parameters in the sampling density for $y$. 
To define a conditional product partition model, we separate the parameters $\{ (\mu^{(x)}_\ell, \sigma^{(x)}_\ell) \}$ in $\{ q_\ell(\cdot) \}$ from their corresponding parameters in the sampling model. As with the PPMx, they are integrated out of the similarity function, which contributes in the usual way to $p(\rho \mid \bx)$.

In the sampling model, $\{ (\mu^{(x)}_\ell, \sigma^{(x)}_\ell) \}$ center and scale the observed covariates in a locally linear predictor. Because they are i) not used to model a quantity of interest, and ii) dissociated from the model for partitions, we replace these parameters with plug-in values that are functions of $\rho$ and $\{ \bxso_j \}$. Perhaps the simplest choice is to use empirical means and standard deviations of $\{x_{i\ell} : i \in \CC_{j\ell} \}$. We elect to use a Bayes plug-in estimator to encourage smoother transitions between distinct partitions, especially across small clusters. This repurposing of parameters from the joint model parallels that of parameters in similarity functions for PPMx models. In similarity functions, probability density functions act as similarity weight kernels and not as models for data. In our sampling model, the plug-in values for $\{(\mu^{(x)}_\ell, \sigma^{(x)}_\ell) \}$ serve to dynamically specify local, cluster-specific covariates. Consequently, the linear predictor in the sampling model relates covariates to the response \textit{only} relative to a given partition $\rho$.

The conditional model with projected (marginalized) and dynamically centered likelihood is given below. For $i = 1,\ldots,m$, $\ell = 1, \ldots, p$, and $j = 1,\ldots,k_m$, we have 
\begin{align}
    \label{eq:fullmodspec}
    \begin{split}
        y_i \mid \bm{\mu}^\star, \bbetas, \bm{\sig}^\star, c_i = j &\simind \mathcal{N}\Big(\mus_j + \sum_{\ell \in \OO_i} \betas_{j\ell} \, z_{i\ell}, \, \sigsqs_j + \sum_{\ell \notin \OO_i} {\betas_{j\ell}}^2 \Big) \, , \\ 
        (\mus_j, \bbetas_j, \sigs_j) \mid \mu_0, \sigma_0 &\simiid \mathcal{N}(\mu_0, \sigma_0^2) \times p(\bbetas_j \mid \sigs_j) \times \mathcal{U}(0, \, a_\sigma) \, , \\ 
        (\mu_0, \sigma_0) &\sim \mathcal{N}(m_0, v^2) \times \mathcal{U}(0, \, a_{\sigma_0}) \, , \\
        p(\rho_m = \{S_1, \ldots, S_{k_m} & \}  \mid \bx^{o},M,\bxi) \propto \prod_{j=1}^{k_m} c(S_j\mid M) \, \tg(\bx^{\star o}_j\mid\bxi),
    \end{split}
\end{align}
where $z_{i\ell} = (x_{i\ell} - \hat{\mu}^{(x)}_{j\ell})/ \hat{\sigma}^{(x)}_{j\ell}$, and
\begin{align}
    \label{eq:plugin_unit}
    \hat{\mu}^{(x)}_{j\ell} = \frac{\nu \mu_{0}^{(x)} + n^o_{j\ell} \bar{x}^o_{j\ell}}{\nu + n^o_{j\ell}}, \quad 
    { \hat{\sigma}^{(x) 2 }_{j\ell} } = \frac{ \nu_s s_0^{(x)2} + \sum_{i \in \CC_{j\ell}} (x_{i\ell} - \bar{x}^o_{j\ell})^2 + \frac{\nu n^o_{j\ell}}{\nu + n^o_{j\ell}} \left( \bar{x}^o_{j\ell} - \mu_{0}^{(x)} \right)^2 }{ \nu_s + n^o_{j\ell}} \, ,
\end{align}
with $n^o_{j\ell}$ denoting the cardinality of $\CC_{j\ell}$ and $\bar{x}^o_{j\ell} = \sum_{i \in \CC_{j\ell}} x_{i\ell} / n^o_{j\ell}$. 
These values correspond to the posterior mean and harmonic mean, for ${\mu}^{(x)}_{j\ell}$ and ${{\sigma}^{(x) 2 }_{j\ell}}$ respectively, under a normal-scaled-inverse-chi-squared prior with prior guesses $\mu_{0}^{(x)}$ and $s_0^{(x)2}$. 
We set $\nu = \nu_s = 1$ so that the plug-in values correspond to Bayes estimates under unit-information priors. The values of $\mu_{0}^{(x)}$ and $s_0^{(x)2}$ need not match the analogous hyperparameters in $\bxi$.  
We use $\mathcal{U}(a,b)$ to denote a uniform distribution on the interval $(a,b)$. 

In what follows we will refer to the model described in \eqref{eq:fullmodspec} and \eqref{eq:plugin_unit} as the variable-dimension local regression model or VDLReg. 
By mimicking the marginalization behavior of a local, jointly Gaussian model, 
the projected sampling model in VDLReg provides a coherent bridge across missingness patterns that borrows strength between them, in the spirit of Figure \ref{fig:simpleMarg}. See Sections \ref{sec:faithful} and \ref{sec:illustration} (of the Supplementary Materials) for illustrations of this behavior.

\subsection{Prior for coefficients}
When considering a prior for the coefficients $\{ \betas_{j\ell} \}$, we first note that setting all coefficients equal to zero recovers the original VDReg model. 
Projection in the sampling model part of \eqref{eq:fullmodspec} inflates variances by the sum of squares of the coefficients indexed by missing covariates. While this aids in both propagating uncertainty due to missingness and maintaining model coherency and parsimony across missingness patterns, care must be taken to appropriately balance uncertainty and precision in model predictions. We therefore apply priors that aggressively shrink $\{ \betas_{j\ell} \}$ toward zero.

We use for $p(\bbetas_j \mid \sigs_j)$ in \eqref{eq:fullmodspec} the Dirichlet--Laplace prior of \citet{bhattacharya2015dirlap}. The prior for $\bbetas_j$ can be expressed as the marginal distribution arising from a global-local scale mixture of Gaussians,
\begin{align}
    \label{eq:DirLap}
    \begin{split}
        \bbetas_{j} \mid \psi_{j\ell}, \phi_{j\ell}, \tau_j, \sigs_j &\simind \mathcal{N}(\bm{0}, \,  {\sigs_j}^2 \, \tau_j^2 \, \bm{D}), \quad \bm{D}_{\ell \ell'} =   \psi_{j\ell} \, \phi_{j\ell}^2 \, 1_{(\ell = \ell')},  \\ 
        \psi_{j\ell} \sim \mathcal{E}(1/2), \quad 
        (\phi_{j1}, \ldots, \phi_{jp}) &\sim \mathcal{D}(p^{-1}, \ldots, p^{-1}) , \quad \tau_j  \sim \mathcal{E}((2\tau_0)^{-1}),
    \end{split}
\end{align}
where $\mathcal{E}(1/2)$ denotes an exponential distribution with mean 2 and $\mathcal{D}$ denotes a Dirichlet distribution. Here, the $\{ \phi_{j\ell} \}$ allocates mass from the global (across $\ell = 1, \ldots, p$) scale parameter $\tau_j$ among the $p$ coefficients with both interdependence and sparsity, while $\sigs_j$ calibrates the prior to the scale of the response. This augmentation resides entirely on the second line of \eqref{eq:fullmodspec}, as all parameters in the set $\{ (\psi_{j1}, \ldots, \psi_{jp}), (\phi_{j1}, \ldots, \phi_{jp}), \tau_j \}$ are indexed by cluster.

\subsection{Prediction and local regression}
\label{sec:mod_pred}

The partition and sampling model components of  \eqref{eq:fullmodspec} yield a predictive model for a new observation $y_{m+1}$ with covariate vector $\bxo_{m+1}$ and observation pattern $\OO_{m+1}$. Given a partition $\rho_m$ of the first $m$ observations, allocation of observation $m+1$ to any of the $k_m$ clusters (or to a new, unoccupied cluster) follows a discrete distribution with probabilities given by
\begin{align}
    \label{eq:predAlloc}
    \begin{split}
        \Pr(c_{m+1} = h \mid \bxo_{m+1}, \OO_{m+1}, \rho_m ) \propto
        \begin{cases}
        \displaystyle
            \frac{ c(S_h \cup \{m+1\}) \, \tg( {\bxso}_h \cup \bxo_{m+1} ) }{ c(S_h ) \, \tg( {\bxso}_h ) } & \text{for } h = 1, \ldots, k_m \, , \\
            c(\{m+1\}) \, \tg( \bxo_{m+1} ) & \text{for } h = k_m + 1\, ,
        \end{cases}
    \end{split}
\end{align}
which collect into the normalized vector $\bm{w} = (w_1, \ldots, w_{k_m + 1})$ with $\bm{1}'\bm{w} = 1$. Formally, these weights are also functions of $\{ \bxo_i, \OO_i : i = 1, \ldots, n \}$ and parameters governing $c(\cdot)$ and $\tg(\cdot)$. Given 
$\{ (\mus_j, \bbetas_j, \sigs_j) \} \in \bths$, we marginalize over cluster allocation $c_{m+1}$ to obtain a mixture formulation for the predictive density
\begin{align}
    \label{eq:predDens}
    p(y_{m+1} \mid \bxo_{m+1}, \OO_{m+1}, \by, \bx, \rho_m, \bths) = \sum_{j = 1}^{k_m + 1} w_j \, \mathcal{N}\Big( y_{m+1} ; \,  \mus_j + \sum_{\ell \in \OO_{m+1}} \betas_{j\ell} z_{(m+1)\ell}, \, \sigsqs_j + \sum_{\ell \notin \OO_{m+1}} {\betas_{j\ell}}^2 \Big)
\end{align}
and corresponding mean surface
\begin{align}
    \label{eq:predMean}
    \mathbb{E}(y_{m+1} \mid \bxo_{m+1}, \OO_{m+1}, \by, \bx, \rho_m, \bths) = \sum_{j = 1}^{k_m + 1} w_j \, \Big( \mus_j + \sum_{\ell \in \OO_{m+1}} \betas_{j\ell} z_{(m+1)\ell} \Big) \, .
\end{align}
Both expressions can be integrated with respect to prior or posterior distributions over model parameters $(\rho, \bths)$ to obtain the usual posterior predictive density, \\ $p(y_{m+1} \mid \bxo_{m+1}, \OO_{m+1}, \by, \bx) = \int p(y_{m+1} \mid \bxo_{m+1}, \OO_{m+1}, \by, \bx, \rho_m, \bths) \, \diff ( p( \rho_m, \bths \mid \by, \bx) )$, and corresponding expectation.

The predictive density in \eqref{eq:predDens} reveals a locally linear regression model in which both weights and conditional means depend on the observed covariate vector $\bxo_{m+1}$. This yields a mean predictive surface in \eqref{eq:predMean} that is generally nonlinear. Furthermore, predictive densities flexibly adapt to any missingness pattern in $\OO_{m+1}$.

Note that the mixture-component means in \eqref{eq:predDens} are \textit{not} defined directly as functions mapping covariate values to the response. The linear predictors in \eqref{eq:predDens} and \eqref{eq:predMean} are defined relative to the partition of observations and values of covariates for grouped observations through the $\{ z_{i\ell} \}$. While the $\{ \mus_j \}$ and $\{ \sigsqs \}$ parameters retain absolute interpretations for their clusters, the coefficient $\{ \bbetas_j \}$ vectors provide only cluster-specific gradients. Thus, $\rho_m$ and $\bm{x}$ are necessary to create a mapping from $x$ to $y$ within the proposed framework.

Predictive inference with samples from the posterior distributions of $\rho_m$ and $\bths$ computationally follow the procedure outlined above. 
At each iteration, an allocation for a new observation, $c_{m+1}$, is drawn from a discrete distribution with probabilities in $\bm{w}$ given by \eqref{eq:predAlloc}. Given $c_{m+1}$, a value for $y_{m+1}$ is drawn from the corresponding mixture component in \eqref{eq:predDens}, with one modification: the $\{ z_{(m+1)\ell} \}$ use values of $\{ \hat{\mu}^{(x)}_{c_{m+1} \ell} \}$ and  $\{ \hat{\sig}^{(x)}_{c_{m+1} \ell} \}$ that are calculated with respect to $\rho_m$ and $\bx$, and do \textit{not} include $\bxo_{m+1}$. 
Predictions with singleton and small clusters will be sensitive to this choice, which departs slightly from the true predictive distribution implied by \eqref{eq:fullmodspec}. 
We acknowledge that this runs counter to calculation of predictive cluster membership weights in \eqref{eq:predAlloc}, which incorporates the new covariates. 
However, we find it appealing to fix the functional form of component kernel means to be defined by covariates from the training sample and \textit{not} modified by the new covariate values. 
When making multiple predictions (e.g., over a grid of predictor values), we customarily treat each point as the  $(m+1)$th observation with no reference to other points at which prediction is sought.

\subsection{Effect of missingness on co-clustering weights}
\label{sec:clustweights}

In the PPMx framework, similarity functions allow covariates to influence the probability distribution over partitions, most commonly through a concentration metric \citep{PPMxMullerQuintanaRosner}. To understand the effect of missing a covariate on clustering probabilities in both the proposed model and that of VDReg, consider the case of two observations, each with two covariates, $x_1$ and $x_2$. Without loss of generality, we fix $(x_1, x_2) = (0,0)$ for the first observation and study the probability that the two observations are clustered together a priori, i.e., $\Pr(\rho_2 = \{1, 2\})$, as a function of $x_1$ and $x_2$ for the second observation, including the possibility that $x_2$ is missing.

\begin{figure}[tb]
    \centering
    \includegraphics[width=6.5in]{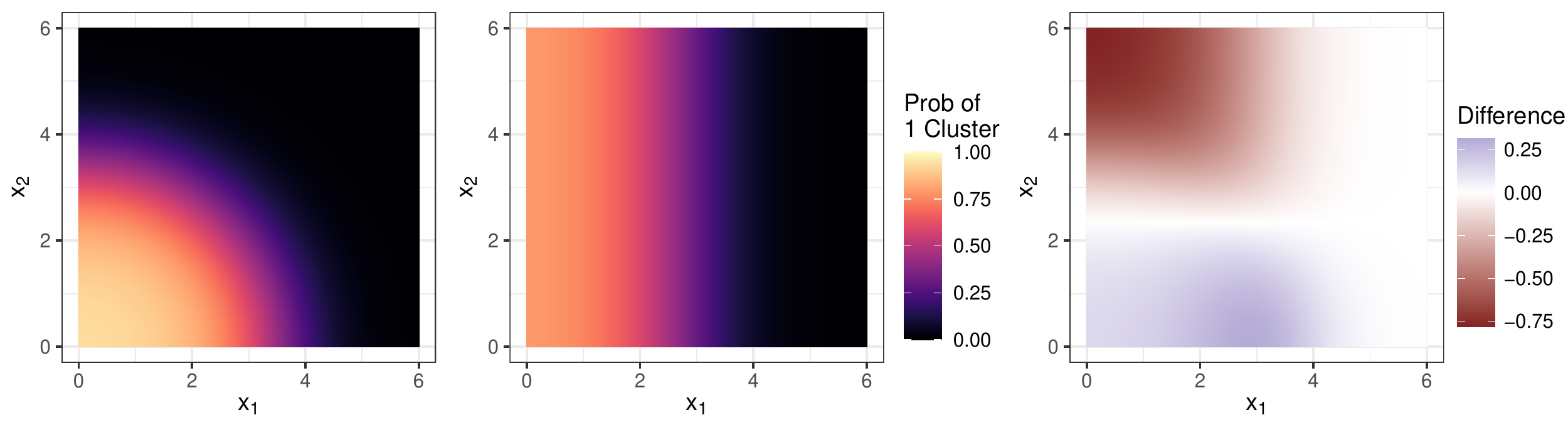} \\
    \includegraphics[width=6.5in]{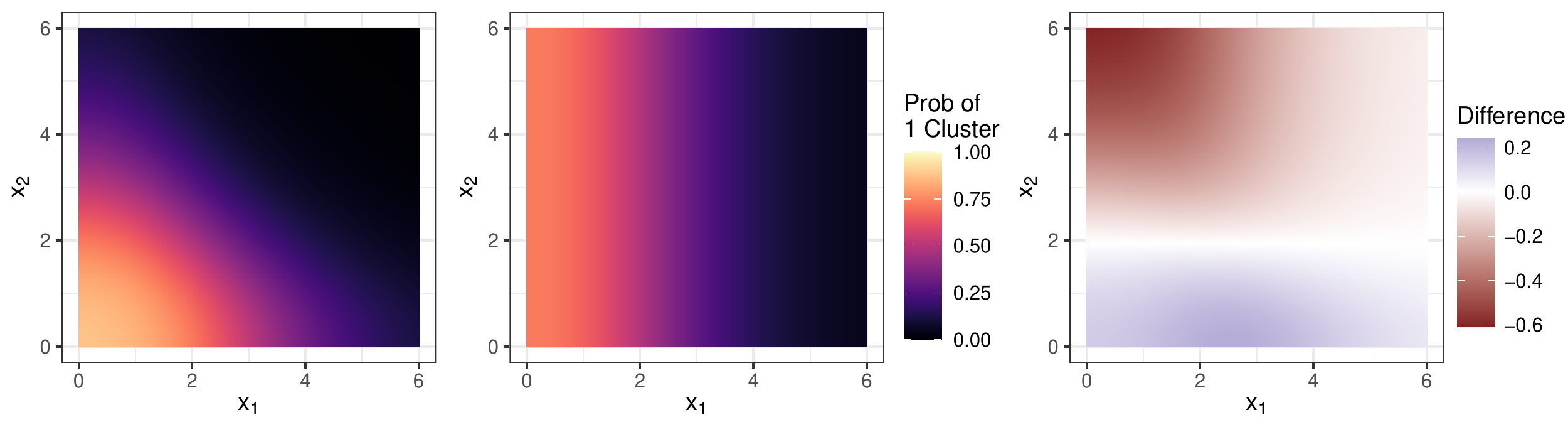}
    \caption{Illustration of the projection property of similarity functions. Prior probability that two observations co-cluster under various values of covariates with the normal-normal (top) and normal-normal-inverse-gamma (bottom) similarity functions. The covariate vector for the first observation is $(0,0)$, while the $x_1$ and $x_2$ axes provide covariate pairs for the second observation. Both covariates are observed in the left panel, $x_2$ is missing in the center panel, and the difference in co-clustering probabilities between these two scenarios is given in the right panel.}
    \label{fig:clustweights}
\end{figure}

Figure \ref{fig:clustweights} depicts the co-clustering probability as a function of $(x_1, x_2)$ for the second observation under two similarity functions and cohesion precision $M = 1$. In the normal-normal (NN) case, we use $\tg(\bxso_j) = \prod_{\ell=1}^2 \int \prod_{i \in \mathcal{C}_{j\ell}} \mathcal{N}(x_{i\ell}; t, 1) \, \mathcal{N}(t; 0, 5^2) \diff t$, and in the normal-normal-inverse-gamma case (NNIG), we use \\ $\tg(\bxso_j) = \prod_{\ell=1}^2 \int \int \prod_{i \in \mathcal{C}_{j\ell}} \mathcal{N}(x_{i\ell}; t, s^2) \, \mathcal{N}(t; 0, 10 s^2) \, \mathcal{IG}(s^2; 2.5, 2.5 ) \diff t \diff s^2$, where $\mathcal{IG}(\cdot; a,b)$ indicates an inverse-gamma density with mean $b/(a-1)$.

Probability contours under the NN and NNIG similarities are qualitatively distinct. 
Designating $x_2$ as missing in the second observation reveals the projection (marginalization) behavior of both similarity functions, which automatically self adjust in a consistent manner to any missing pattern. When both $x_1$ and $x_2$ are missing, the probability that the two observations share one cluster derives entirely from the cohesion function (and equals $0.5$).

\subsection{Guidance for selecting hyperparameters}
\label{sec:hypers}

We recommend centering and scaling all data as a preprocessing step to facilitate interpretation and specification for hyperparameters. Here we build intuition about the roles of hyperpareters and make recommendations for default values.

Consider first the similarity function $\tg(\bx^{\star o}_j \mid \bxi)$, which is usually taken as the product of location and/or scale densities (see \citealp{PPMxMullerQuintanaRosner, ppmxMissing}). 
If all covariates share a common scale, and we lack prior knowledge of the functional response to specific covariates, we make all $\bxi_j = \bxi$ common. 
Owing to its adaptability, we favor using 
\begin{align}
\label{eq:simNNSIX2}
\tg(\bxso_j \mid \bxi) = \prod_{\ell=1}^p \int \int \prod_{i \in \mathcal{C}_{j\ell}} \mathcal{N}(x_{i\ell}; \, t, \, s^2) \, \mathcal{N}(t; \, \tilde{\mu}_0, \, s^2/ \tilde{\kappa}) \, \mathcal{IG}(s^2; \, 0.5 \, \tilde{\nu}, \, 0.5 \, \tilde{\nu} \, \tilde{s}_0^2 ) \diff t \diff s^2 \, , 
\end{align}
which is NNIG parameterized as scaled-inverse-chi-square with $\bxi = (\tilde{\mu}_0, \tilde\kappa, \tilde\nu, \tilde{s}_0^2)$. We refer to this similarity as NNSI$\chi^2$. The location parameter $\tilde{\mu}_0$ centers the ``prior," and we typically take it to be 0 if the covariates are centered. The parameter $\tilde\kappa$ scales the prior variance of $\mu$ with respect to $\sigma^2$, the ``bandwidth" parameter of the auxiliary model for $x_{i\ell}$, and we default to 0.1. The remaining hyperparameters are the most important and influential to similarity, and thus clustering behavior. Scale parameter $\tilde\nu$ acts as a prior effective sample size (against ``likelihood" sample size $\lvert \mathcal{C}_{j\ell} \rvert$; defaults to 4), determining the strength of concentration around $\tilde{s}_0^2$, the prior harmonic-mean bandwidth. In the context of similarity, $\tilde{s}_0^2$ directly influences the width of the marginal density kernels in $\tg(\bxso_j \mid \bxi)$. Using wider kernels encourages smooth/fuzzy transitions between fewer clusters, while narrow kernels lead to local, sharp transitions between more clusters. If covariates are prescaled to unit variance, we recommend defaulting to $\tilde{s}_0^2 \in (0.1^2, 0.5^2)$. The number of clusters also increases as the cohesion concentration parameter increases, and we default at $M=1$.

The other hyperparemeter that strongly influences model behavior is $a_\sigma$, which places a hard upper boundary on cluster-specific error standard deviations. This bound can be used to eliminate local posterior modes that are supported by the model, but are of no practical utility, e.g., a single cluster with flat mean and large error. 
Using small $a_\sigma$, and thereby low noise error, forces model flexibility to compensate and account for complexities in the data, often resulting in more clusters. This strategy can improve point prediction at the risk of overfitting, and is generally not recommended if inferential goals include predictive density estimation. If the response is prescaled to unit variance, values of $a_\sigma$ below 0.2 assume a high signal-to-noise ratio, while values in the 0.3 to 0.6 range can be used to accommodate noisier data.

Finally, Supplemental Section \ref{sec:hypers-appx} explores the role of the global scale parameter $\tau_0$, which influences shrinkage of $\{\bbetas_j\}$. We recommend using values below 0.4, and default to setting $\tau_0 = 0.1$.

Regardless of the strategy used to specify the model, we emphasize the importance of assessing the fit and exploring multiple settings of hyperparameters. As with any nonparametric regression tool, a wide variety of behaviors lies within the scope of these hyperparameters. Different settings impose a variety of assumptions regarding the relationship between covariates and response, and the response distribution.

\subsection{On adding covariates to the likelihood}
\label{sect:indicator}

The PPMx includes an explicit covariate dependence in the prior, which has been shown to be beneficial 
for predictions in various contexts, including in particular, when missing data are present \citep{ppmxMissing}. It was however
surprising and to a certain extent counter-intuitive to find examples where adding them to the likelihood (sampling)
model not only did not improve predictions but in fact produced
worse results. Section \ref{sec:flatbetter} discusses possible causes of this behavior and gives an example using real data. 
We thus found it useful, in the context of data analysis, to explore methods that help detecting whether the addition of covariates to the likelihood would increase
predictive performance. In other words, we need to be able to
detect linear patterns on a local (rather than global) level.

We approach this problem with a simple and easy to implement procedure based on the {\em model based clustering} (MBC) method developed in \cite{fraley-raftery:02}. The procedure first estimates a partition of the data on the response and covariates jointly, then fits cluster-specific linear regressions, and summarizes the resulting statistics (e.g., p-value of the global F test, coefficient of determination) as an indicator of local linearity. 
Details for the procedure appear in Section \ref{sec:flatbetter}, and its use is exemplified in Section \ref{sec:boston}.

\section{Implementation} 
\label{sec:computation}

Here we present an estimation and computation strategy for Markov chain Monte Carlo (MCMC) sampling of the joint posterior distribution. 
We briefly describe more standard updates and focus attention on modified algorithm for updating latent allocation variables. Sampling from predictive distributions was described in Section \ref{sec:mod_pred}. We conclude with notes on run time.

\subsection{MCMC algorithm}
\label{sec:MCMC}

The model outlined in \eqref{eq:fullmodspec} and augmented with \eqref{eq:DirLap} admits a full joint posterior density reported in Section \ref{sec:fulljointpost}. We use the following notation:
$\bm{\eta} = (\mu_0, \sigma_0)$ denotes baseline hyperparameters; 
$\bths = \{ \bths_j \}$ collects all cluster-specific parameters belonging to the sampling model with $\bths_j = \{ {\mu}^\star_j, {\sigma}^\star_j, \bbetas_j, \bm{\psi}_j, \bm{\phi}_j, \tau_j \}$; 
$\bm{\psi}_j = (\psi_{j1}, \ldots, \psi_{jp})$ and $\bm{\phi}_j = (\phi_{j1}, \ldots, \phi_{jp})$ control shrinkage of $\bbetas_j$; 
and 
$c_i \in \{1, \ldots, k_m\}$ indicates latent cluster membership, such that $\{ c_i : i = 1, \ldots, m \}$ fully specifies the partition $\rho_m$. 
We employ a Gibbs sampling scheme that cycles through the following block full conditional distributions: $ [ \bths \mid \rho_m, \bm{\eta}, \by, \bxo]$, $[ \bm{\eta} \mid \rho_m, \bths, \by, \bxo]$, and $[\rho_m \mid \bths, \bm{\eta}, \by, \bxo]$.

\textbf{Update for $\bths$}: 
The updates for cluster-specific means, $\{ \mus_j \}$, are conjugate under rearrangement of the sampling model. Setting ${y}^{c}_i = y_i - \sum_{\ell \in \OO_i} \betas_{j \ell} z_{i\ell}$ for all $i \in S_j$ yields the full conditional $p(\mus_j \mid -) \propto \mathcal{N}(\mus_j \mid \mu_0, \sigma_0^2) \prod_{i \in S_j} \mathcal{N}\big(y^c_i \mid \mus_j,\, {\sigs}^2_j + \sum_{\ell \notin \OO_i} {\betas}^2_{j\ell} \big) $. The uniform prior and observation-specific variance inflation render the updates for $\{ \sigs_j \}$ non-conjugate. We employ a slice sampler \citep{neal2003slice}.

The marginalized sampling model in the first line of \eqref{eq:fullmodspec} precludes conjugate updates for $\{ \bbetas_j \}$, as the coefficients appear in observation means or variances determined by patterns in $\{ \OO_i \}$. Given $\sigs_j$ and the augmentation hyperparameters in $\{ \bm{\psi}_j, \bm{\phi}_j, \tau_j \}$, the prior for $\bbetas_j$ is multivariate Gaussian, facilitating our use of an elliptical Slice sampler \citep{murray2010elliptical}. Given $\bbetas_j$, the updates for $\{ \bm{\psi}_j, \bm{\phi}_j, \tau_j \}$ are blocked Gibbs steps, outlined in Section 2.4 of \citet{bhattacharya2015dirlap}.

\textbf{Update for $\bm{\eta}$}: 
The full conditional for $\mu_0$ is standard conjugate normal, updated using $\{ \mus_j : j = 1, \ldots, k_m \}$. The uniform prior on $\sigma_0$ necessitates a nonstandard update, and we again employ a slice sampler.

\textbf{Update for $\rho_m$}: 
We update the partition by drawing each observation's cluster allocation successively. 
The specific full conditional for the allocation $c_i$ is given as
\begin{align}
    \label{eq:fullcond_alloc}
    \begin{split}
        \Pr(c_i = h \mid - ) \propto
        \begin{cases}
        \displaystyle
            \frac{ c(S_h^{-i} \cup \{i\}) \, \tg( {\bxso}^{-i}_h \cup \bxo_i ) }{ c(S_h^{-i} ) \, \tg( {\bxso}^{-i}_h ) }  \, p(\by \mid c_i = h, - ) & \text{for } h = 1, \ldots, k_m^{-i} \, , \\
            c(\{i\}) \, \tg( \bxo_i ) \, p(\by \mid c_i = h, - ) & \text{for } h = k_m^{-i} + 1\, ,
        \end{cases}
    \end{split}
\end{align}
where all terms with $-i$ are computed excluding observation $i$ and $p(\by \mid c_i = h, - )$ is the product of $m$ univariate normal densities represented in the first line of \eqref{eq:fullmodspec}, evaluated as though $c_i = h$. In the case of proposing a new cluster, a new $\bths_h$ is drawn and used in the sampling density. Expression \eqref{eq:fullcond_alloc} is modified from its equivalent full conditional in \citet{ppmxMissing}, which uses the sampling density of $y_i$ only. This is because the values of all $\bm{z}_{i'} = (z_{i'1}, \ldots, z_{i'p})$, for $i' = 1, \ldots, m$, are affected when all clusters are considered for $c_i$. The full conditional therefore requires two evaluations of the likelihood for {\em every} observation (once with $c_i$ in the cluster and once without).

To reduce the computational burden of updating $c_i$, we instead implement the following modification to Algorithm 7 of \citet{neal:2000} that utilizes a Metropolis-Hastings move with proposals that importantly {\em exclude} the sampling density. The algorithm first randomly selects (with user-specified probability that we take as 0.5) among two proposal types: i) a singleton-to-group or group-to-singleton move, and ii) a move between two currently occupied clusters.

If the first proposal type is selected and observation $i$ is currently a singleton, the proposal is to join a different existing cluster. The proposal distribution utilizes prior full-conditional weights. Let $\mathcal{H} = \{1, \ldots, k_m\} \setminus c_i$ denote the indices of alternate clusters with at least one member. Then the distribution for the proposed value, $c^*$, is given as
\begin{align}
    \label{eq:alg7_1a}
        \Pr(c^* = h ) &\propto
        \displaystyle
            \frac{ c(S_h \cup \{i\}) \, \tg( {\bxso}_h \cup \bxo_i ) }{ c(S_h ) \, \tg( {\bxso}_h ) }  \quad \text{for } h \in \mathcal{H} \,.
\end{align}
Let $\varsigma_{\text{a}}^*$ denote the sum over all $h \in \mathcal{H}$ of the terms in \eqref{eq:alg7_1a}. 
The probability of accepting $c^*$ is given as $\min(1, A^*_{1\text{a}})$, where
\begin{align}
    \label{eq:accpt_alg7_1a}
    A^*_{1\text{a}} = \frac{ \varsigma_{\text{a}}^* \, \prod_{i' \in S_{c^*} \cup \{i\}} p(y_{i'} \mid c_{i'} = c^*) }{ c(\{i\}) \, \tg( \bxo_i ) \, p(y_i \mid c_i) \, \prod_{i' \in S_{c^*}} p(y_{i'} \mid c_{i'} = c^*) } \, ,
\end{align}
and the $p(y_{i'} \mid c_{i'})$ are the univariate densities represented in the first line of \eqref{eq:fullmodspec}. Note that these density values do not cancel in \eqref{eq:accpt_alg7_1a} because those in the numerator are calculated with observation $i$ in the cluster while the same densities in the denominator exclude observation $i$ from the cluster, resulting in distinct $\{z_{i'\ell}\}$.

If the first proposal type is selected and observation $i$ currently belongs to a cluster with at least one other observation, then a singleton for $i$ is proposed, together with a new $\bths_{k_m+1}$, with probability 1. The proposal $c^* = k_m+1$ is accepted with probability $\min(1, A^*_{1\text{b}})$, where
\begin{align}
    \label{eq:accpt_alg7_1b}
    A^*_{1\text{b}} = \frac{ c(\{i\}) \, \tg( \bxo_i ) \, p(y_i \mid c^* ) \, \prod_{i' \in S_{c_i} \setminus \{i\} } p(y_{i'} \mid c_{i'} = c_i) }{ \varsigma_{\text{b}}^* \, \prod_{i' \in S_{c_i} \cup \{i\}} p(y_{i'} \mid c_{i'} = c_i) } \, ,
\end{align}
and
\begin{align}
    \label{eq:alg7_1b_wgts}
        \varsigma_{\text{b}}^* & = \sum_{h = 1}^{k_m}
        \displaystyle
            \frac{ c(S_h^{-i} \cup \{i\}) \, \tg( {\bxso}_h^{-i} \cup \bxo_i ) }{ c(S_h^{-i} ) \, \tg( {\bxso}_h^{-i} ) } \,,
\end{align}
and all terms with $-i$ are computed excluding observation $i$. Again, the densities in the numerator and denominator of \eqref{eq:accpt_alg7_1b} are distinct because they depend on the complete membership of the cluster through $\{z_{i'\ell}\}$.

If the second proposal type is selected and observation $i$ is a singleton, do nothing. Otherwise, let $\omega_h$ denote the $h$th summand in \eqref{eq:alg7_1b_wgts}, for $h = 1, \ldots, k_m$, and propose $c^*$ from a discrete distribution over $\{ 1, \ldots, k_m \} \setminus c_i $ with corresponding probabilities $\{ \omega_h / \sum_{h' \ne c_i } \omega_{h'} \}$. The proposal is accepted with probability $\min(1, A^*_2)$, where
\begin{align}
    \label{eq:accpt_alg7_2}
    A^*_{2} = \frac{ (\sum_{h \ne c_i} \omega_h) \, (\prod_{i' \in S_{c^*} \cup \{i\} } p(y_{i'} \mid c_{i'} = c^*) ) \prod_{i' \in S_{c_i}^{-i} } p(y_{i'} \mid c_{i'} = c_i)   }{ (\sum_{h \ne c^*} \omega_h) \, (\prod_{i' \in S_{c^*}^{-i} } p(y_{i'} \mid c_{i'} = c^*) ) \prod_{i' \in S_{c_i} \cup \{i\}  } p(y_{i'} \mid c_{i'} = c_i)  } \, .
\end{align}

If the proposal is accepted, $c_i$ is set to $c^*$, otherwise it remains unchanged. Before updating individual allocations, the sampler randomly permutes the order in which the elements in $\{ c_i : i=1,\ldots,m \}$ will be updated in the current iteration of the encompassing Gibbs sampler.

Note that calculation of $A^*_{1\text{a}}$, $A^*_{1\text{b}}$, and $A^*_2$ importantly requires evaluating densities for all observations in only two clusters instead of all $k_m$. The Metropolized proposal distributions encourage movement among indicators \citep{liu1996gibbs} while avoiding evaluation of sampling densities. We examine the effect of this choice on computational complexity in Section \ref{sec:complexity}.

\subsection{Benchmarking run time}
\label{sec:benchmark}

Table \ref{tab:runtime} reports benchmark timing for MCMC on simulated data of varying sizes with 20\% missing values.  
Step data were generated using four clusters each with a constant mean response, while cluster-specific means in the linear data include covariates. Model indicates the mean specification for the sampling model. 
VDLReg runs were fit with the {\tt ProductPartitionModels} package (version 0.8.2; \citealp{ProductPartitionModels:package}) in \textit{Julia} (version 1.8.3; \citealp{Julia-2017}). VDReg runs fit with {\tt ProductPartitionModels} used a simplified version of the proposed $\rho$ update. 
We further include run times on VDReg using the {\tt ppmSuite} package (version 0.2.4; \citealp{ppmSuite:2022}) in \textit{R} \citep{Rlanguage}, called through \textit{Julia} using the {\tt RCall} package \citep{RCall:package}. 
Reported times are medians of 10 successive samples of 1,000 iterations for the {\tt ProductPartitionModels} implementation, and 10 independent samples for the {\tt ppmSuite} implementation (each starting from iteration 1 and resulting in longer times). 
All benchmarking was performed using the {\tt BenchmarkTools} package \citep{BenchmarkTools.jl-2016} on a 2021 MacBook Pro laptop with Apple M1 Pro chip. 

\begin{table}[tb]
    \centering
    \begin{tabular}{l l l l r r }
    \toprule
    & & & & \multicolumn{2}{c}{$p$} \\
    \cmidrule{5-6}
     Data & $m$  & Package & Model & 5 & 10 \\
        \midrule
Step & 100 & ProductPartitionModels & VDLReg & 7.35 $\pm$ 0.42 & 9.54 $\pm$ 0.12 \\
 & & ProductPartitionModels & VDReg & 4.93 $\pm$ 0.45 & 5.82 $\pm$ 0.31 \\
 & & ppmSuite & VDReg & 1.05 $\pm$ 0.02  &  1.47 $\pm$ 0.19 \\
\midrule
Step & 300 & ProductPartitionModels & VDLReg & 36.10 $\pm$ 0.68 & 47.38 $\pm$ 1.51 \\
 & & ProductPartitionModels & VDReg & 12.13 $\pm$ 1.30 & 20.11 $\pm$ 2.16  \\
 & & ppmSuite & VDReg & 2.73 $\pm$ 0.05 & 5.61 $\pm$ 0.32 \\
 \midrule
Linear & 100 & ProductPartitionModels & VDLReg & 7.48 $\pm$ 1.70 & 10.23 $\pm$ 0.38 \\
 & & ProductPartitionModels & VDReg & 6.00 $\pm$ 0.38 & 8.65 $\pm$ 0.36 \\
 & & ppmSuite & VDReg & 1.26 $\pm$ 0.05 & 1.99 $\pm$ 0.05 \\
\midrule
Linear & 300 & ProductPartitionModels & VDLReg & 35.80 $\pm$ 1.12 & 47.11 $\pm$ 2.52 \\
 & & ProductPartitionModels & VDReg & 23.36 $\pm$ 2.75 &  25.87 $\pm$ 4.71 \\
 & & ppmSuite & VDReg & 5.25 $\pm$ 0.30 & 6.71 $\pm$ 0.38 \\
 \bottomrule 
    \end{tabular}
    \caption{Median run time, in seconds, for 1,000 iterations of MCMC at various model specifications, data type, sample sizes ($m$), and number of covariates ($p$). Plus-minus values report one standard deviation.}
    \label{tab:runtime}
\end{table}

It is clear from Table \ref{tab:runtime} that the dynamically specified likelihood (sampling) model in VDLReg adds substantial computational overhead that at times will not be justified by predictive performance gains like those demonstrated in Sections \ref{sec:simulations} and \ref{sec:applications}. 
We intend for Table \ref{tab:runtime} to serve as a quick reference for data scenarios we consider appropriate for use of VDLReg and not as an comprehensive treatment of algorithm performance. 
MCMC sampling speed for these models will also vary with data complexity, which manifests in the PPMx framework as the number of clusters. 

\section{Simulation Study}
\label{sec:simulations}

We conduct a simulation study to illustrate VDLReg's performance in handling incomplete covariate vectors.  The simulation is based on creating synthetic datasets with 300 training and 300 testing observations. The procedure for data generation mimics 
that of 
the second simulation in \cite{ppmxMissing} which is based on the procedure of \cite{friedman:1991}.  We provide a few details for sake of completeness. 

Fix $p=10$ and create covariate values using $x_{i1}, \ldots, x_{ip} \simiid \mathcal{U}(0,1)$.  Response values $y_i$  are generated using $y_i = f(\bm{x}_i) + \epsilon_i$, where
\begin{align*}
f(\bm{x}_i) = 10\sin(\pi x_{i1}x_{i2}) + 20(x_{i3} - 0.5)^2 + 10x_{i4} + 5x_{i5}.
\end{align*}
Notice that $x_{i6}, \ldots, x_{i10}$ are noise covariates as they do not contribute to the response value. We consider two different $ \epsilon_i$ terms.  The first is the \textit{iid} case with $\epsilon_i \simiid \mathcal{N}(0, 1)$. The second error terms depend on $\bm{x}_i$ such that $\epsilon_i \simind \mathcal{N}(0, \exp(x_{i1}))$.  As an aside, it is worth noting that the generated data sets do not originate from clusters that are explicitly covariate informed as defined in the PPMx.

We consider four levels of missing rates (0\%, 10\%, 25\%, 50\%).  Missing values in the covariates are inserted under the missing at random (MAR) and the missing not at random (MNAR) paradigms. 
Generating both types of missing is facilitated using the {\tt ampute} function found in the {\tt mice} R-package (\citealt{micePack}). For MNAR, the {\tt ampute} function  is used for each covariate with the missing probabilities being a function of the covariate value (see \citealt{schouten&lugtig&vink} for specific details regarding the function used to produce probability of missing). The {\tt ampute} function is also applied separately to each covariate for the MAR case where each covariate entry is equally likely to be classified as missing.

In summary, we generate 100 data sets under simulation truths that vary the following factors: (A) {\em type of missing} (MAR or MNAR), (B) {\em missing fraction} (0\%, 10\%, 25\%, 50\%), and
(C) {\em heteroscedasticity} (yes, no). For all fits in the simulation study, the data were \emph{not} centered and scaled.

\subsection{Comparison metrics}

We compare methods with multiple metrics, noting that using incomplete covariates can lead to conditional heteroscedasticity and, in some cases, introduce multimodality to a regression response distribution. The mixture structure of the PPMx-based predictive distributions gives these models a distinct advantage over competitors in this regard, which we highlight in this simulation study and in Section \ref{sec:applications}. The common default of mean squared error can be difficult to interpret, or entirely inappropriate, in the the presence of multimodal predictive distributions.

To describe the metrics that we employ, let $\{ (y_{i}, \bx_{i}) : i = 1, \ldots, m'\}$ denote the $m' = 300$ out-of-sampling testing observations in each synthetic dataset. We used the following metrics to compare methods. 

\noindent{\textbf{Mean squared prediction error (MSPE)}}: Let $\hat{y}_i$ denote the model-based point prediction for $y_i$. Bayesian implementations use the posterior mean of the predictive distribution for $\hat{y}_i$. MSPE is calculated as ${m'}^{-1}\sum_{i=1}^{m'} (y_i - \hat{y}_i)^2$.

\noindent{\textbf{Predictive deviance}}: Let $\mathcal{L}(y_i)$ denote the log-likelihood evaluation (for any model) at $y_i$, and let $\hat{\mathcal{L}}(y_i)$ denote a Monte Carlo estimate of the  posterior mean of the log-likelihood. In the VDLReg model, we use as $\mathcal{L}$ the log of the mixture of normal densities given in \eqref{eq:predDens}. The same calculation for a VDReg model is equivalent to \eqref{eq:predDens} with all $\betas_{j\ell}$ fixed at 0. For each method, the deviance is then $-2{m'}^{-1}\sum_{i=1}^{m'} \hat{\mathcal{L}}(y_i)$.

Instead of calculating log-likelihood using point estimates of parameters, which requires two steps in PPMx models, 
we derive point estimates of log-likelihood from their posterior distributions. Posterior mean likelihood is incomparable to a likelihood evaluation using point estimates of parameters due Jensen's inequality. We therefore use predictive deviance as a comparison metric only among Bayesian methods.

\noindent{\textbf{Goodness of fit (GoF)}}: Methods that yield a (Monte Carlo estimate of the) posterior predictive distribution for $y_i$ further admit a quantile residual, $q_i \in (0,1)$, taken as the quantile corresponding with $y_i$, i.e., $q_i = \Pr(Y \le y_i)$ with respect to the predictive distribution \citep{dunnSmyth1996qresid}. A successfully fitting model would then produce a set $\{ q_i \}$ that are uniformly distributed. We use as our goodness-of-fit metric the Kolmogorov-Smirnov test statistic for uniformity of $\{ q_i \}$.

\subsection{Methods}

The methods listed below are included in our simulation study and data analyses to provide context to VDLReg's performance.  Throughout this and Section \ref{sec:applications}, we use the default settings of the available software employed to fit the methods listed here.

\noindent{\textbf{BARTm}}: This method using regression trees was extended by \citet{kapelner2015BARTm} to admit missing covariates in splitting decisions, and is implemented in the {\tt bartMachine} package \citep{bartMachinePackage} in \texttt{R} \citep{Rlanguage}. Posterior predictive samples of $y_i$ and posterior samples of error variance are extracted to calculate the deviance and goodness-of-fit.

\noindent{\textbf{Random Forest}}: This approach first uses the {\tt missForest} package to impute missing covariate values using random forests \citep{Stekhoven2012, missForestPackage}. We then fit a random forest to the imputed data \citep{randomForestPackage}.

\noindent{\textbf{PSM}}: Pattern submodels are fit using the method of \citet{Mercaldo2020patternSub} and implemented with code accompanying the article that is available on GitHub.

\noindent{\textbf{MI}}: The \texttt{mi} package \citep{miPackage} is used to generate 10 $\bx$ matrices  with multiple imputation. Then a Bayesian linear model is fit to each imputed set using the {\tt rstanarm} package that calls STAN in the background \citep{rstanarmPackage}. This is done to extract posterior predictive samples of $y_i$ and posterior samples of error variance to calculate deviance/GoF.

\noindent{\textbf{VDReg, VDLReg}}: Model settings for both VDReg and VDLReg are as follows. We used cohesion precision $M=1$, NNSI$\chi^2$ similarity with $\bxi 
= (\tilde{\mu}_0, \tilde\kappa, \tilde\nu, \tilde{s}_0^2)
=  (0.5,\ 0.1,\ 10,\ 0.2^2)$, 
$m_0 = \bar{y}$, 
$v = 2 \, \text{stdev}(\by)$, 
$a_{\sigma_0} = 5 \, \text{stdev}(\by)$, 
$\tau_0 = 0.1$, and $a_\sigma = 2$. 
While preferred settings vary between VDLReg and VDReg, as well as among missing covariate rates, we selected one common setting among a few alternatives that yields generally strong performance throughout. 
We ran MCMC for 100,000 iterations, with 50,000 burn-in, and thinned to every 50th sample.

\subsection{Simulation Results}

\begin{figure}[hb!]
\begin{center}
\includegraphics[scale=0.7]{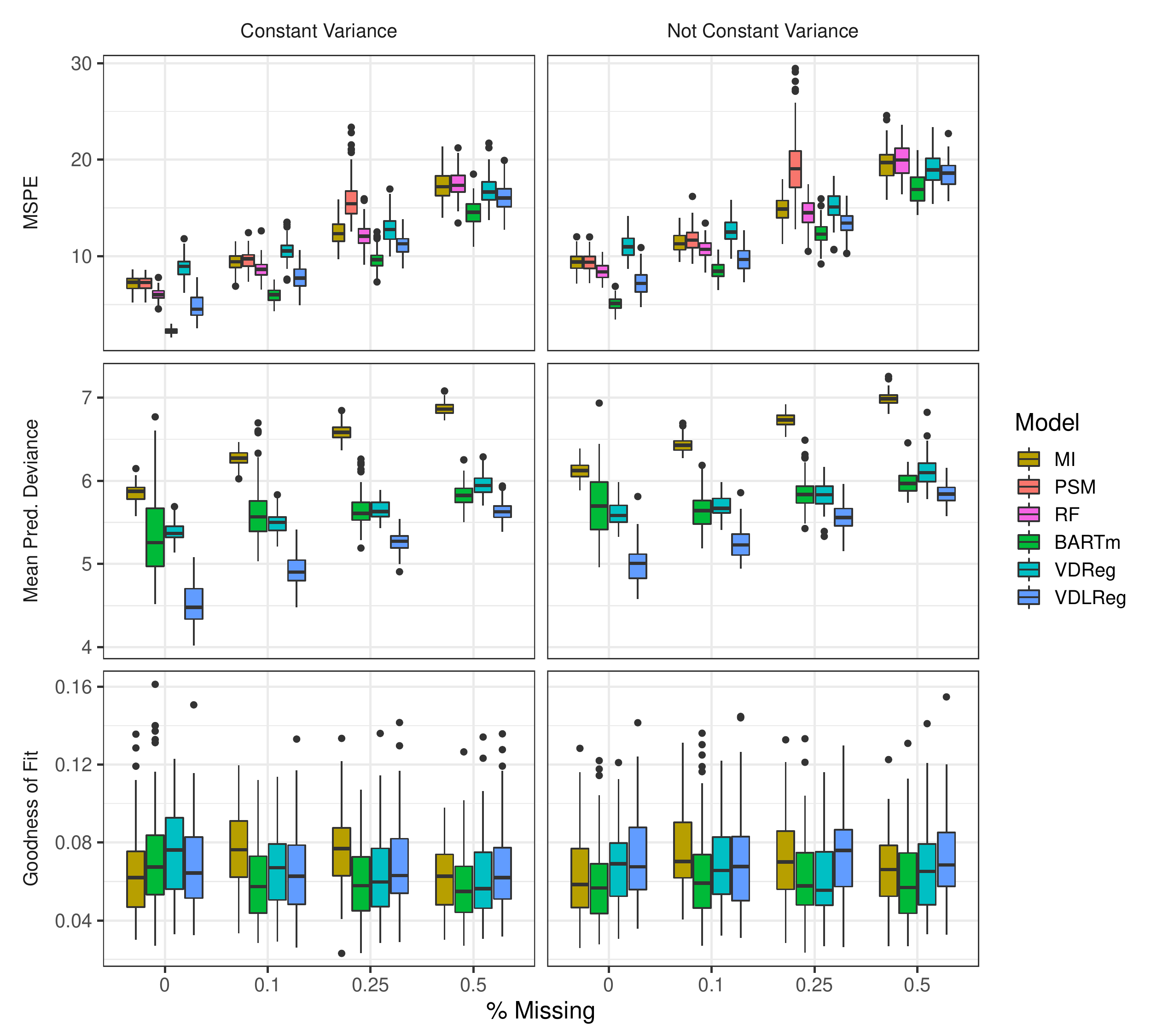}
\caption{Out-of-sample results from the Friedman simulation with MAR data. Lower values indicate better fit with all three metrics.}
\label{fig:friedman_mar}
\end{center}
\end{figure}

We present the MSPE, predictive deviance, and GoF results here for the MAR case. The MNAR results yield similar patterns (see Section \ref{sec:sim-appx}), suggesting a degree of robsustness to missingness assumptions for all methods, with the Friedman data. In-sample metrics are also reported in Section \ref{sec:sim-appx}. Figure \ref{fig:friedman_mar} provides side-by-side box plots for each of the methods, summarizing each metric across replicate data sets. We first note that when 50\% of covariate values are missing, PSM's performance suffers greatly due to the large number of missingness patterns.

With regard to MSPE, it appears that BARTm performs the best, with its advantage over VDLReg diminishing as the missing rate increases. VDLReg can outperform BARTm on larger samples ($m=1000$, not shown) with moderate missing rates. 
PSM degrades most severly as the missing rate increases.  

With regard to predictive deviance, VDLReg performs best, followed by BARTm and VDReg, which perform similarly. The reason for the apparent discrepancy is that MSPE corresponds to a point prediction while the deviance evaluates the predictive density, which can adjust in nonstandard ways when one or more relevant dimensions are missing. 
This may be apparent with more data; in the $m=1000$ case (not shown), BARTm has superior predictive deviance when no covariates are missing, but thereafter usually lags behind VDLReg at nonzero missing rates. All methods performed similarly in GoF; we considered K-S statistics below 0.1 generally acceptable.

\section{Applications}
\label{sec:applications}

We illustrate use of VDLReg and compare methods with two data sets that are popular in the regression literature, 
highlighting different aspects of modeling with VDLReg in each analysis.

\subsection{Boston housing}
\label{sec:boston}

The Boston housing data set aggregates several housing, demographic, social, and environmental variables, compiled from various sources in 1970, to 506 census tracts in Boston \citep{harrison1978}. The data are available through the \texttt{MASS} package in \texttt{R} \citep{ MASSpackage}. We follow \citet{kapelner2015BARTm} and use $p=8$ continuous variables to predict the median value owner-occupied homes (details are given in Supplemental Section \ref{sec:boston-appx}). All variables, including median home value, were centered and scaled using all 506 observations. The original data contain no missing values, which we imposed at varying rates for comparison.

Each of VDLReg, VDReg, BARTm, random forest (RF), PSM, and MI was fit to 100 replicate training sets of size $m=400$. This was repeated on data sets with 10\%, 25\%, and 50\% of covariate values missing completely at random. Out-of-sample test metrics were then computed for each replicate test set of 106 observations, which also contained missing covariates at the selected rates. VDReg and VDLReg used the following settings, which generally perform best among a few alternatives tested with initial runs. We used $M=3$ and NNSI$\chi^2$ similarity with $\tilde{s}_0^2=0.1^2$ to encourage more clusters and higher flexibility, and $a_\sigma=0.4$ to allow a fairly low signal-to-noise ratio. 

\begin{figure}
    \centering
    \includegraphics[scale=0.62, trim={0 0 0 20}, clip]{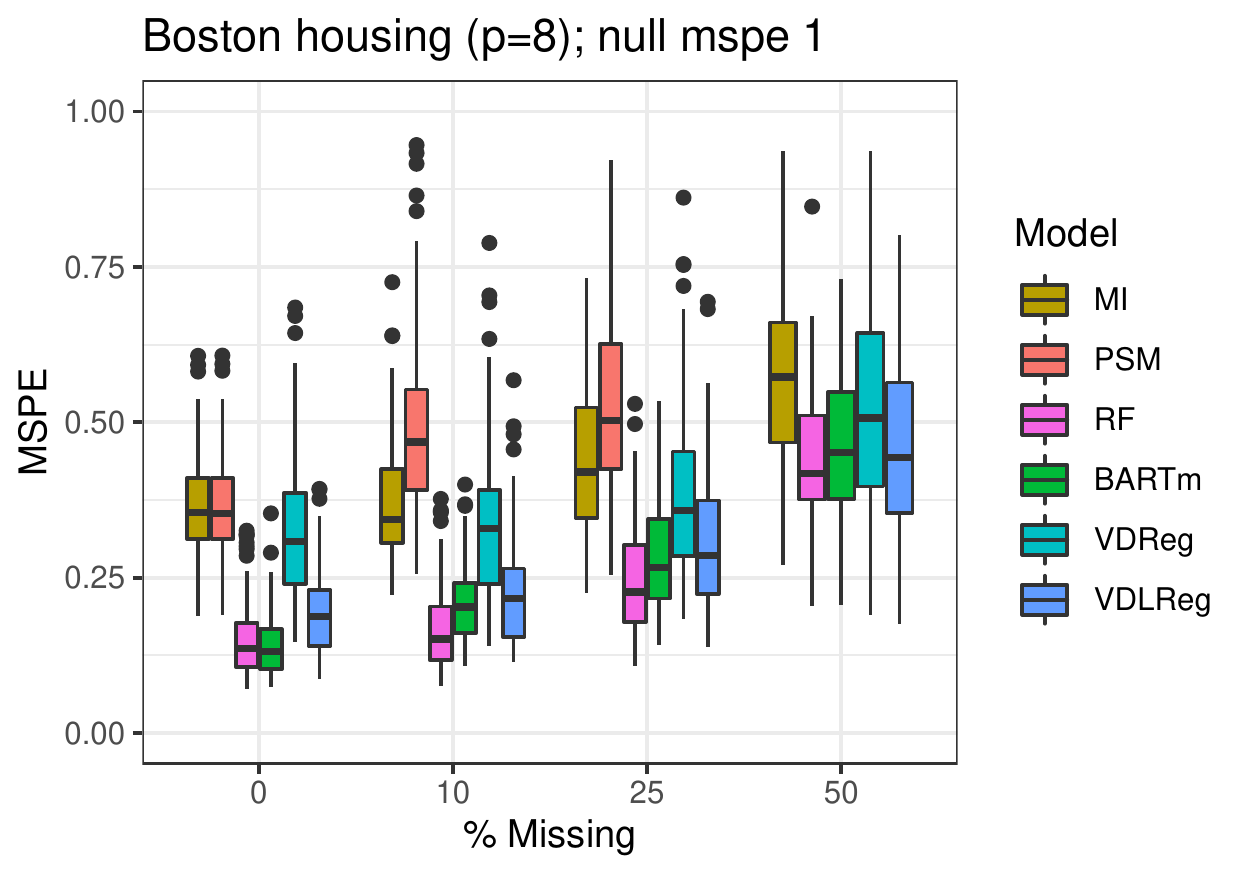}     \includegraphics[scale=0.62, trim={0 0 0 20}, clip]{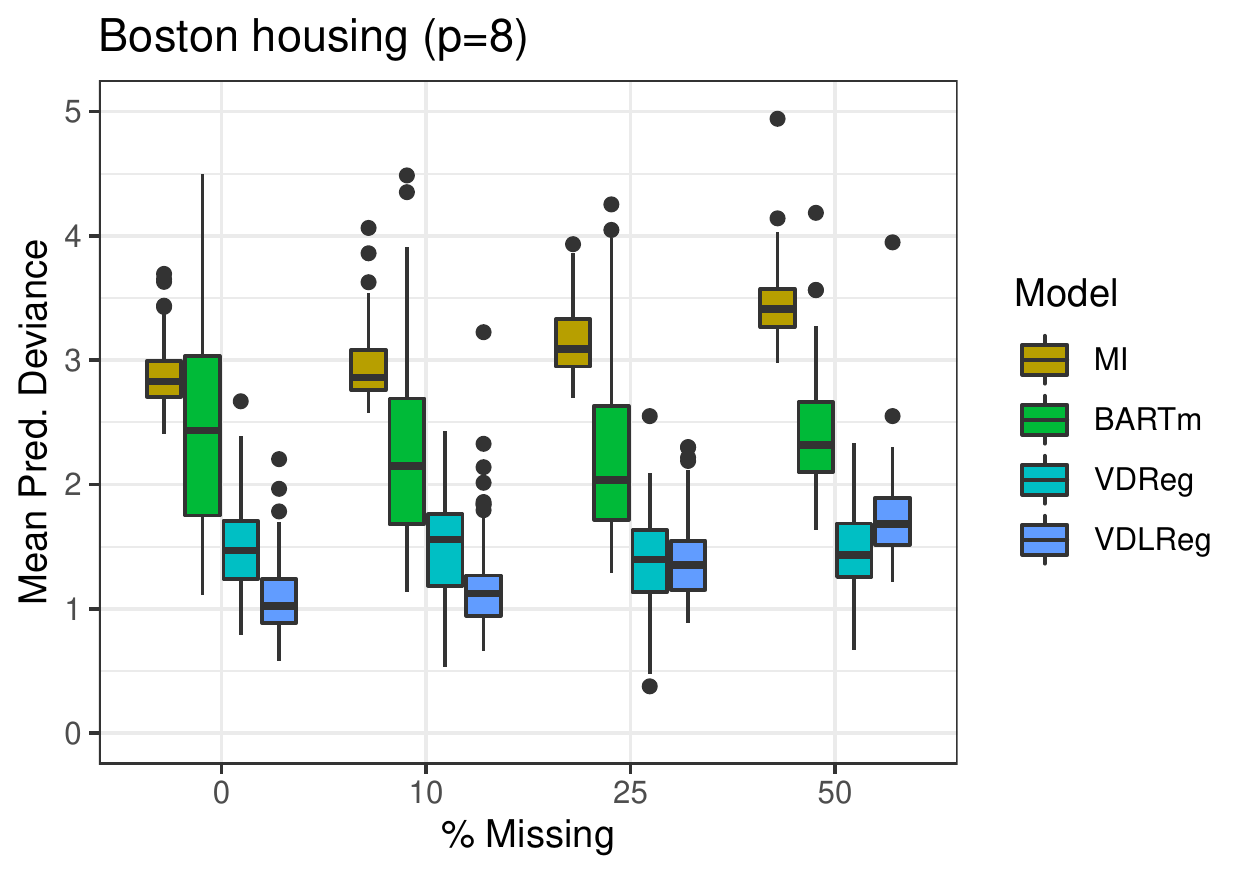}
    \caption{Box plots summarizing mean squared prediction error (left) and mean predictive deviance (right) for model fits to replicated training/test pairs of the Boston housing data under four missing covariate rates.}
    \label{fig:bostonMSPE}
\end{figure}

Figure \ref{fig:bostonMSPE} summarizes MSPE and mean predictive deviance for all model fits. Note that the PSM approach failed for several runs at higher levels of missingness, including all runs at 50\%. From the perspective of point prediction, RF and BARTm are superior on more complete data, while VDLReg is more competitive at higher missing rates. VDLReg always outperforms VDReg in MSPE, suggesting either nonlinear associations between median home values and the covariates, or more likely, subsets of similar census tracts exhibiting (roughly) linear associations. The linearity test from Section \ref{sect:indicator} corroborates these results with small p-values (weighted avearage 0.0001) and large R-squared values (weighted average 0.81) across nine clusters. VDLReg consistently fits approximately three fewer clusters than VDReg.

PPMx-based models yield the best predictive density performance, with VDLReg beating VDReg at lower missing rates (0\% and 10\%). K-S statistics on out-of-sample quantile residuals (shown in Section \ref{sec:boston-appx}) indicate similar performance between VDReg and VDLReg. BARTm and MI predictive distributions are occasionally overdispersed at higher missing rates.

\subsection{Old Faithful}
\label{sec:faithful}

We next demonstrate the projection and density regression capabilities of VDLReg by predicting the distribution of waiting times between successive eruptions of the Old Faithful geyser in Yellowstone National Park, USA. We use data from \citet{azzalini1990} that track 299 eruptions between August 1 and 15, 1985; the data are also available in the \texttt{MASS} package. The response $(y)$ is waiting time in minutes, and we use two covariates: the duration of the previous eruption (\texttt{d1}), and the waiting time to the previous eruption (\texttt{w1}). We again centered and scaled all variables to facilitate model interpretation and fitting, but report in original units (always minutes). For purposes of illustration, we eliminated durations rounded to whole numbers (about 25\%), most of which represent qualitative nocturnal measurements and treated them as missing values. 

Each of the competing models was fit to 100 replicate training sets of size $m=200$. Out-of-sample test metrics were then computed for each replicate test set of 97 observations (two lags of waiting time were also considered, reducing the sample size; see Section \ref{sec:OldFaithful-appx}). 
VDReg and VDLReg used the following settings, which generally perform best among a few alternatives tested with initial runs. We used $M=1$ and NNSI$\chi^2$ similarity with $\tilde{s}_0^2=0.5^2$ to admit fewer, larger clusters; and $a_\sigma=0.5$ to accommodate substantial noise. 
Performance in point prediction is very similar among VDLReg, VDReg, PSM, and BARTm (shown in Section \ref{sec:OldFaithful-appx}). The PPMx-based models enjoy a clear advantage in predictive deviance.

\begin{figure}[p]
    \centering
    \includegraphics[scale=0.70, trim={0 30 0 0}, clip]{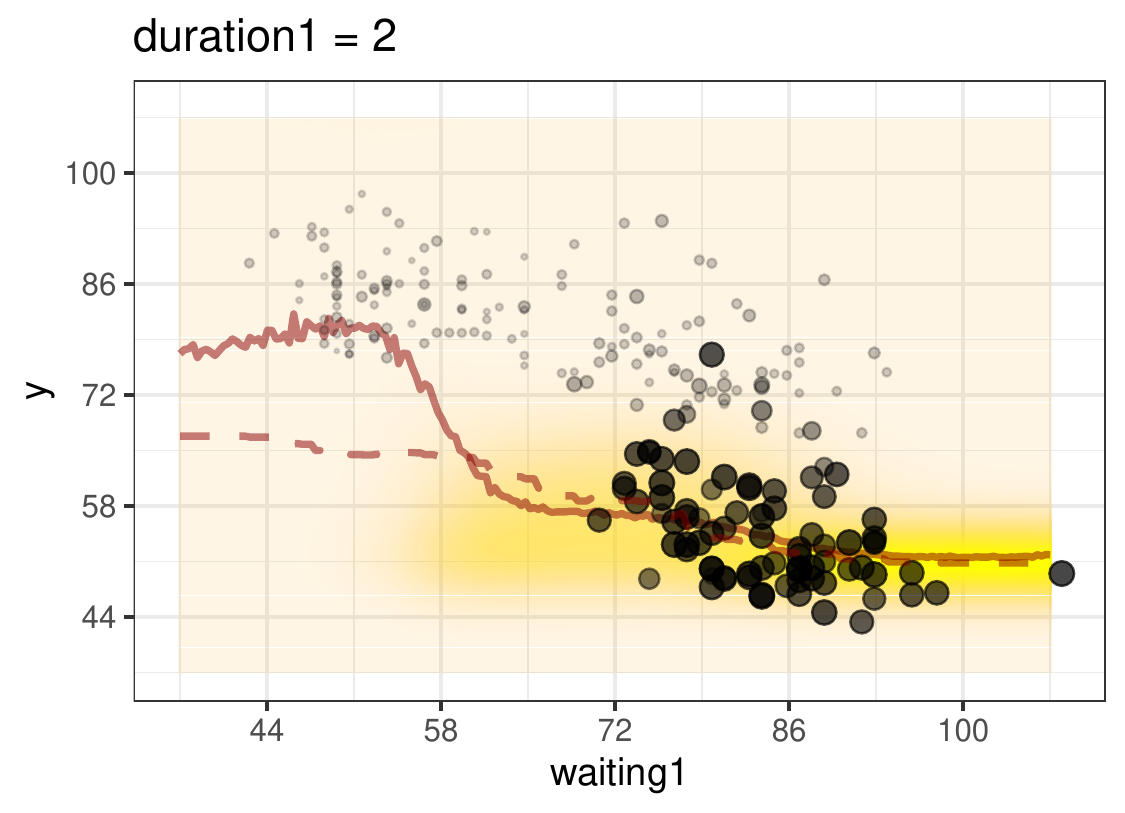}     \includegraphics[scale=0.70, trim={35 30 0 0}, clip]{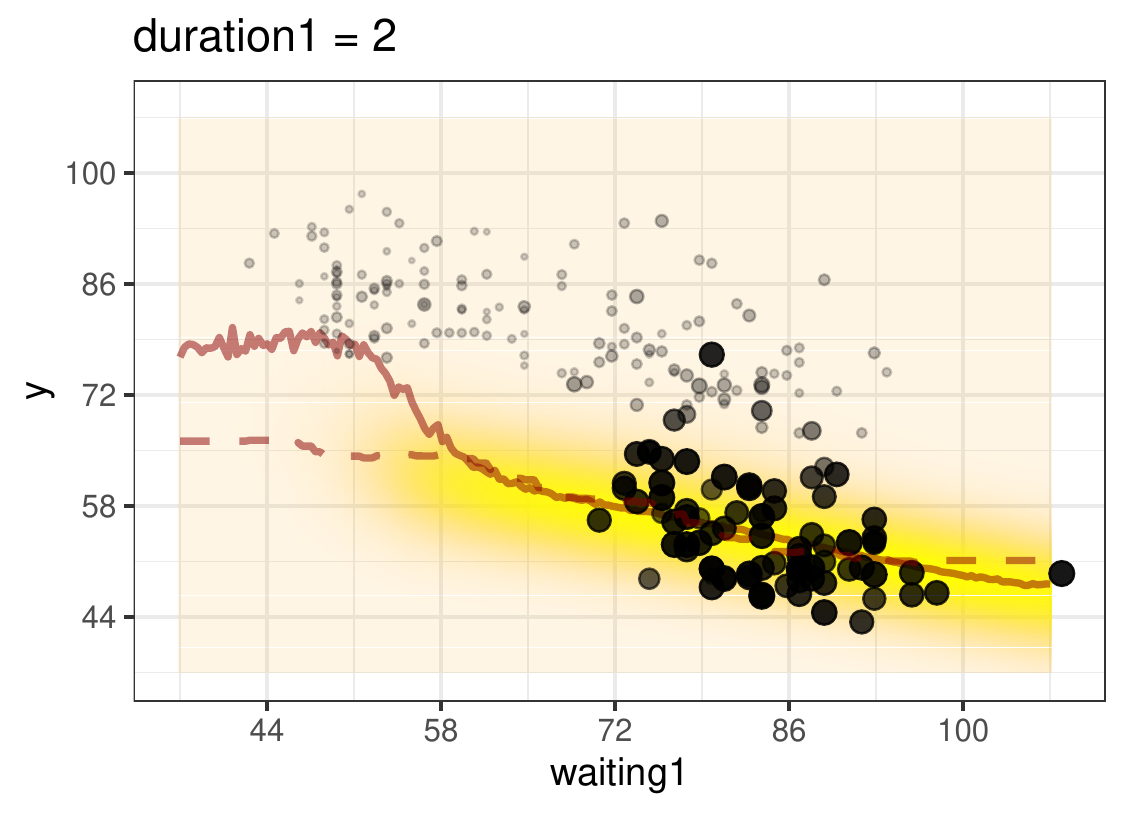}

    \includegraphics[scale=0.70, trim={0 30 0 0}, clip]{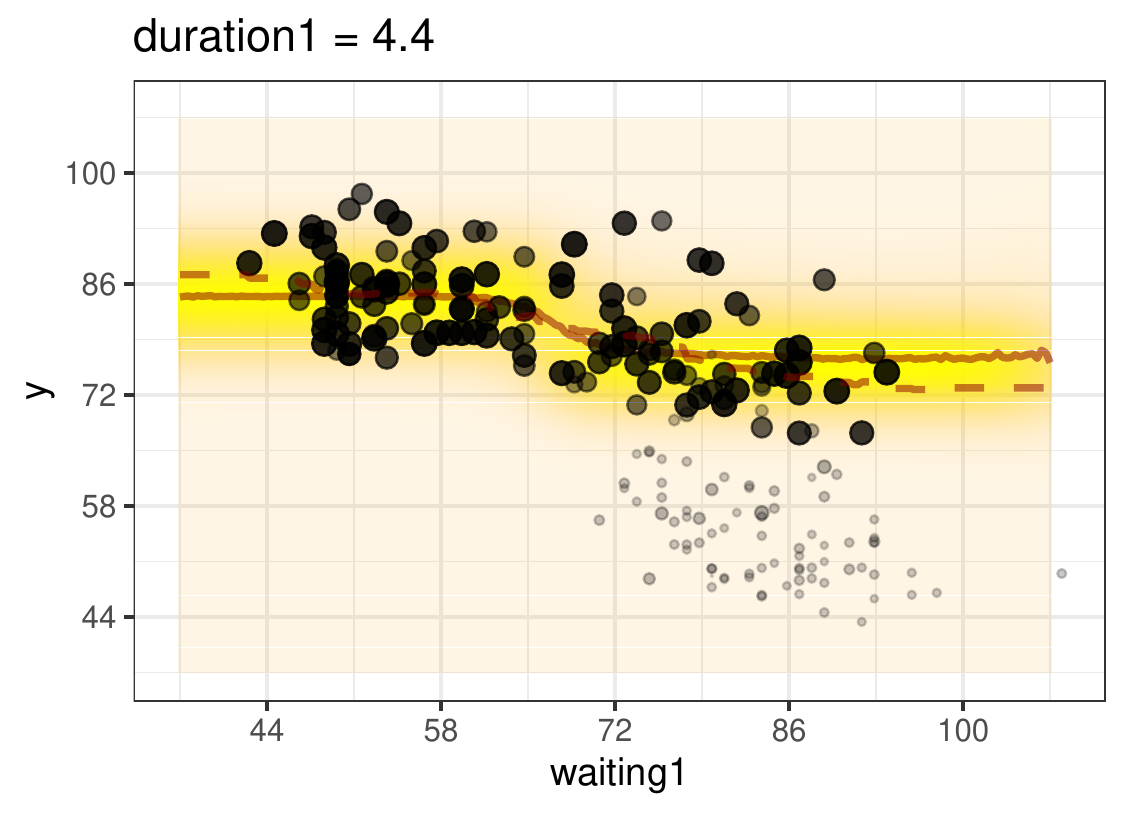}     \includegraphics[scale=0.70, trim={35 30 0 0}, clip]{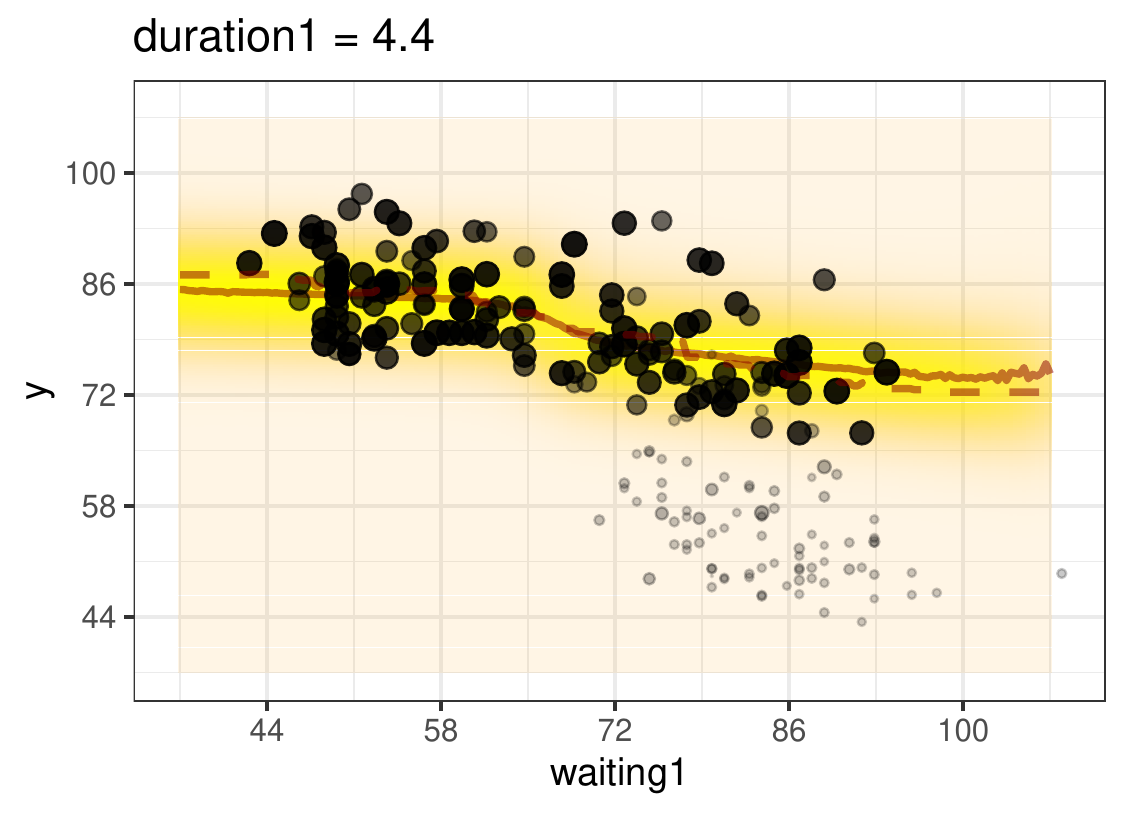}

    \includegraphics[scale=0.70, trim={0 0 0 0}, clip]{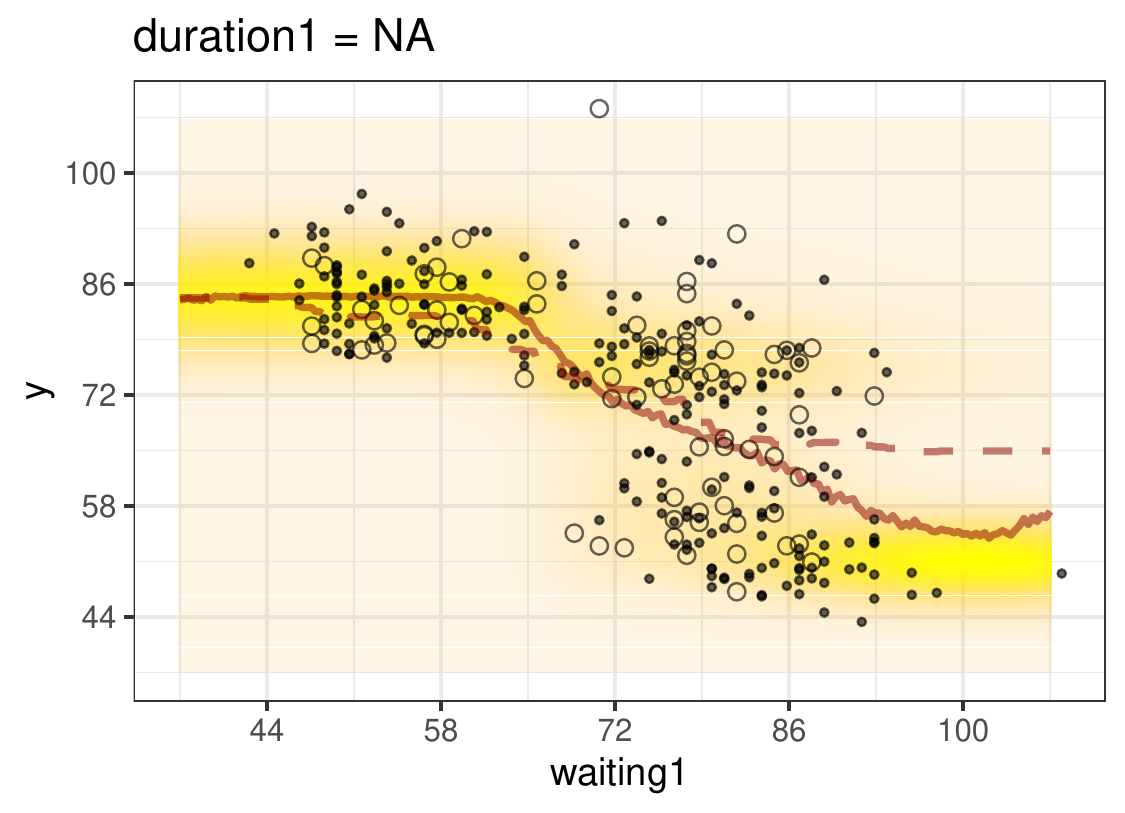}     \includegraphics[scale=0.70, trim={35 0 0 0}, clip]{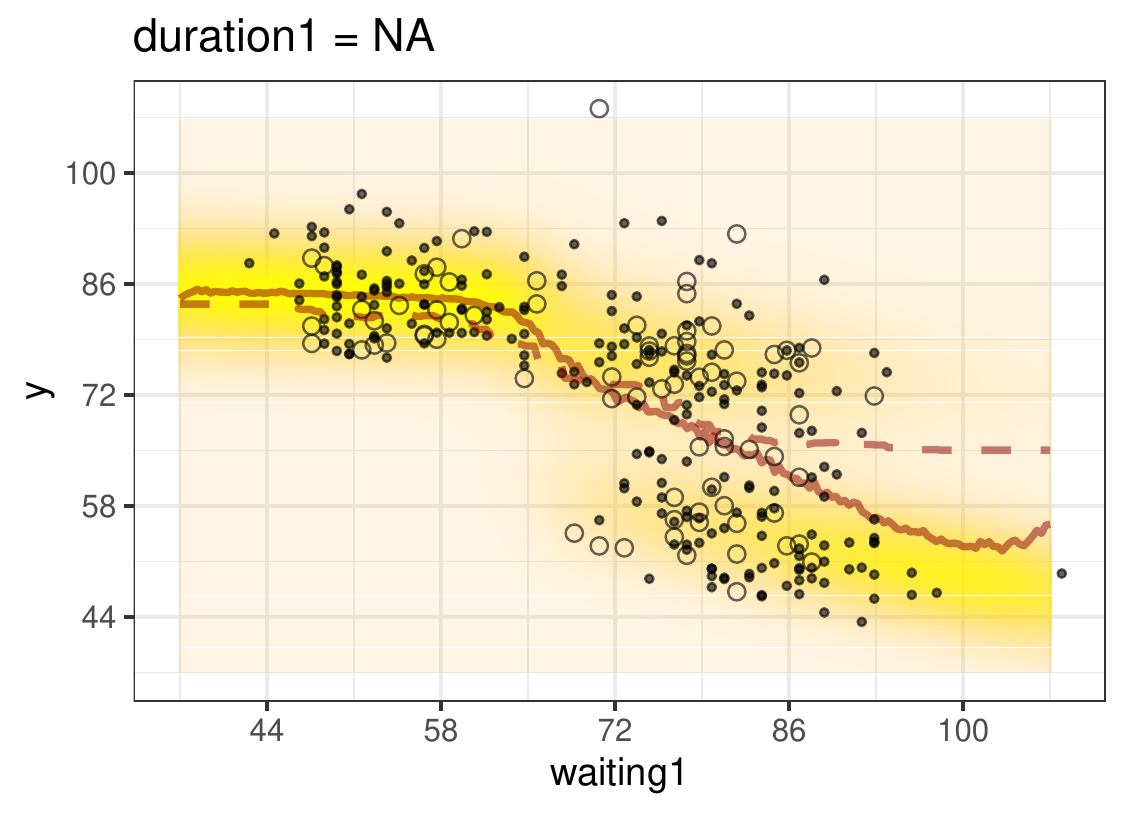}
    \caption{Old Faithful waiting time in minutes ($y$) against the previous waiting time in minutes (\texttt{waiting1}) for different durations of the previous eruption (\texttt{duration1}), including missing values (\texttt{NA}). Darker, larger points indicate observations with \texttt{duration1} values closer to the panel's represented value and circles indicate missing \texttt{duration1}. Left panels correspond with a VDReg fit and the right panels with the VDLReg fit. The solid red curve estimates the mean regression functional; the dashed red curve indicates the same from BARTm. Posterior mean predictive density values over a grid, indicated with color, demonstrate the models' capacity to project (marginalize) over missing values of \texttt{duration1} (bottom panels) and capture resulting bimodality. }
    \label{fig:OldFaithfulPredDens}
\end{figure}

Figure \ref{fig:OldFaithfulPredDens} illustrates the flexible projection property of VDReg (left panels) and VDLReg (right) with scatter plots of eruption waiting times against the previous waiting times, paneled by value of previous eruption duration (rows). 
Posterior predictive densities on a grid, indicated by color, concentrate and project appropriately when \texttt{d1} is missing (bottom panels), capturing bimodality in the response without additional parameters or model structure. All observations contribute to the fit, regardless of missingness pattern.

When \texttt{d1} is observed, the response distribution is unimodal and nonlinear. Here we can appropriately use the regression mean for prediction, as indicated with red curves in Figure \ref{fig:OldFaithfulPredDens}, and evaluate with mean squared error. In this case, BARTm yields similar regression curves (dashed red). 
When \texttt{d1} is missing, point predictions for \texttt{w1} above 65 minutes are unsatisfactory due to the bimodal response distribution. See Section \ref{sec:OldFaithful-appx} for analysis of quantile residuals.

We finally highlight that the slight advantage of VDLReg over VDReg in this example is evident for \texttt{w1} values above 65 minutes in the $\texttt{d1}=2$ minutes case. Because VDReg component means are flat, an extra cluster is necessary to accommodate the negative slope between $y$ and \texttt{w1}. VDLReg alternatively captures this feature with a single cluster using a regression in the component mean.

\section{Discussion}
\label{sec:discussion}

We have developed a method for nonparametric, locally linear regression that accommodates covariate vectors of varying dimensions without imputation. Building on the projection (marginalization) interpretation of VDReg, we have introduced linear covariate dependence to cluster-specific sampling means that i) relieves the burden of capturing features of the regression relationship from random partitions alone, and ii) parsimoniously and appropriately adapts to any missingness pattern while propagating uncertainty from missing covariates in a controlled manner. In contrast with other imputation-free methods for variable-dimension regression, VDReg and VDLReg yield prediction densities composed of mixtures, thereby accommodating multimodal response distributions that can occur when relevant covariates are missing. The method appears robust to violation of the built-in assumption that covariates are missing completely at random. We have further explored the effects of missingness on the random partition mechanism, introduced a tuning-free MCMC algorithm, and proposed a simple and fast screening method for determining whether locally linear sampling means are warranted. In this section we address additional issues relevant to the model and its use, including areas that merit future consideration.

The demonstrations in Sections \ref{sec:simulations} and \ref{sec:applications} employed multiple metrics for comparisons among modeling options. 
We emphasize that which metric is preferred will also depend on the modeling objective at hand, which should be considered with care. Mean squared error can be appropriate in scenarios that call for point prediction. Predictive deviance is appropriate for comparing accuracy of prediction distributions. Applications emphasizing inferential objectives, including detection of subpopulations, should refer to goodness-of-fit metrics (such as the K-S test in Section \ref{sec:simulations}) and analysis. While PPMx-based methods often outperform the other methods in alternate evaluation criteria, one must also weigh these benefits against the substantially increased computational cost of VDLReg.

The framework underlying the proposed VDLReg model is designed for continuous covariates, but does not readily admit categorical or ordinal covariates in the sampling model. This is because marginalization over a missing categorical variable calls for summation over alternate discrete values, complicating the sampling model specification and exponentially slowing computation. The simplest solution is to use a hybrid model with VDLReg specification involving only continuous covariates in the sampling model and all covariates in the similarity functions. \citet{PPMxMullerQuintanaRosner} explore options for similarity functions on discrete-valued covariates.

Construction of similarity functions with products of densities 
restricts the number of covariates 
the PPMx class of dependent random partition models 
can accommodate. 
While VDLReg employs shrinkage in the sampling model and performs competitively in our examples with $p=8$ and $p=10$ covariates, it is very difficult to detect multimodal response distributions with any more than a few covariates. One approach to improving density regression, and performance with other predictive criteria, assumes relevance of a small subset of covariates and employs variable selection in the similarity functions. \citet{quintana2015varsel} propose binary selection/inclusion of similarities in the PPMx, which is itself challenging due to doubly intractable likelihoods complicating inference for similarity hyperparameters. This and similar approaches, and their interaction with variable-dimension covariates, make future development in this area especially appealing.

Scalability to large sample sizes is another priority for future development, especially since the benefits of VDLReg over VDReg seem to be more evident with larger samples. 
The current implementation can admit modeling of hundreds to (low) thousands of observations with few covariates, or up to tens of covariates with hundreds of observations, before becoming prohibitively slow. 
Dynamic scaling of the linear predictors in the sampling model renders the update for latent allocations less efficient than that of other dependent random partition models. Overcoming this bottleneck will require additional innovations in computation, and likely approximation. Work by \citet{guha2010scalable} and \citet{ni2020scalable} offer promising directions for the effort to scale these models for use on larger data sets.


\section*{Acknowledgements}

The authors gratefully acknowledge helpful conversations with Peter M\"{u}ller.

\bigskip
\begin{center}
{\large\bf SUPPLEMENTARY MATERIAL}
\end{center}

\begin{description}

\item[Supplementary Materials:] (included) Illustration of marginalization behavior in VDLReg; fast screening tool for local linearity indicator, with simulations and data illustration; role of the global shrinkage hyperparameter; full posterior; computational complexity; additional simuation results; and additonal details from applications.

\item[VDLocalReg\_examples:] (\url{https://github.com/mheiner/VDLocalReg_examples.git}) R scripts that call the \texttt{ProductPartitionModels} Julia package to fit VDLReg models and recreate examples in the paper.

\end{description}

  \bibliographystyle{jasa3}
\bibliography{reference}

\newpage

\beginsupplement

\section{Illustration of Marginalization/Projection Behavior in VDLReg}
\label{sec:illustration}

To provide intuition for the conditionally specified sampling model in the top line of \eqref{eq:fullmodspec} in the main document, we examine a case with two covariates that can be visualized. We simulated $m=500$ observations from three fixed clusters. Covariates were generated from independent, unit-variance Gaussian distributions with cluster-specific means $(0,0)$, $(-3,-1.5)$, and $(1,3)$, with 25\% of all values randomly selected to be missing. Responses were generated from the sampling model with cluster-specific parameters $\mus_1 = 1.5$, $\mus_2 = 2.5$, $\mus_3 = -5.0$, $\bbetas_1 = (-0.9, 2.0)$, $\bbetas_2 = (-0.3, -1.0)$, $\bbetas_3 = (0.7, 0.0)$, $\sigs_1 = 1.2$, $\sigs_2 = 0.5$, and $\sigs_3 = 0.8$. We fit the proposed model to the data employing a normal-normal similarity function with unit variance. 

\begin{figure}[h!p]
    \centering
    \begin{tabular}{cc}
    \includegraphics[height=2.65in, trim = 17 80 40 13, clip]{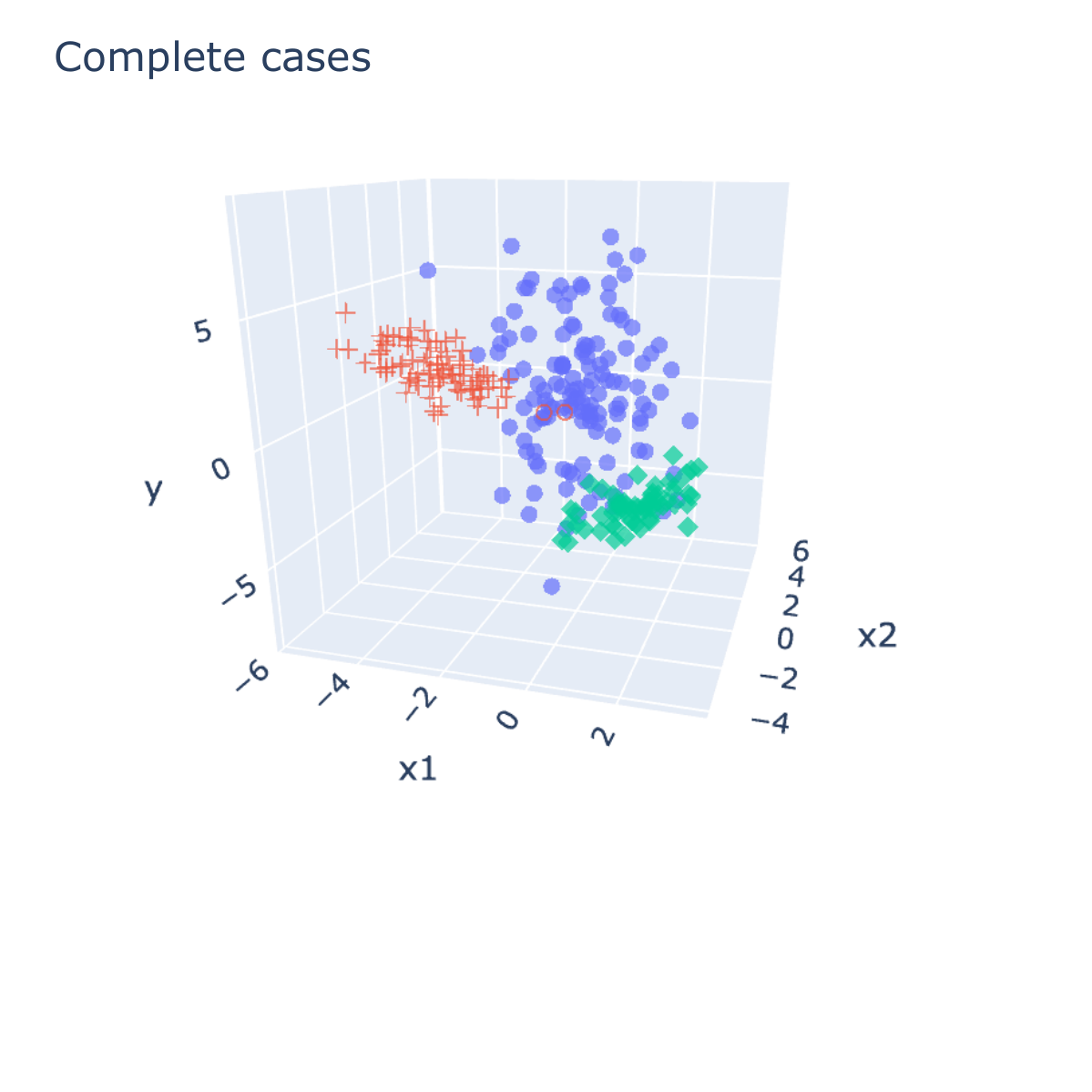} & \includegraphics[height=2.65in, trim = 0 0 0 13, clip]{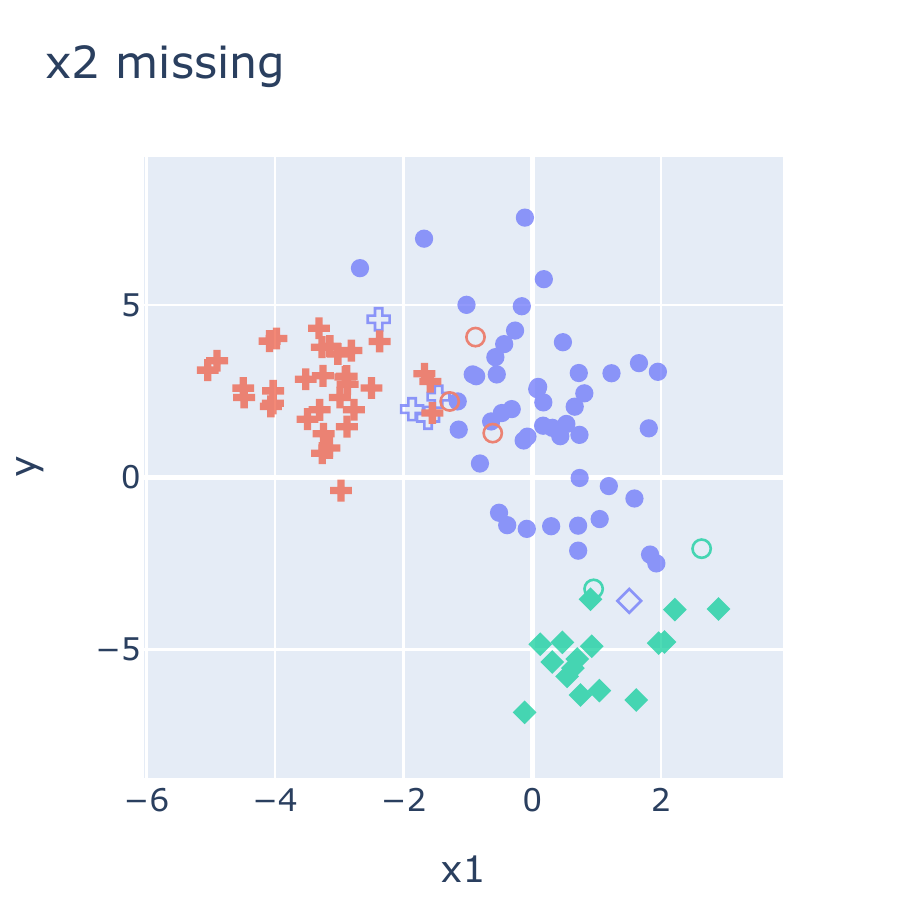} \\
    \includegraphics[height=2.65in, trim = 17 80 40 13, clip]{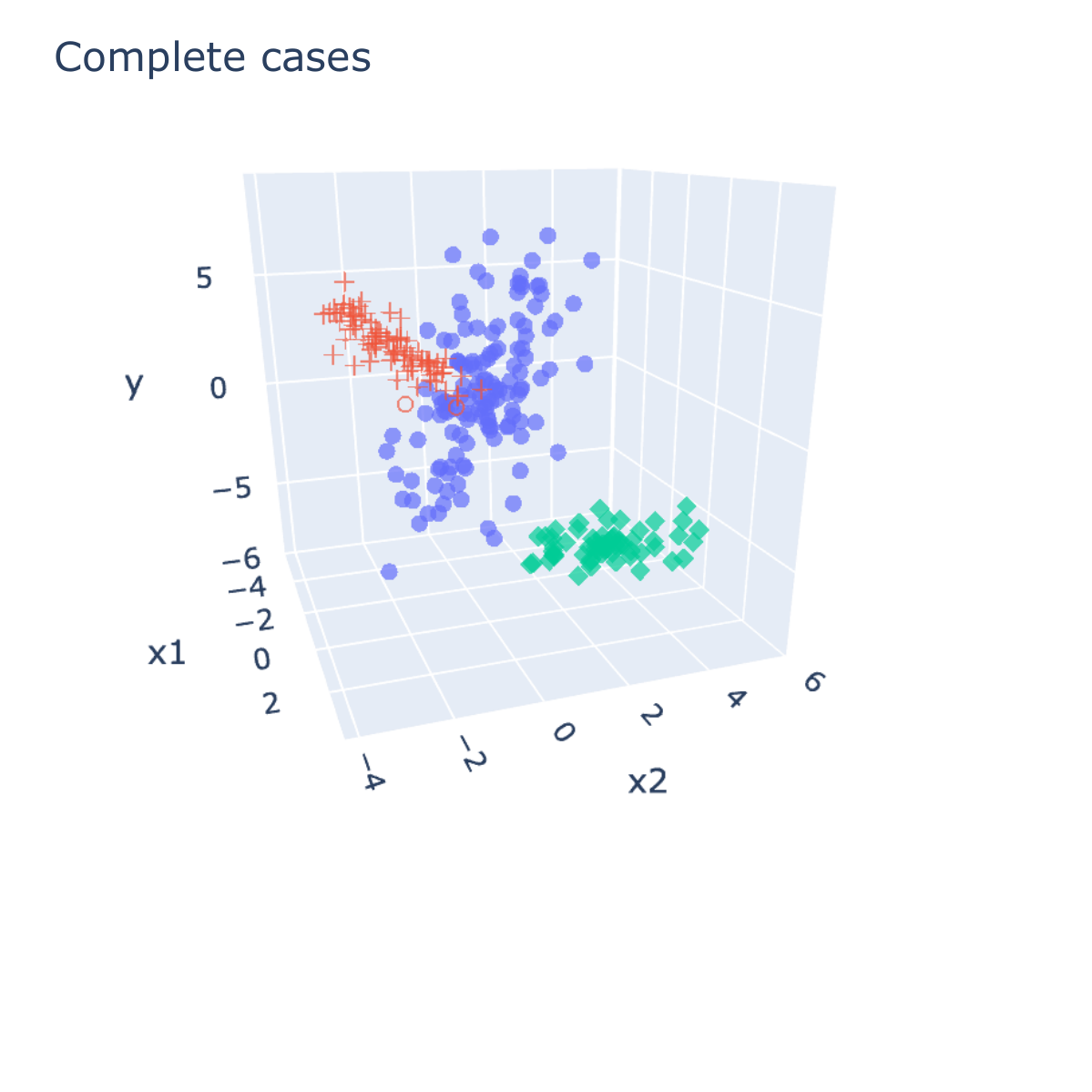} & \includegraphics[height=2.65in, trim = 0 0 0 13, clip]{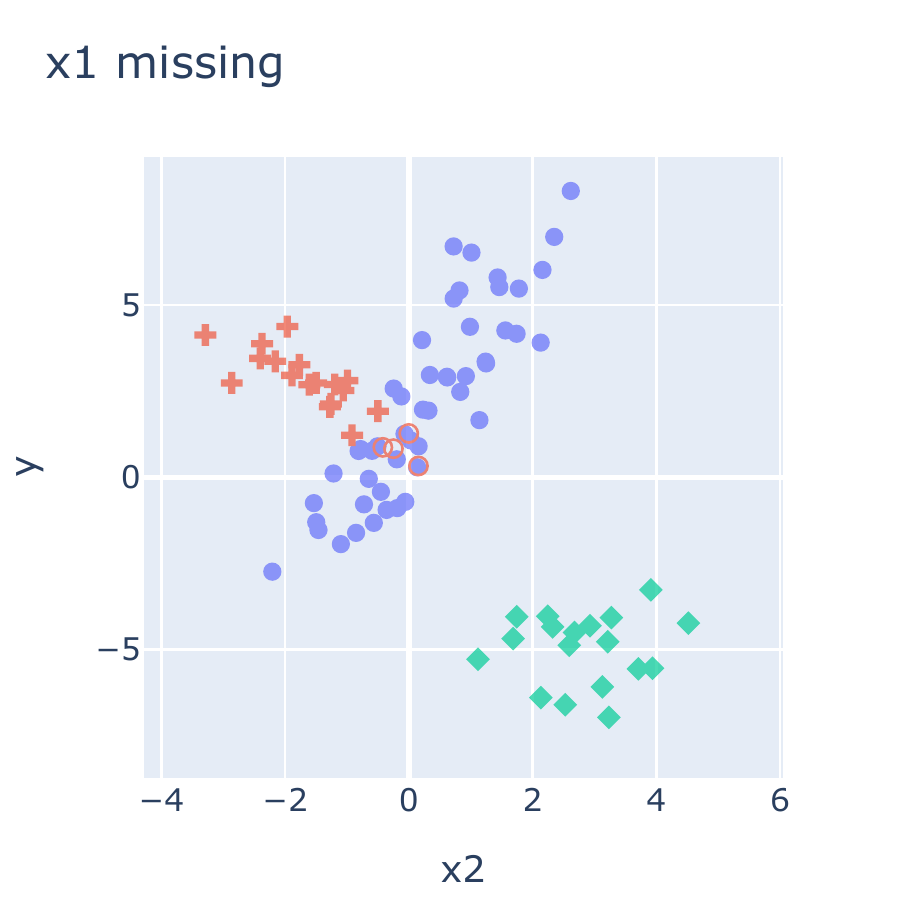} \\
    \includegraphics[height=2.65in, trim = 0 0 0 13, clip]{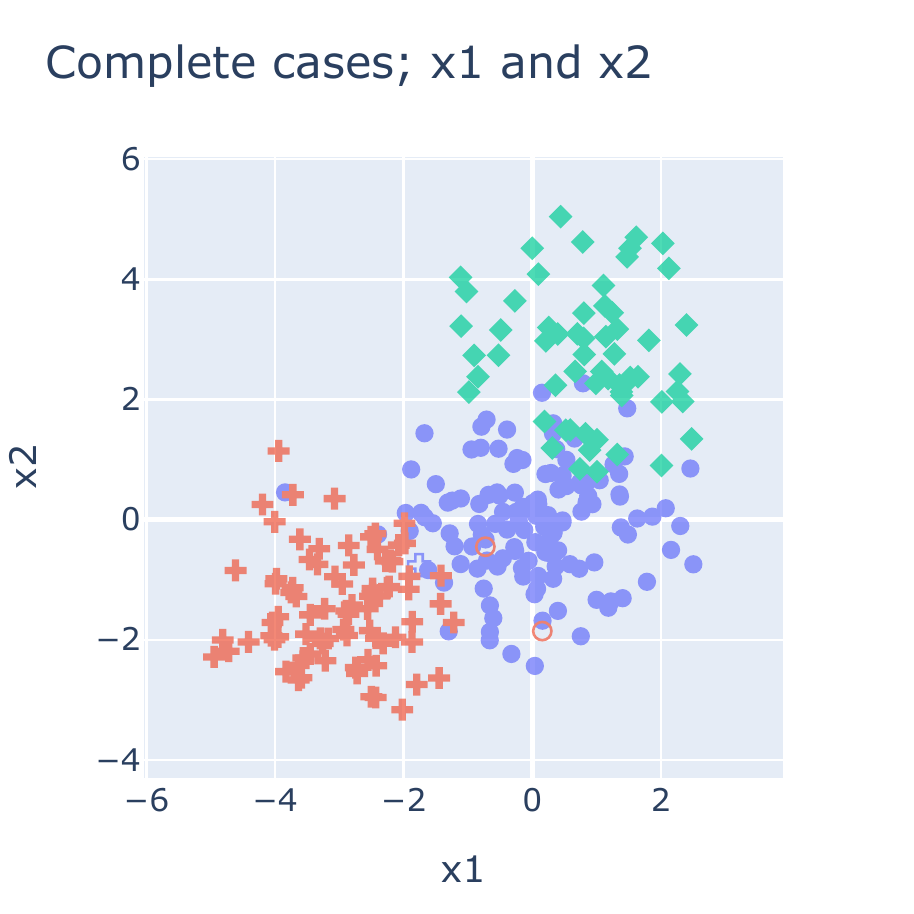} & \includegraphics[height=2.65in, trim = 0 0 0 13, clip]{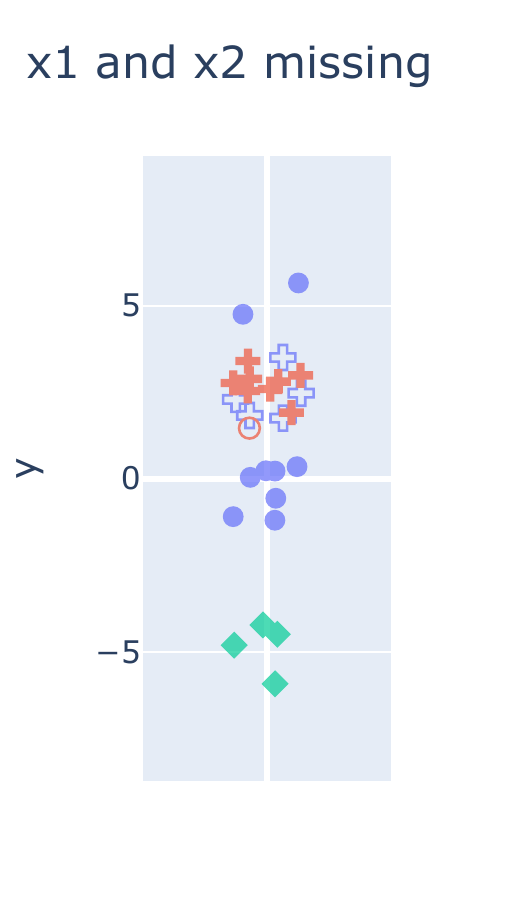} \\
    \end{tabular}
    \caption{Scatter plots of simulated illustration data by missingness pattern. Clusters are identified by color and model-estimated clusters are identified by shape; filled shapes are correctly classified observations and hollow shapes are misclassified.}
    \label{fig:illustration}
\end{figure}

Figure \ref{fig:illustration} plots the simulated data by missingness pattern. Points are colored by true cluster membership, while shape indicates a posterior point estimate of cluster membership, obtained from posterior samples using the SALSO algorithm of \citet{SALSOpackage, dahl2022salso}. Filled shapes indicate correct classification, and hollow shapes flag missclassified observations. The top two rows give comparable perspectives on partially missing and complete cases. When a coordinate is missing, the two-dimensional plot contains all the information provided by the observation.

The response mean functions for clusters 1 and 2 (blue and red, respectively) are planes that depend on both covariates $x_1$ and $x_2$. Cluster 2 depends primarily on $x_2$, which is evident from rotating the axes or comparing the strength of signal among red points in the panel with $x_2$ missing against those in the panel with $x_1$ missing. When $x_{i,2}$ is missing for observation $i \in S_2$, the sampling distribution assumes a mean value for $x_{i,2}$ (i.e., $z_{i,2} = 0$) and the variance is inflated by ${\betas_{2,2}}^2$ (in this case, approximately 1) to marginalize over uncertainty in the value of the missing $x_{i,2}$. Even when both $x_1$ and $x_2$ are missing, projection in the sampling model yields an appropriate response density that assists with classification between clusters 1 and 2, with correct classifications to cluster 1 (blue) occurring above and below the plausible cluster 2 range.

By mimicking the marginalization behavior of a local, jointly Gaussian model, 
the projected sampling model provides a coherent bridge across missingness patterns that borrows strength between them. For example, the data in both panels of the top row of Figure \ref{fig:illustration} contribute to the same cluster-specific parameters, not requiring a separate model specification for the case when $x_2$ is missing. This parsimony is also helpful when the observed covariates alone are less informative about cluster membership. When $x_2$ is missing, for example, it is more difficult to distinguish between clusters 1 (blue) and 3 (green). In this case, the sampling model furnishes information critical to discriminating membership.

\section{On Adding Covariates to the Likelihood}
\label{sec:flatbetter}

VDReg is a special case of VDLReg when the coefficients in the sampling model are fixed at zero, and the default settings of the latter model aggressively shrink toward the former. There are nevertheless situations in which fitting the simpler model is preferred. VDLReg also requires more care and attention in the fitting process. For example, given the same hyperparameter settings, the VDLReg model will often fit a plane through two separated clusters of points, trading variance for bias. Furthermore, under a fixed computing budget, the \texttt{ppmSuite} implementation of VDReg enables fits to data with more samples and covariates (see Table \ref{tab:runtime} in the main document). It is therefore useful to perform a preliminary assessment with any given data set prior to fitting a VDLReg model.

\subsection{A local linearity indicator}
\label{sec:linearIndicator}

We propose the following procedure as a fast screening tool for locally linear behavior. It is based on the {\em model based clustering} (MBC) method developed in \cite{fraley-raftery:02}, as implemented
in the \texttt{mclust} package \citep{mclust}. 
As a preliminary step,
eliminate all data points with missing entries, to achieve
$\tilde{m}$ complete observations. Let $\tilde{\bm{x}}=\{\tilde{\bm{x}}_i=(y_i,\bm{x}_i):\, \text{for $i=1,\ldots,\tilde{m}$}\}$ 
denote the complete data. Let $\tilde{p}$ denote the covariate vector dimension (note that $\tilde{p}$ may or may not
coincide with $p$), which thus contains $\tilde{m}$ vectors of dimension $\tilde{p}+1$. Carry out the following steps:

\begin{description}
	\item[1. Clustering:] Use \texttt{mclust} with $\tilde{\bm{x}}$ as input data to obtain an estimate of the number of clusters $\tilde{k}$
	and the corresponding partition $\tilde{\rho}=(\tilde{S}_1,\ldots,\tilde{S}_{\tilde{k}})$.
	\item[2. Regression:] For each $j \in \{1,\ldots,\tilde{m}\}$ such that $|\tilde{S}_{j}|>\tilde{p}+2$, compute the
	ordinary least squares estimator and appropriate measure of linearity $\tilde{q}_{j}$ associated to this ``local'' regression model. Possible options for $\tilde{q}_{j}$ include the p-value of the significance test for the corresponding regression coefficients (excluding the intercept), the coefficient of determination $R^2$, and its adjusted version 
 $$\bar{R}^2=1-\left(1-R^2\right) \frac{|\tilde{S}_{j}|-1}{|\tilde{S}_{j}|-\tilde{p}}.$$ 
	\item[3. Combination:] Compute and return
	\begin{equation}\label{eq:q*}
		\tilde{q}^*=\frac{\sum\limits_{j:\,|\tilde{S}_{j}|>\tilde{p}+2} |\tilde{S}_{j}| \, \tilde{q}_{j}}
	{\sum\limits_{j:\,|\tilde{S}_{j}|>\tilde{p}+2}  |\tilde{S}_{j}|}.
	\end{equation}
\end{description}

The quantity $\tilde{q}^*$ represents the average of selected measures as described earlier, weighted by cluster size, for those clusters with
sizes beyond a minimum level. Small values of $\tilde{q}^*$ suggest the presence of a local regression effect when the selected measure is the p-value, and the same applies when larger, closer to 1 values of $R^2$ for $\bar{R}^2$ are observed. 
We note here that \texttt{mclust} uses a joint covariance model for the $\tilde{\bm{x}}_i$ observations that does not necessarily parallel the VDLReg construction. Nevertheless, $\tilde{q}^*$ is easy and cheap to compute, which makes it a useful indicator for the purpose of 
detecting local linearities.

\subsection{Tests on the local linearity indicator}
\label{sec:indicator-simulation}

The local linearity indicator introduced above provides a simple procedure to detect potentially important regression effects at the cluster level. To test this procedure we carried out a small simulation experiment. We considered data sets with $n=200$ observations each, all of them with 4 clusters and three covariates, $x_1$, $x_2$ and $x_3$, defined as follows: $x_1$ is a sequence of equally spaced values ranging from $-2$ to $2$, $x_2$ is a uniform random sample on $[-3,3]$ and $x_3$ is a uniform random sample on $[0,1]$. Four clusters were considered in all cases, defined by those values of $x_1$ in the sub-intervals $[-2,-1)$, $[-1,0)$, $[0,1)$, and $[1,2)$, respectively.
Three different scenarios were considered, as indicated in Table~\ref{tab:IndicatorScenarios}.

In all scenarios, clustering is determined by values of $x_1$, but the type of covariate dependence changes with clusters.  Furthermore, the true generating distribution ignores $x_3$ in Scenario 1, ignores $x_1$ in Scenario 3, and does not use covariates in Scenario 2. In addition, we have
included quadratic (i.e., nonlinear) covariate dependence in Scenarios 1 and 3.

\begin{figure}[p]
\centering
\includegraphics[width=6in,height=6in]{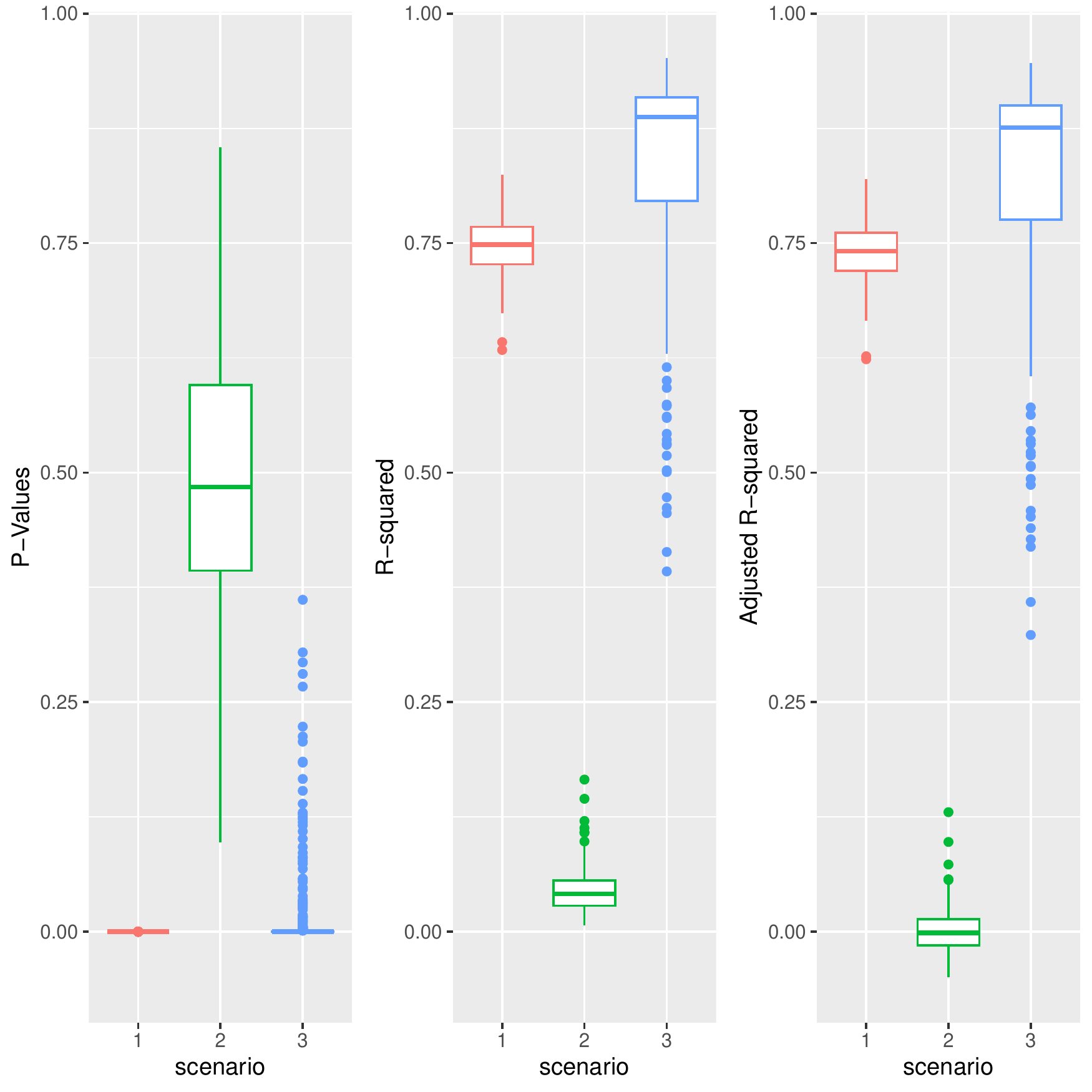}
\caption{Box plots summarizing performance of the local linearity indicator across scenarios. Reults using p-values, $R^2$ and adjusted $R^2$ coefficients are shown in the left, middle and right panels, respectively.}
	\label{fig:Indicator}
\end{figure}

For each scenario we generated 500 replications and computed indicator
\eqref{eq:q*} based on weighted averages of (i) p-values; (ii) $R^2$ coefficients; and (iii) adjusted $R^2$ coefficients. The results are summarized in Figure~\ref{fig:Indicator}, and they are generally in agreement that in Scenarios 1 and 3 there is indeed a substantial local dependence that may justify adopting a linear regression in the likelihood. Results from Scenario 1 are much less varied across repeated simulations than for Scenario 3, reflecting the fact that in the latter case, clusters are constructed from a covariate ($x_1$) that plays no role in the sampling generating mechanism, unlike the former case. On the other hand, in Scenario 2 results suggest that no local linear regression effect was detected, advising against including covariates in the sampling model.

\begin{table}[h]
\begin{center}
\begin{tabular}{c|l|l|l}
\multicolumn{1}{c|}{Cluster} & \multicolumn{1}{c|}{Scenario 1} & 
\multicolumn{1}{c|}{Scenario 2} & \multicolumn{1}{c}{Scenario 3} \\ \hline
1 & $y\sim \mathcal{N}(1+10x_1^2+ \frac{1}{2}x_2,10^2)$ & $y\sim \mathcal{N}(20,10^2)$ & 
$y\sim \mathcal{N}(1+10x_2^2+ \frac{1}{2}x_3,10^2)$ \\
2 & $y\sim \mathcal{N}(2+10x_1^2-x_2,6^2)$ & $y\sim \mathcal{N}(-20,6^2)$ & 
$y\sim \mathcal{N}(2+10x_2^2-x_3,8^2)$\\
3 & $y\sim \mathcal{N}(-1-20 x_1^2+2 x_2,10^2)$ & $y\sim \mathcal{N}(30,10^2)$ & 
$y\sim \mathcal{N}(-1-20x_2^2+ 2x_3,10^2)$\\
4 & $y\sim \mathcal{N}(-2+20x_1^2-2 x_2,8^2)$ & $y\sim \mathcal{N}(-30,8^2)$ & 
$y\sim \mathcal{N}(-2+20x_2^2- 2x_3,8^2)$ \\ \hline
\end{tabular}
\end{center}
\caption{Definition of simulation scenarios to test the proposed local linearity indicator}\label{tab:IndicatorScenarios}
\end{table}

\newpage
\subsection{Illustration with SUPPORT data}

\citet{Mercaldo2020patternSub} demonstrate the pattern submodel approach using data from the Study to Understand Prognoses and Preferences for Outcomes and Risks of Treatments (SUPPORT; \citealp{knaus1995support}), modeling survival of 9,105 hospitalized adults. The data are publicly available through \citet{mercaldoGITHUB}. Following \citet{Mercaldo2020patternSub}, we use as response a physiology score (\texttt{sps}) that was derived in part from the following $p=10$ covariates: partial pressure of oxygen in the arterial blood (\texttt{pafi}), mean blood pressure (\texttt{meanbp}), white blood count (\texttt{wblc}), albumin (\texttt{alb}), APACHE III respiration score (\texttt{resp}), temperature (\texttt{temp}), heart rate per minute (\texttt{hrt}), bilirubin (\texttt{bili}), creatinine (\texttt{crea}), and sodium (\texttt{sod}). The data contain 3,842 complete cases; among other missing covariates, each of \texttt{pafi}, \texttt{alb}, and \texttt{bili} have 1,000$+$ missing values. 

Although weighted p-values from the local linearity indicator applied to the full data set 
are small, 
weighted R-squared values fall below 0.2 when $p=6$ and $p=10$ (see below), suggesting only marginal gain from locally linear predictors. We find this surprising given that the response is a nonlinear derived product of the covariates. However, the data also omit a categorical covariate known to contribute to \texttt{sps} \citep{Mercaldo2020patternSub}, adding further noise to empirical regression.

Because neither implementation of VDReg accommodates data of this size, we repeatedly subsampled to training sets of $m=500$ and test sets of $m'=1000$ observations. We favored inclusion of cases with incomplete data, using sampling weights proportional to one more than the number of missing covariates. Each of the competing models was fit to 60 replicate training sets using all covariates listed above, as well as a subset identified by imputed random forest: \texttt{pafi}, \texttt{meanbp}, \texttt{alb}, \texttt{hrt}, \texttt{bili}, and \texttt{crea}. VDReg and VDLReg used $M=1$ and NNSI$\chi^2$ similarity with $\tilde{s}_0^2=0.5^2$ to admit fewer, larger clusters; and $a_\sigma=0.9$ to accommodate substantial noise. BARTm, RF, PSM, and MI again used default settings.

\begin{figure}[t!]
    \centering
    \includegraphics[width=3.1in, trim={0 0 0 20}, clip]{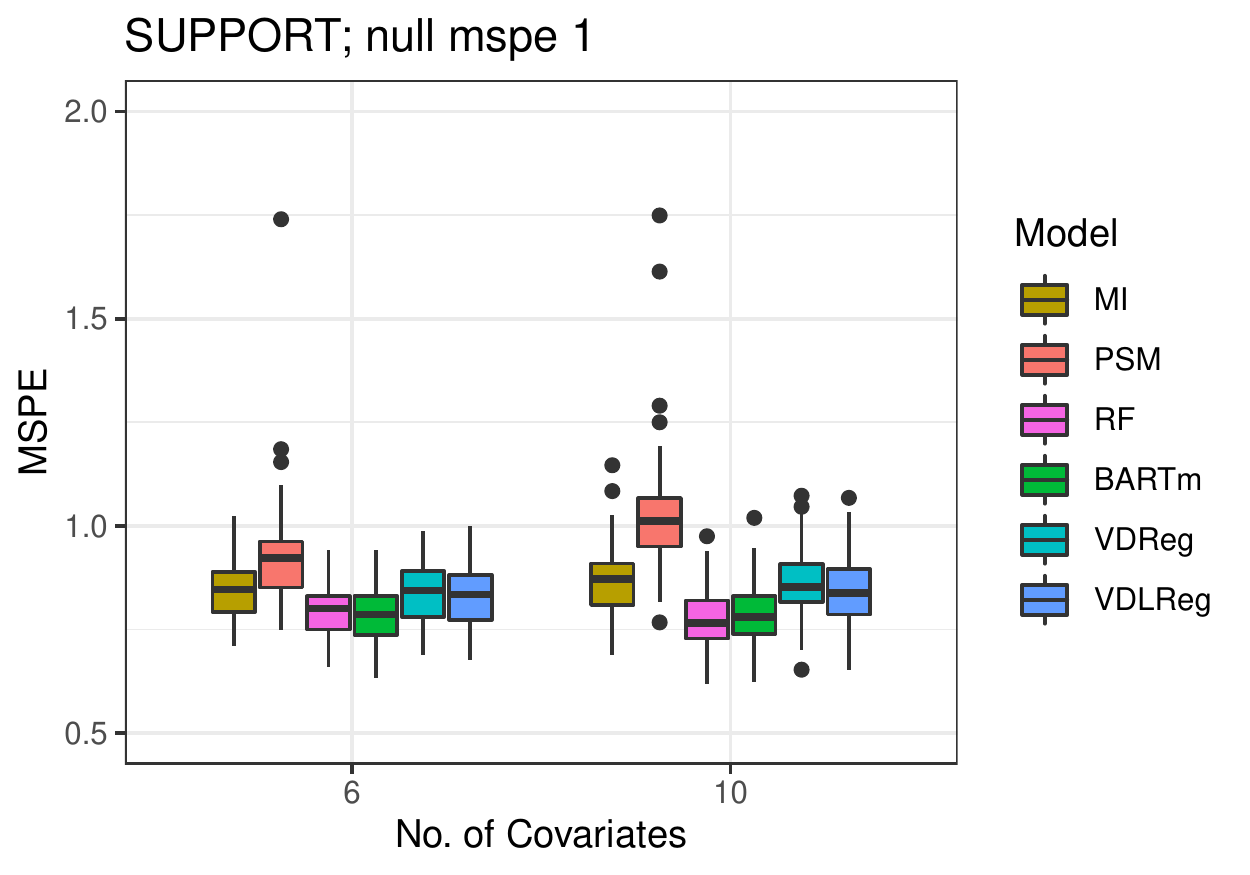}         \includegraphics[width=3.1in, trim={0 0 0 20}, clip]{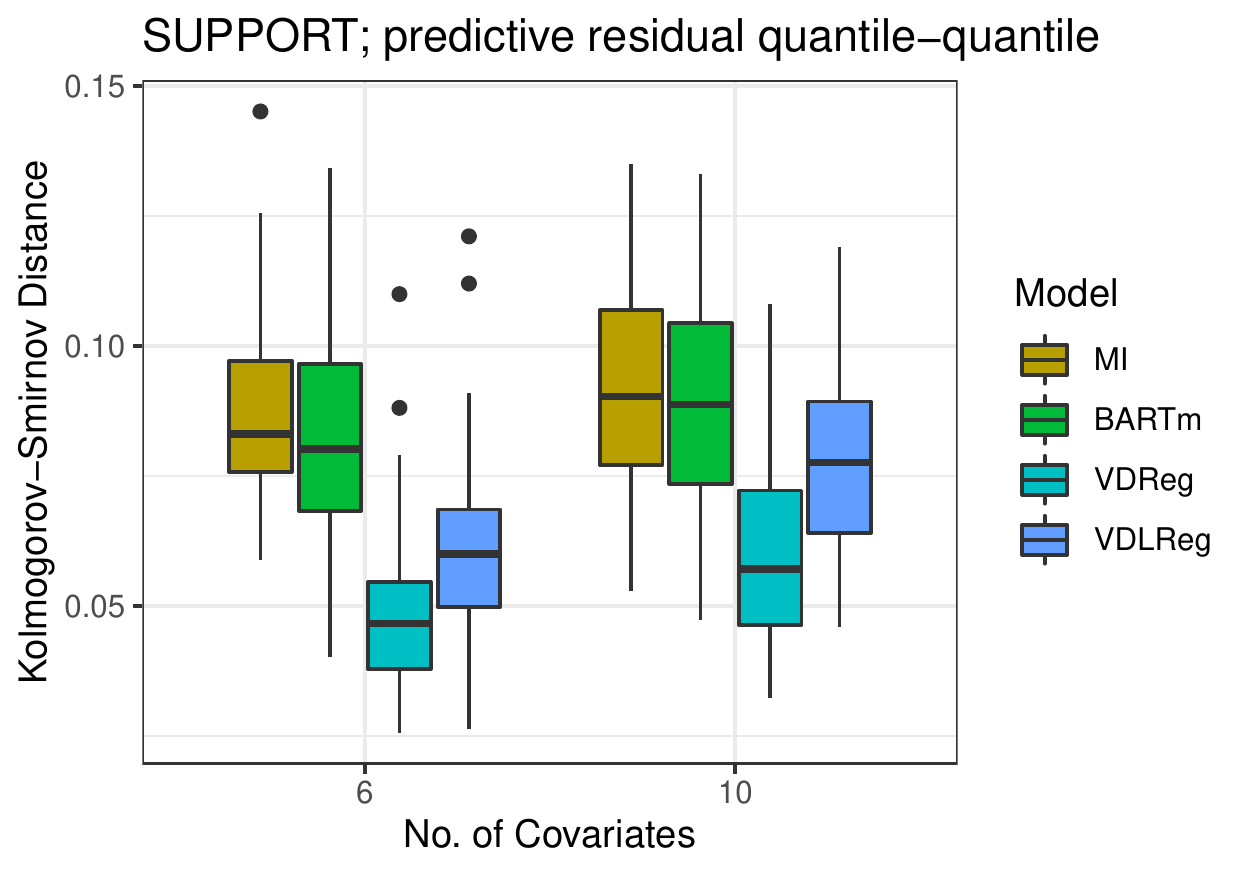} \\
    \includegraphics[width=3.1in, trim={0 0 0 20}, clip]{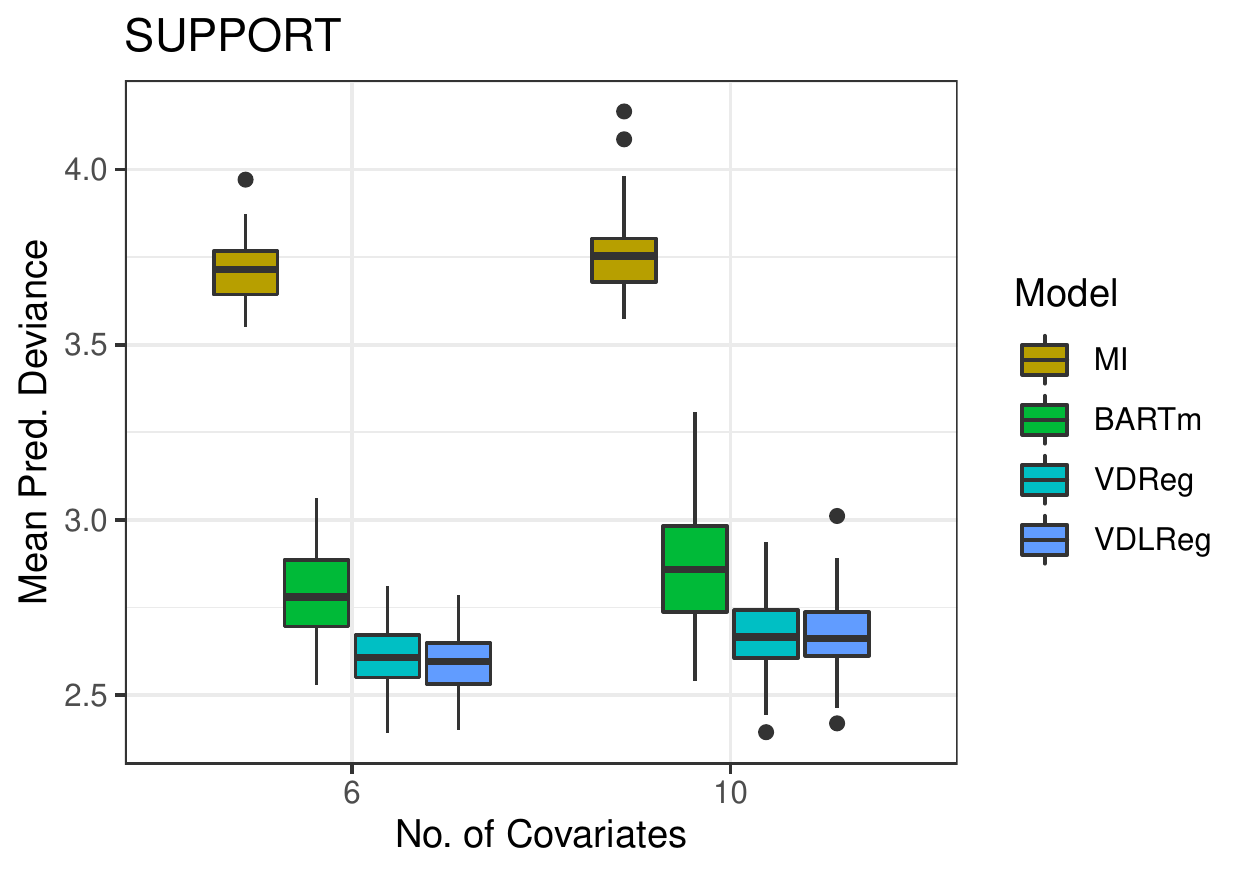}
    \includegraphics[width=3.1in, trim={0 0 0 20}, clip]{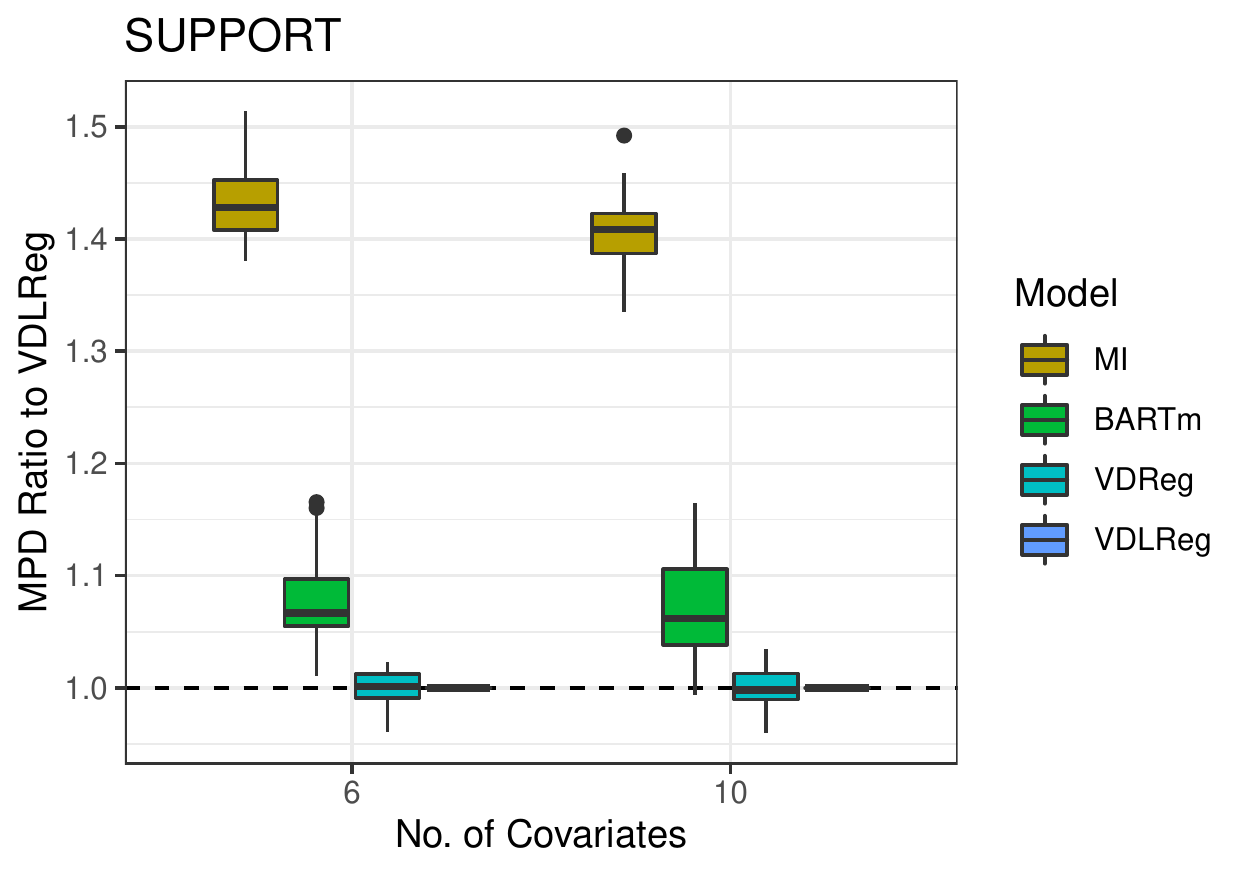}
    \caption{Box plots summarizing mean squared prediction error (top left), K-S distance of predictive quantile residuals from uniformity (top right) and mean predictive deviance (raw, bottom left; ratio, bottom right) for model fits to replicated training/test pairs of subsampled SUPPORT data.}
    \label{fig:supportMSPE}
\end{figure}

Figure \ref{fig:supportMSPE} reports out-of-sample MSE, deviance, and K-S goodness-of-fit statistics for the SUPPORT fits. BARTm and random forest lead at point prediction. 
VDLReg has slightly lower MSPE than VDReg, but both show little advantage over imputed linear models. Preference between the two in predictive deviance depends on the covariates included and 
VDReg appears to have the best-calibrated predictive residuals. 
Without clear advantages, we would find the added computational cost of VDLReg difficult to justify in this scenario. However, it may be that an alternate set of hyperparameter values could improve VDLReg performance relative to that of VDReg.

\section{Role of the global shrinkage hyperparameter}
\label{sec:hypers-appx}

To understand the role of $\tau_0$, the hyperparameter informing the global scale of $\bm{\beta}$, consider the multivariate normal distribution corresponding with the joint model in \eqref{eq:normal_reg_marg},
\begin{align}
    \begin{pmatrix}
    y \\ \bx
    \end{pmatrix} \sim \mathcal{N}\left( 
    \begin{pmatrix}
        \mu \\ \bm{\mu}^{(x)}
    \end{pmatrix}, \, 
    \begin{pmatrix}
        {\sigma^{(y)}}^2 & \sigma^{(y)} \bm{\varrho}' \bm{D}^{(x)} \\
        \sigma^{(y)} \bm{D}^{(x)} \bm{\varrho} & {\bm{D}^{(x)}}^2
    \end{pmatrix}
    \right) \, ,
\end{align}
parameterized with correlation vector $\bm{\varrho} = (\varrho_1, \ldots, \varrho_p)$ and $\bm{D}^{(x)} = \diag(\sigma_1^{(x)}, \ldots, \sigma_p^{(x)})$. The conditional variance of $y$ is $\sigma^2 = {\sigma^{(y)}}^2 (1 - \sum_{\ell=1}^p \varrho_\ell^2)$, which corresponds with the error variance. The conditional mean of $y$ is $\mu + \bm{\beta}' \bm{z}$, where $\bm{z} = {\bm{D}^{(x)}}^{-1} (\bm{x} - \bm{\mu}^{(x)})$ and $\bm{\beta} = \sigma^{(y)}\bm{\varrho}$ are parameterized as \eqref{eq:normal_reg_marg} and the sampling model in \eqref{eq:fullmodspec}. Substituting the error variance into the expression for $\bm{\beta}$ yields a relation that is useful for interpretation and specification: $\bm{\beta} = \sigma (1 - 
\sum_{\ell=1}^p \varrho_\ell^2)^{-1} \bm{\varrho}$. The prior for coefficients in \eqref{eq:DirLap} is scaled by the error standard deviation, leaving the global scale of the shrinkage prior, $\tau_0$, to correspond with a function of the correlations, $(1 - 
\sum_{\ell=1}^p \varrho_\ell^2)^{-1} \bm{\varrho}$. In one dimension, this function is approximately equal to $\varrho$ for correlations of magnitude below 0.4, after which it diverges to $-\infty$ and $+\infty$ near the respective boundaries, $-1$ and $1$. 
This suggests using a value of $\tau_0 < 1$, and we have found that $\tau_0 = 0.1$ performs well across a variety of scenarios.

\section{Full Posterior}
\label{sec:fulljointpost}

The model outlined in \eqref{eq:fullmodspec} and augmented with \eqref{eq:DirLap} admits a full joint posterior density 
\begin{align}
    \label{eq:fullpostdens}
    \begin{split}
        p(\rho = \{S_1, \ldots, S_{k_m}\}, \bm{\eta}, \bths \mid \by, \bxo) \propto & \: \mathcal{N}(\mu_0 \mid m_0, v^2) \times \text{Unif}(\sigma_0 \mid 0, \, a_{\sigma_0}) \, \times \\ & \prod_{j=1}^{k_m} \left[ c(S_j\mid M) \, \tg(\bx^{\star o}_j\mid\bxi) \right] \, \times \\ 
        & \prod_{j=1}^{k_m} \Big[ \mathcal{N}(\mus_j \mid \mu_0, \sigma_0^2) \times \text{Unif}(\sigs_j \mid 0, \, a_\sigma) \, \times \\
        & \qquad \mathcal{N}(\bbetas_j \mid \bm{0}, \, {\sigs_j}^2 \, \tau_j^2 \, \bm{D}(\bm{\psi}_j, \bm{\phi}_j)) \, \times \\
        & \qquad \text{Dirichlet}(\bm{\phi}_j) \times \text{Exp}(\tau_j) \times \prod_{\ell=1}^p \text{Exp}(\psi_{j\ell}) \Big] \, \times \\
        & \prod_{i=1}^m \mathcal{N}\left(y_i \mid \mus_{c_i} + \sum_{\ell \in \OO_i} \betas_{{c_i}\ell} z_{i\ell}, \, \sigsqs_{c_i} + \sum_{\ell \notin \OO_i} {\betas_{{c_i}\ell}}^2 \right) \, ,
    \end{split}
\end{align}
where $\bm{\eta} = (\mu_0, \sigma_0)$, $\bths = \{ \bths_j \}$ with $\bths_j = \{ {\mu}^\star_j, {\sigma}^\star_j, \bbetas_j, \bm{\psi}_j, \bm{\phi}_j, \tau_j \}$, $c_i \in \{1, \ldots, k_m\}$, $\bm{\psi}_j = (\psi_{j1}, \ldots, \psi_{jp})$, $\bm{\phi}_j = (\phi_{j1}, \ldots, \phi_{jp})$, and $\bm{D}(\bm{\psi}_j, \bm{\phi}_j)_{\ell \ell'} = \psi_{j\ell}\phi_{j\ell}^2 \, 1_{(\ell = \ell')}$. Note that $z_{i\ell} = (x_{i\ell} - \hat{\mu}^{(x)}_{j\ell}) / \hat{\sigma}^{(x)}_{j\ell}$ also depends on $c_i$. We employ a Gibbs sampling scheme, described in Section \ref{sec:computation}, that cycles through the following block-full conditional distributions based on \eqref{eq:fullpostdens}: $[\rho_m \mid \bths, \bm{\eta}, \by, \bxo]$, $ [ \bths \mid \rho_m, \bm{\eta}, \by, \bxo]$, and $[ \bm{\eta} \mid \rho_m, \bths, \by, \bxo]$.

\section{Computational Complexity}
\label{sec:complexity}

The update for latent allocations is easily the most computationally demanding and time consuming step of the Gibbs sampler. As noted in the description of the update for $\rho_m$, the full conditional distribution for $c_i$ in standard PPMx models requires the sampling (likelihood) contribution of $y_i$ only, whereas VDLReg requires sampling contributions from all observations potentially involved in a change in $c_i$. Algorithm 8 of \citet{neal:2000} considers all clusters as candidates, thus requiring likelihood evaluation for all observations, repeated $m$ times and resulting in $O(m^2p)$ complexity. Because these full conditional distributions are heavily influenced by $\bx$ \citep{wade2014enrichedDP, page:2018}, our Metropolis proposals in \eqref{eq:alg7_1a} involve only cohesion and similarity weights, reducing the complexity to approximately $O(mp(\bar{k} + \bar{m}_c))$, where $\bar{k}$ is an average number of clusters during the complete scan, and $\bar{m}_c$ generically refers to an average cluster size involved in a proposal. In a typical Gibbs scan, $\bar{k} \bar{m}_c$ may be close to $m$, but we often have $\bar{k} + \bar{m}_c < m$. The extremes of all singletons ($\bar{k} = m$) and one large cluster ($\bar{m}_c = m$) each yields the same complexity as the original Algorithm 8.

\section{Additional Simulation Results}
\label{sec:sim-appx}

\begin{figure}[htbp]
\begin{center}
\includegraphics[scale=0.7]{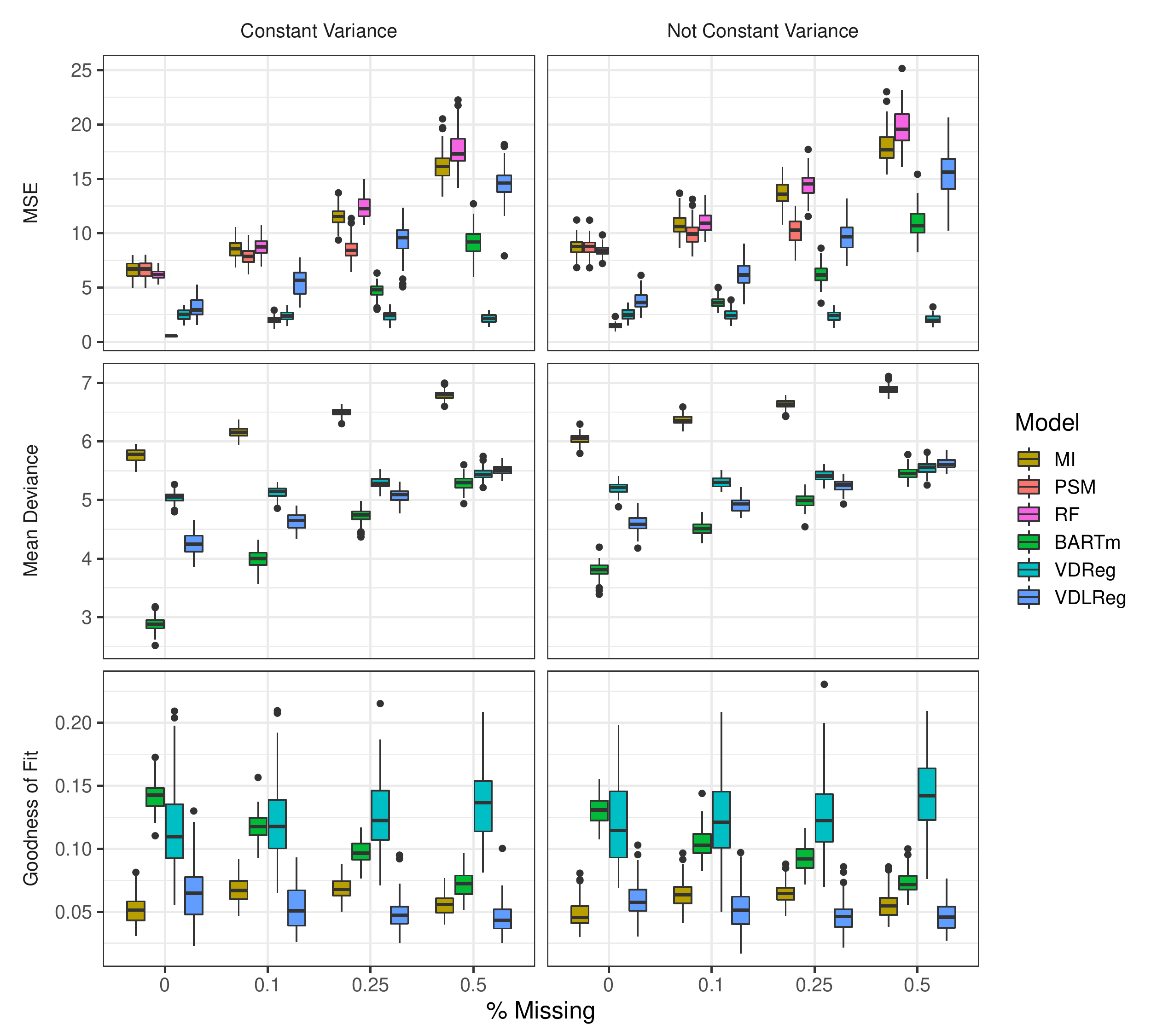}
\caption{In-sample results from the Friedman simulation with MAR data.}
\label{fig:friedman_mar_is}
\end{center}
\end{figure}

\begin{figure}[htbp]
\begin{center}
\includegraphics[scale=0.7]{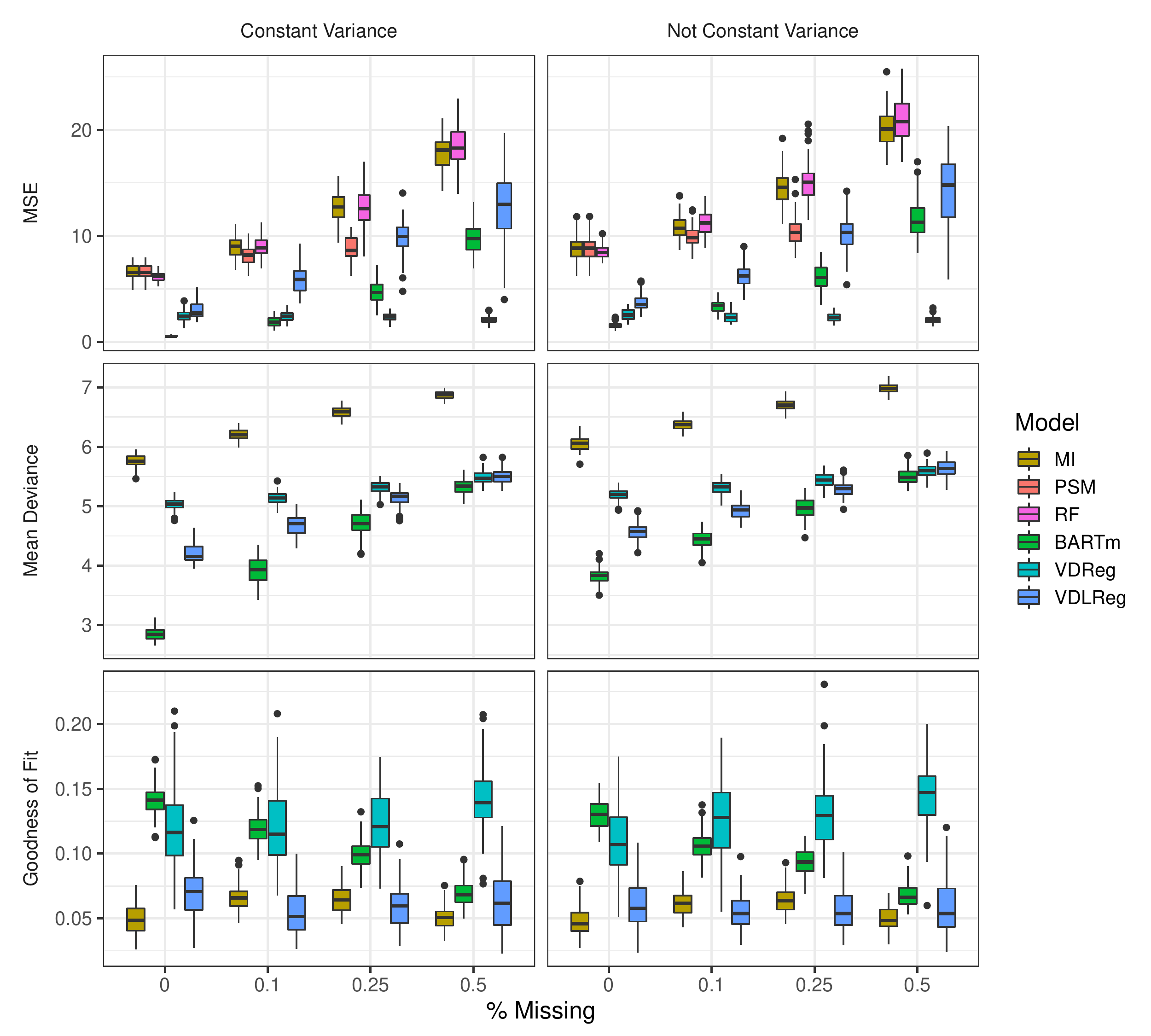}
\caption{In-sample results from the Friedman simulation with MNAR data.}
\label{fig:friedman_mnar_is}
\end{center}
\end{figure}

We include, for completeness, additional results from the simulation study on the Friedman data in Section \ref{sec:simulations}. Figures \ref{fig:friedman_mar_is} and \ref{fig:friedman_mnar_is} depict in-sample results, analogous to Figure \ref{fig:friedman_mar}, for the MAR and MNAR scenarios, respectively. As with the out-of-sample results, patterns between MAR and MNAR are nearly identical. Regarding MSPE, VDReg and BARTm appear to struggle the most with generalizing out of the sample. BARTm also suffers substantial loss in deviance moving out of sample. Surprisingly, VDReg and BARTm quantile residuals indicate poor calibration in the sample, with improvement out of the sample.

\begin{figure}[htbp]
\begin{center}
\includegraphics[scale=0.7]{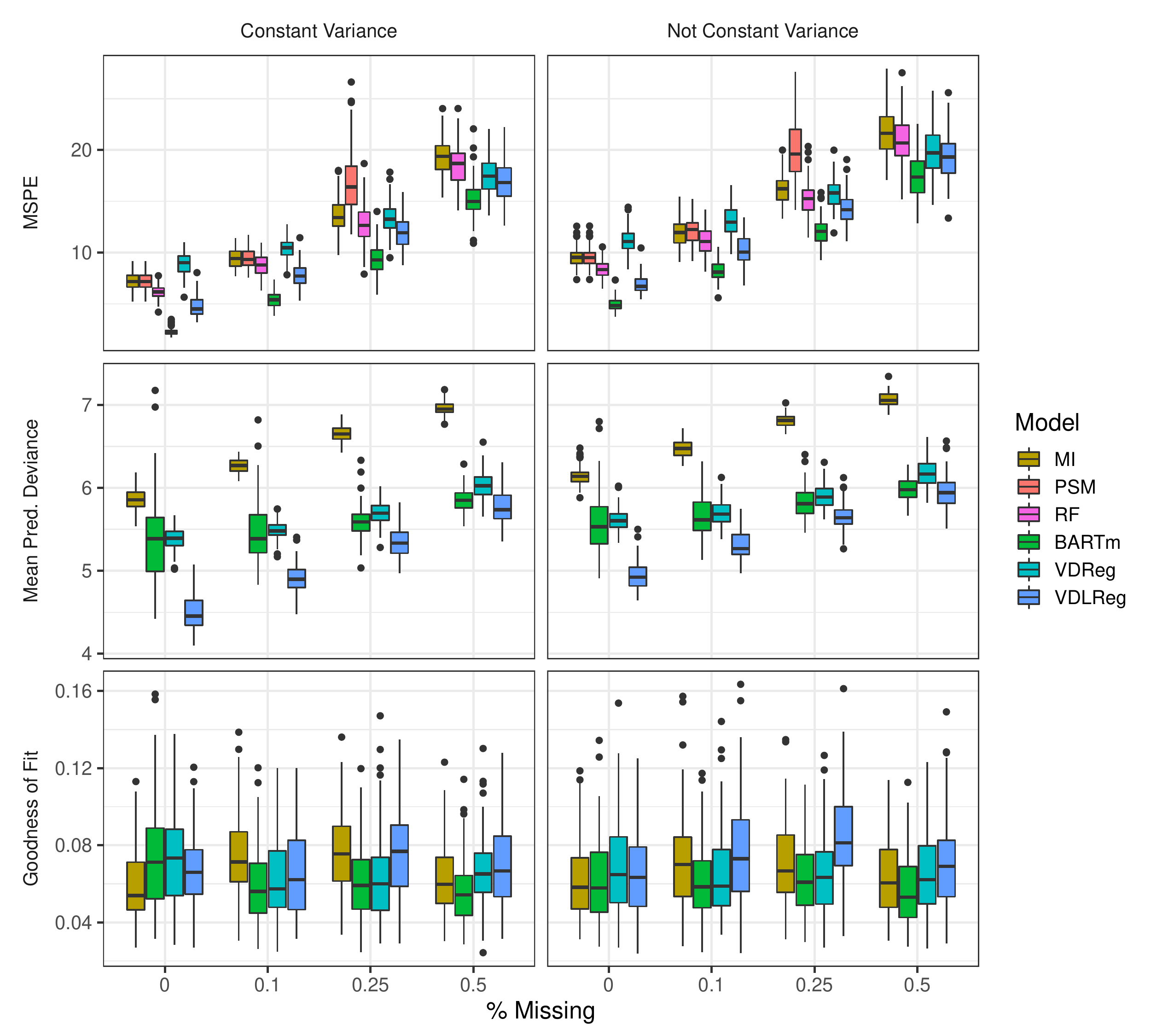}
\caption{Out-of-sample results from the Friedman simulation with MNAR data.}
\label{fig:friedman_mnar}
\end{center}
\end{figure}

Figure \ref{fig:friedman_mnar} reports out-of-sample results for the MNAR scenario. As noted in the main manuscript, the patterns are visually nearly indistinguishable from those of Figure \ref{fig:friedman_mar}. VDLReg typically fits between two and three clusters (fewer in 50\% missing case), while VDReg typically fits between four and six clusters (more in 50\% missing case) across all data scenarios.

\newpage

\section{Additional Details on Applications}
\label{sec:applications-appx}

\subsection{Boston housing}
\label{sec:boston-appx}

The eight covariates used in the Boston housing analysis are average number of rooms per dwelling (\texttt{rm}), per capita crime rate by town (\texttt{crim}), percent of population with lower income status (\texttt{lstat}), nitrogen oxides concentration in parts per 10 million (\texttt{nox}), full-value property-tax rate per \$10,000 (\texttt{tax}), proportion of owner-occupied units built prior to 1940 (\texttt{age}), proportion of non-retail business acres per town (\texttt{indus}), and index of accessibility to radial highways (\texttt{rad}).

\begin{figure}[hb!]
    \centering
    \includegraphics[scale=0.75, trim={0 0 75 20}, clip]{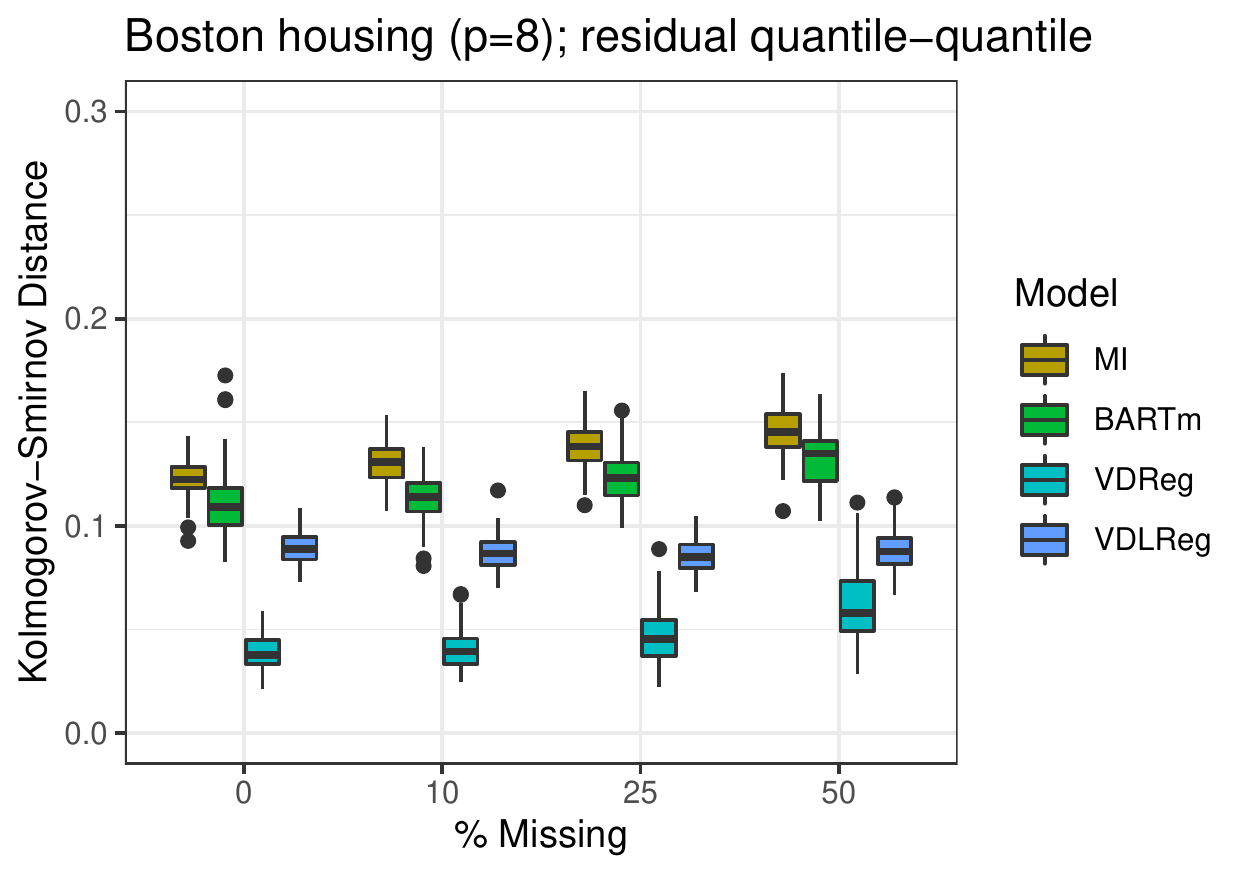}\includegraphics[scale=0.75, trim={17 0 0 20}, clip]{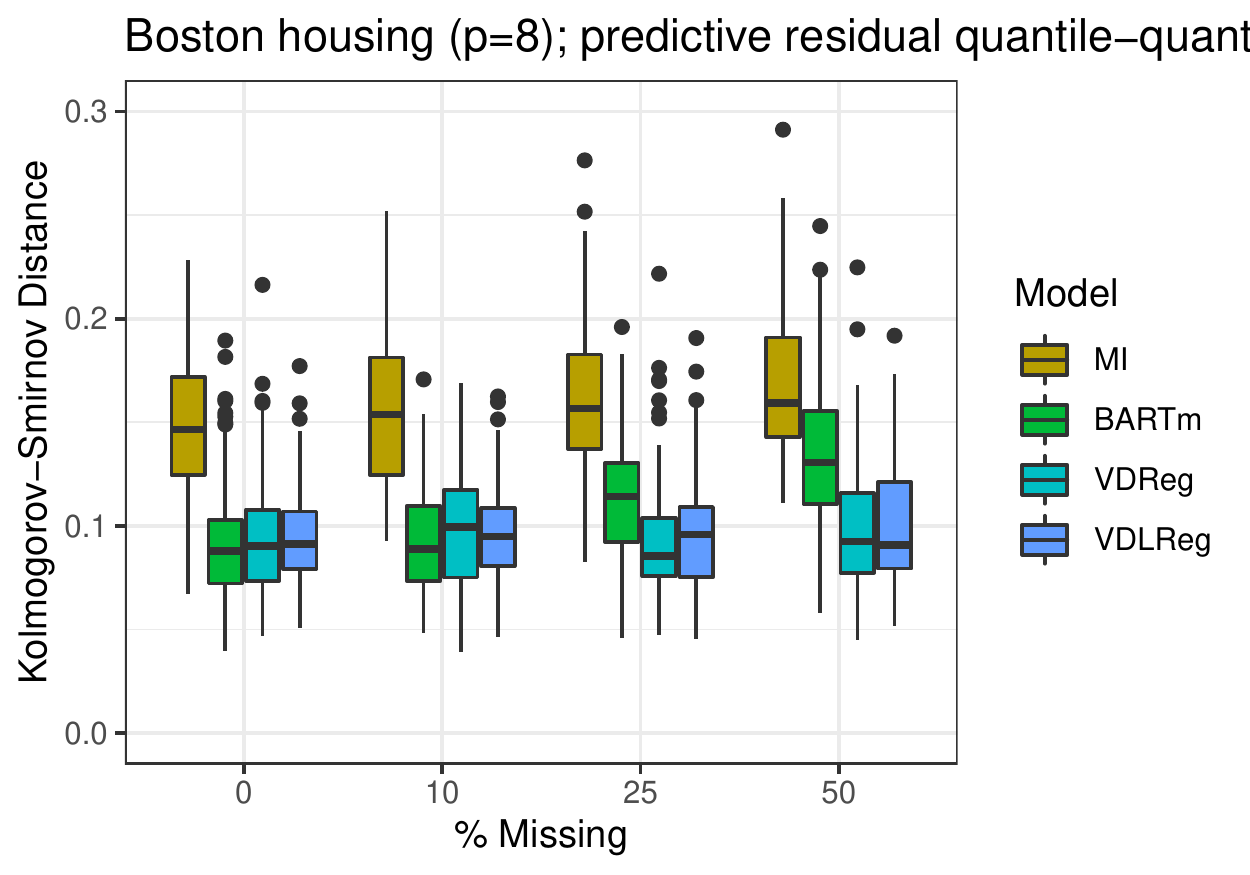}
    \caption{Box plots of Kolmogorov-Smirnov test statistics for uniformity of in-sample (left) and out-of-sample (right) quantile residuals from various fits of the Boston housing data.}
    \label{fig:KS-Boston}
\end{figure}

\begin{figure}[h]
    \centering
    \includegraphics[scale=0.7, page=3, trim={0 35 10 0}, clip]{figures/predQtile_competitors_Boston_pMiss25_ii100_220813.pdf}\includegraphics[scale=0.7, page=1, trim={30 35 10 0}, clip]{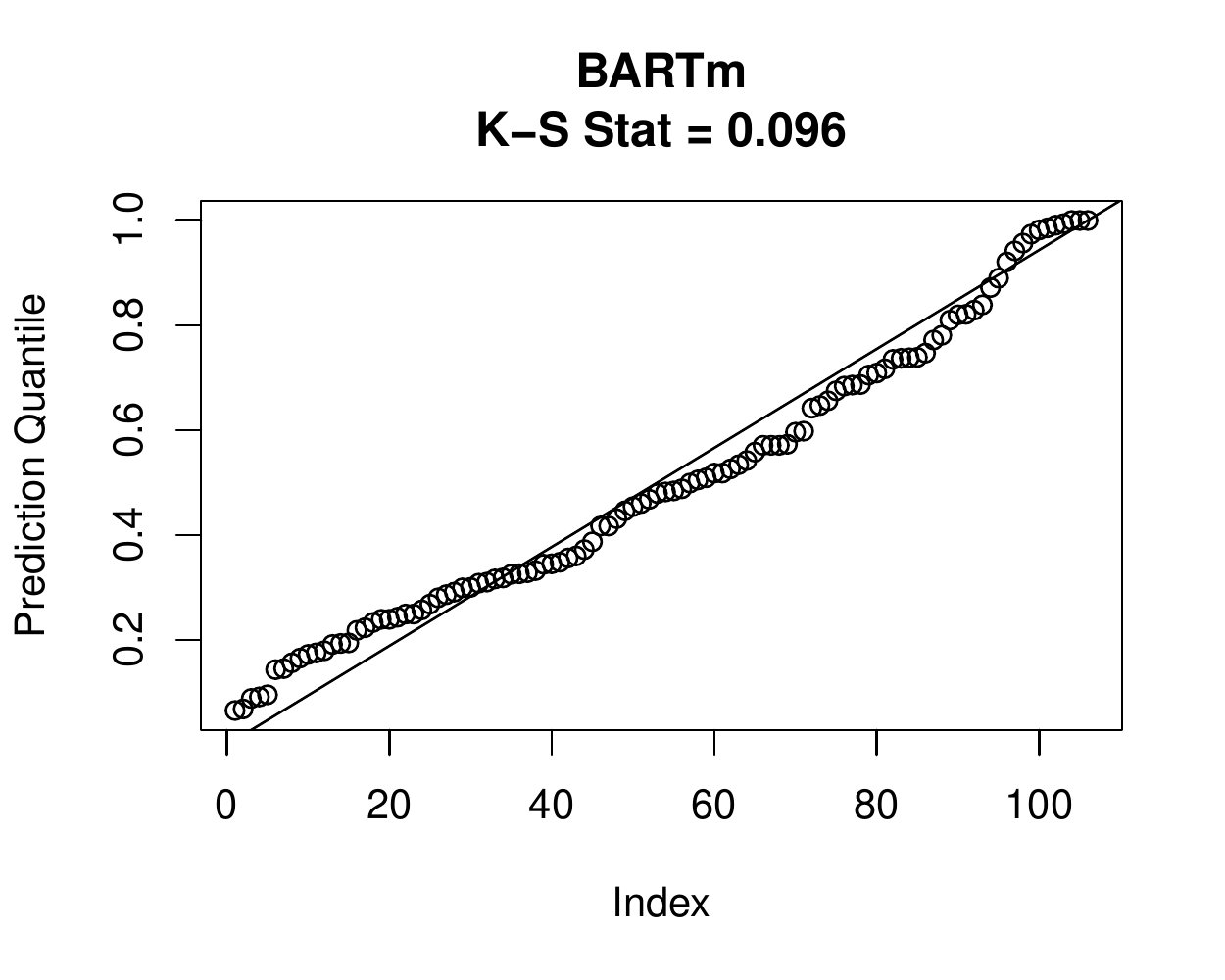}
    \includegraphics[scale=0.7, page=1, trim={0 0 10 0}, clip]{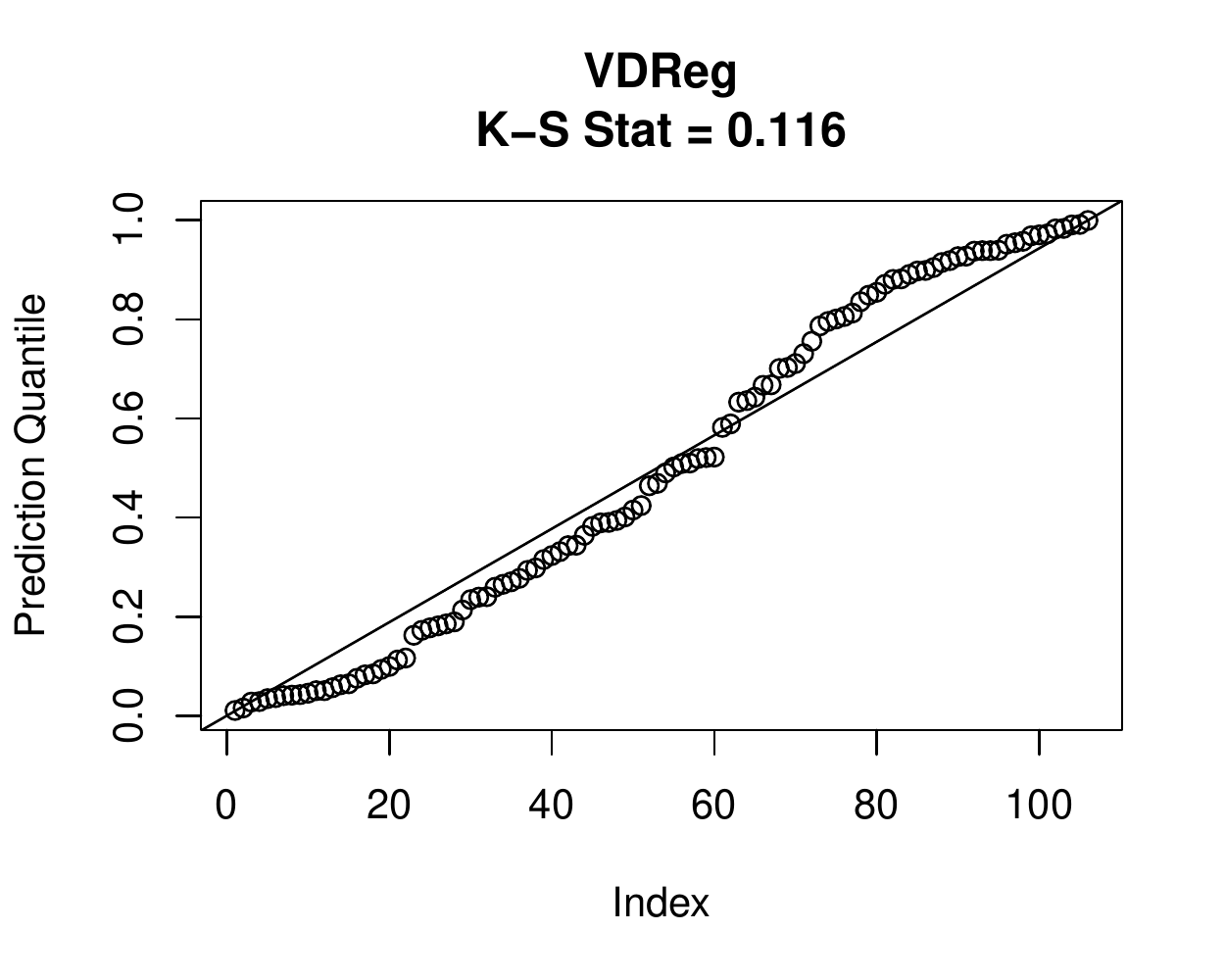}\includegraphics[scale=0.7, page=1, trim={30 0 10 0}, clip]{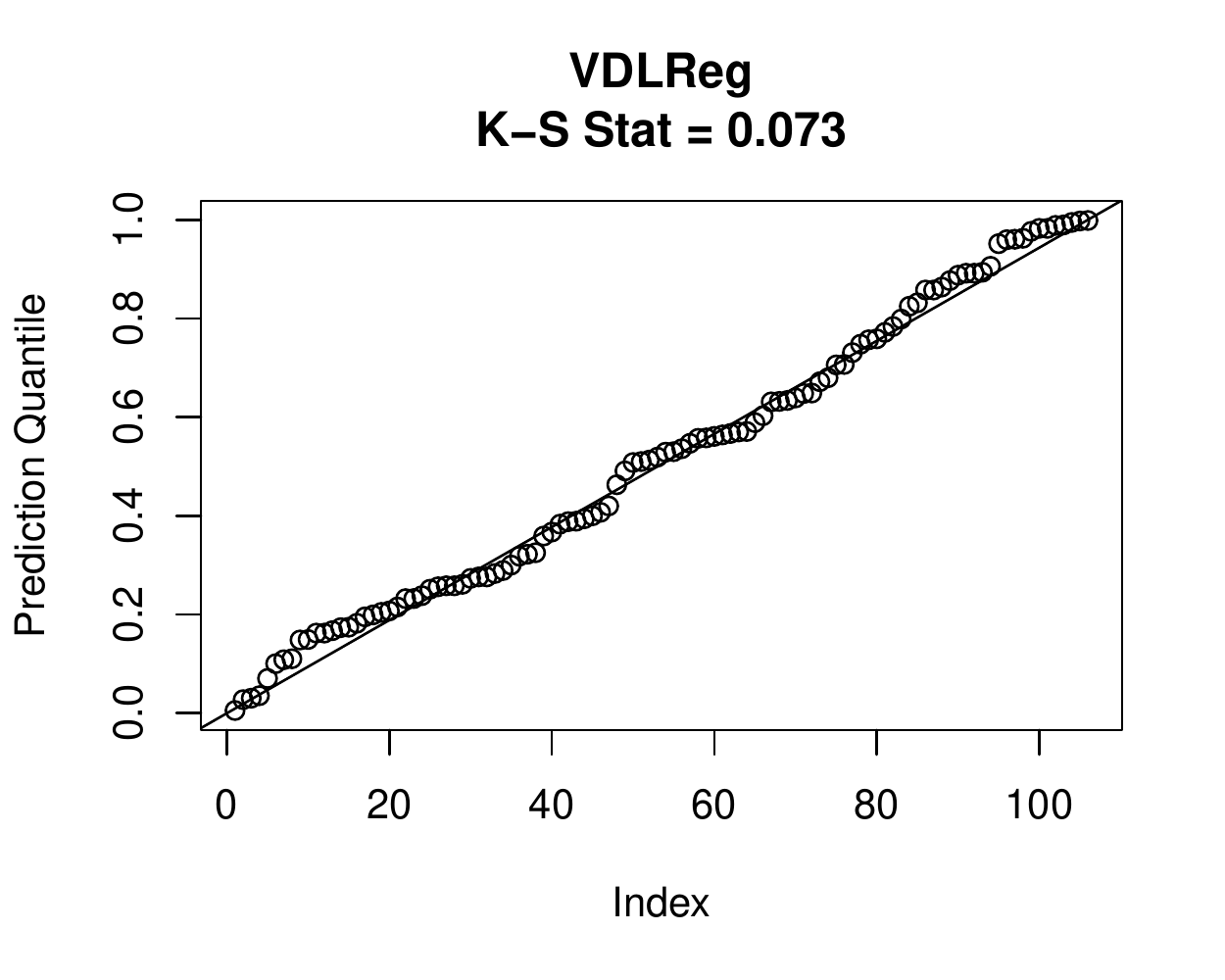}
    
    \caption{Ordered predictive quantile residuals for four fits to one replicate test set of the Boston housing data with 25\% of covariate values missing.}
    \label{fig:Qresid-Boston}
\end{figure}

Figure \ref{fig:KS-Boston} summarizes Kolmogorov-Smirnov test statistics on quantile residuals for the Boston housing data. High values, especially above 0.1, indicate poorly calibrated predictive distributions. VDLReg and BARTm perform consistently in and out of the training sample with this metric (in fact, BARTm improves out of sample), while VLReg generalizes less well. Figure \ref{fig:Qresid-Boston} shows a typical example of the contributing predictive quantile residuals for one test data set when 25\% of the covariate values are missing. In this case, the MI and BARTm predicitive distributions appear to be overdispersed, while the VDReg predictive distribution is underdispersed.

\newpage

\subsection{Old Faithful}
\label{sec:OldFaithful-appx}

We consider up to three covariates: the duration of the previous eruption (\texttt{d1}) in minutes, the waiting time to the previous eruption (\texttt{w1}), and the waiting time to the eruption preceding the previous (second lag, \texttt{w2}). All 100 training/test runs were repeated at three combinations of covariates: (\texttt{w1}, \texttt{w2}), (\texttt{w1}, \texttt{d1}), and (\texttt{w1}, \texttt{w2}, \texttt{d1}).
 
\begin{figure}
    \centering
    \includegraphics[width=3.1in, trim={0 0 0 20}, clip]{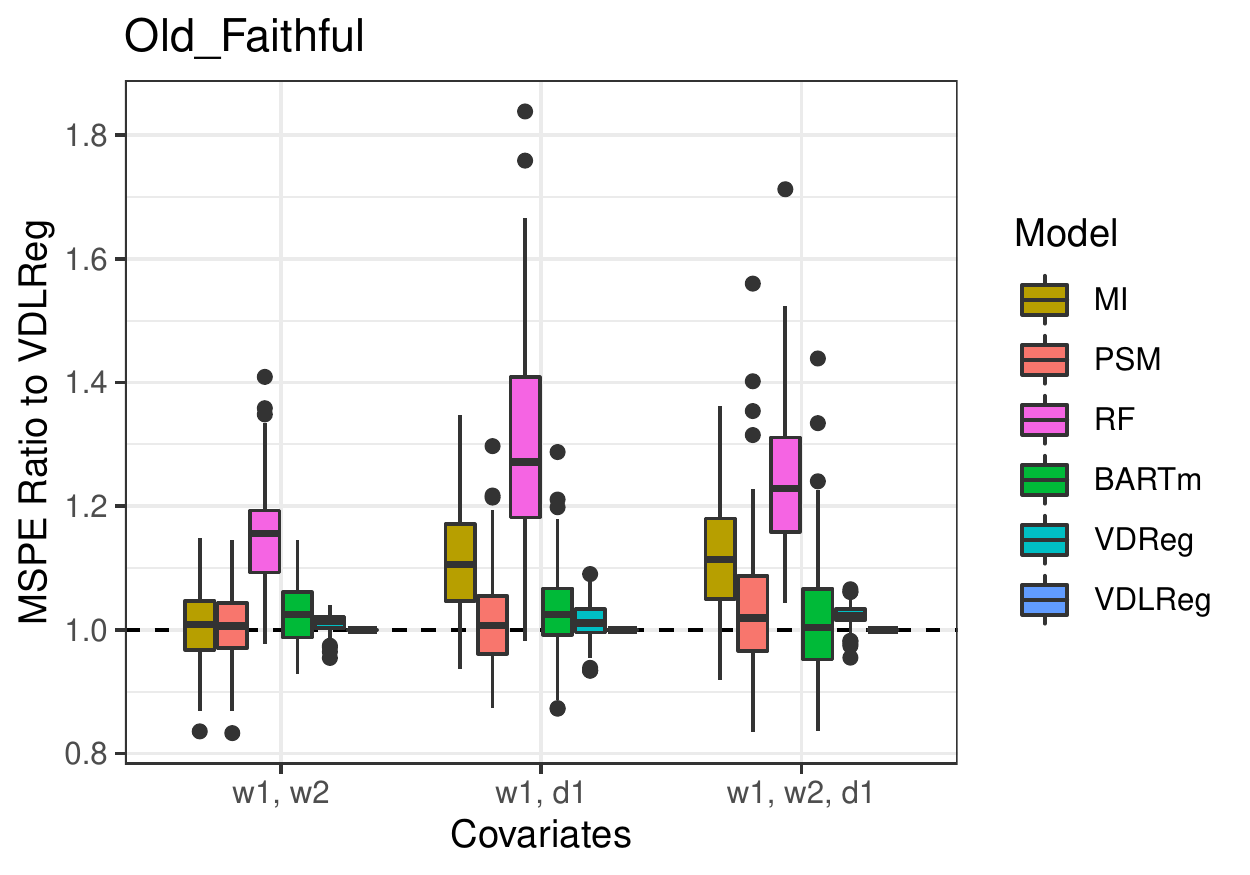} \includegraphics[width=3.1in, trim={0 0 0 20}, clip]{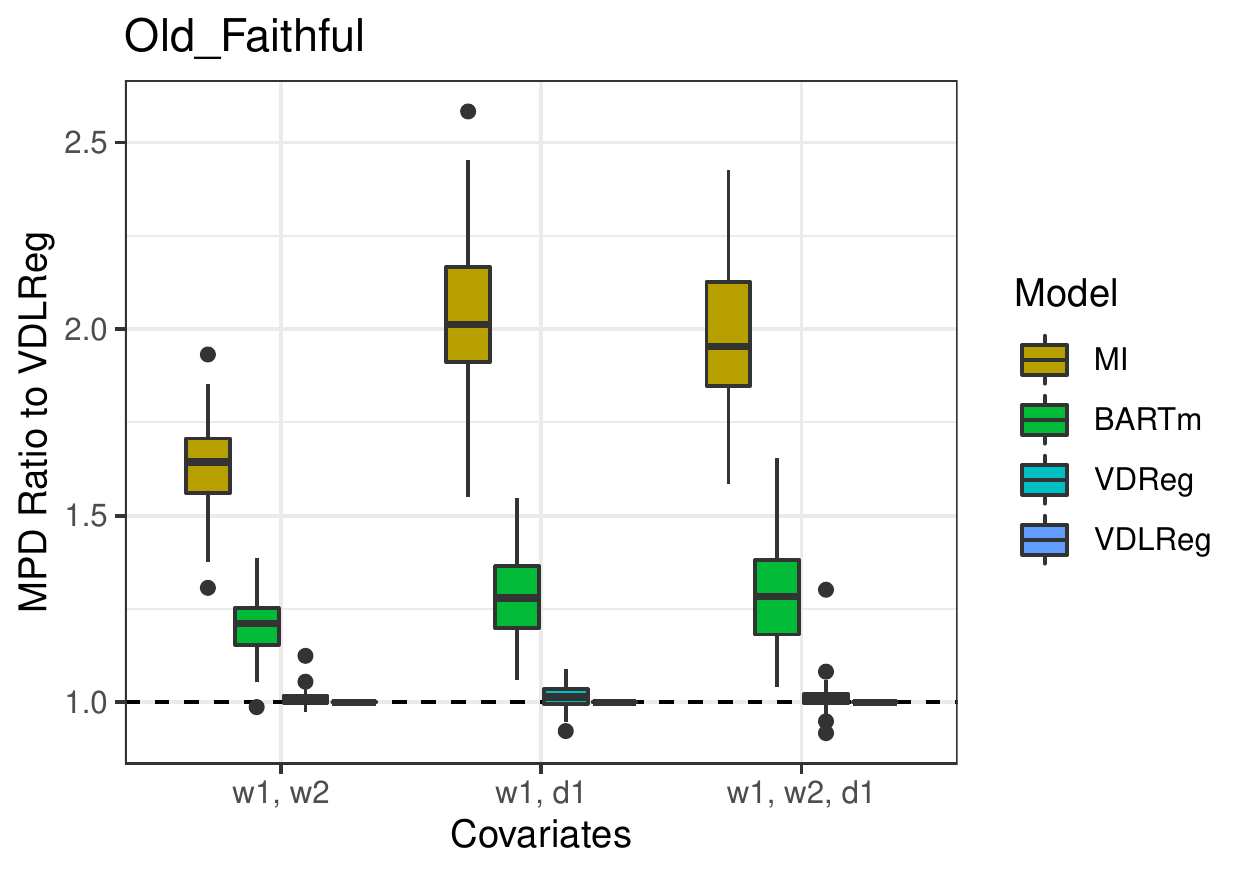}
    \caption{Box plots summarizing ratios of mean squared prediction error (left) and mean predictive deviance (MPD, right) for model fits relative to the VDLReg fit on the same replicated training/test pair of the Old Faithful data. Three covariate configurations are considered.}
    \label{fig:OldFaithfulMSPE}
\end{figure}

Figure \ref{fig:OldFaithfulMSPE} reports MSPE and mean predictive deviance with ratios against the VDLReg fit for the Old Faithful analysis. VDLReg is generally slightly preferred in MSPE, which becomes clear only when comparing methods within replicate data sets. PSM also performs well in MSPE, presumably because there are only two distinct missingness patterns when \texttt{d1} is present. The PPMx-based models enjoy a clear advantage in predictive deviance, possibly aided by their ability to capture bimodality in the response distribution in certain regions of the covariate space. Including \texttt{d1} as a covariate improves predictive accuracy.

\begin{figure}[h]
    \centering
    \includegraphics[scale=0.75, trim={0 0 75 20}, clip]{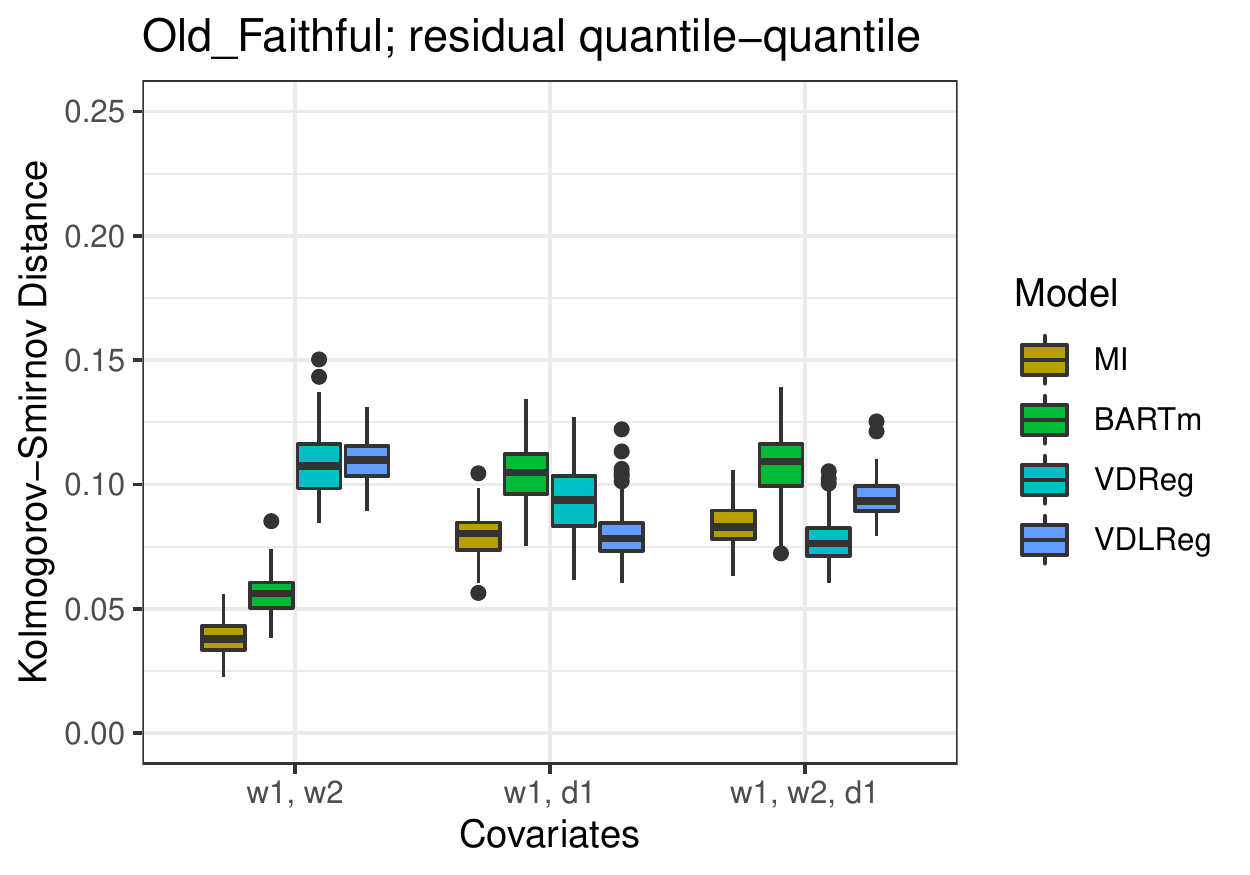}\includegraphics[scale=0.75, trim={17 0 0 20}, clip]{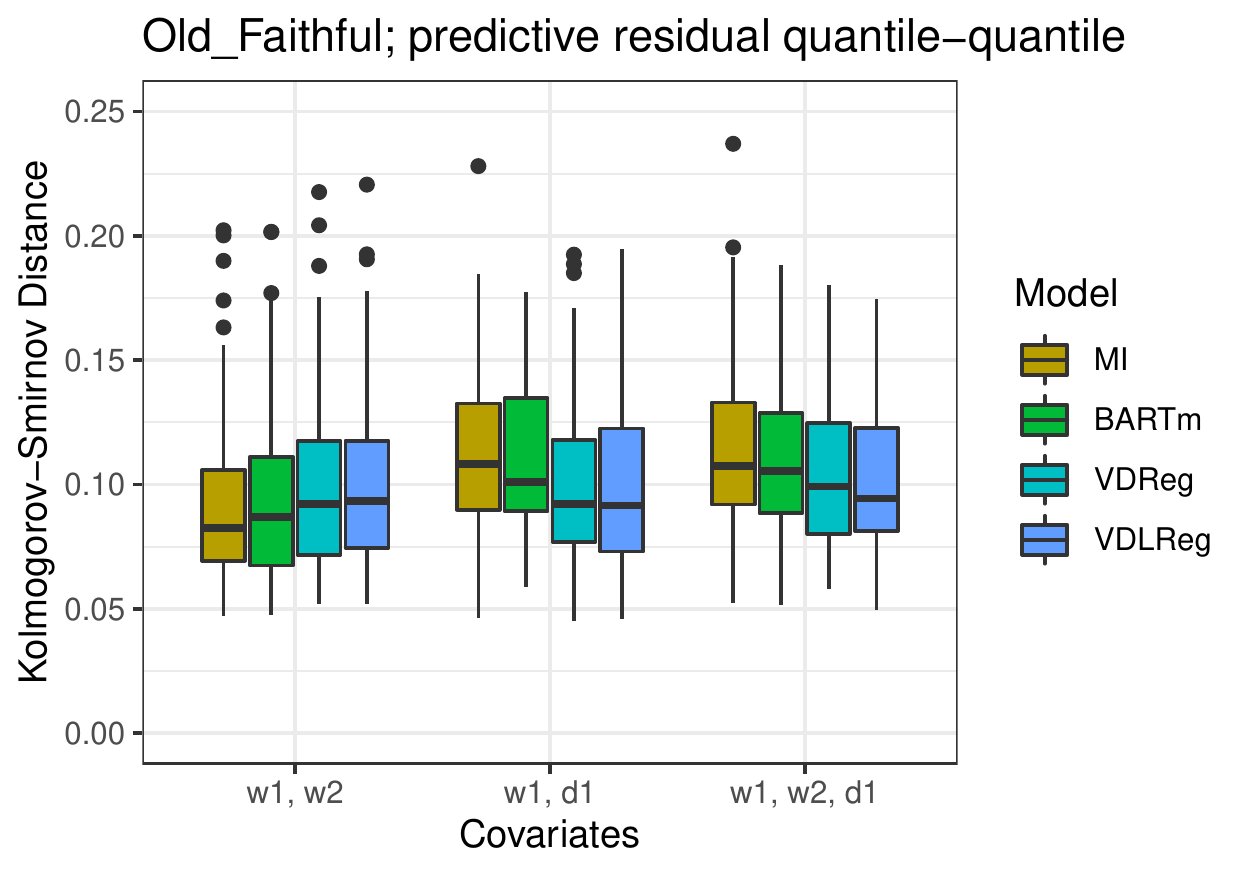}
    \caption{Box plots of Kolmogorov-Smirnov test statistics for uniformity of in-sample (left) and out-of-sample (right) quantile residuals from various fits of the Old Faithful data.}
    \label{fig:KS-OldFaithful}
\end{figure}

\begin{figure}[h]
    \centering
    \includegraphics[scale=0.7, page=3, trim={0 35 10 0}, clip]{figures/predQtile_competitors_OldFaithful_puse2_ii1_220816.pdf}\includegraphics[scale=0.7, page=1, trim={30 35 10 0}, clip]{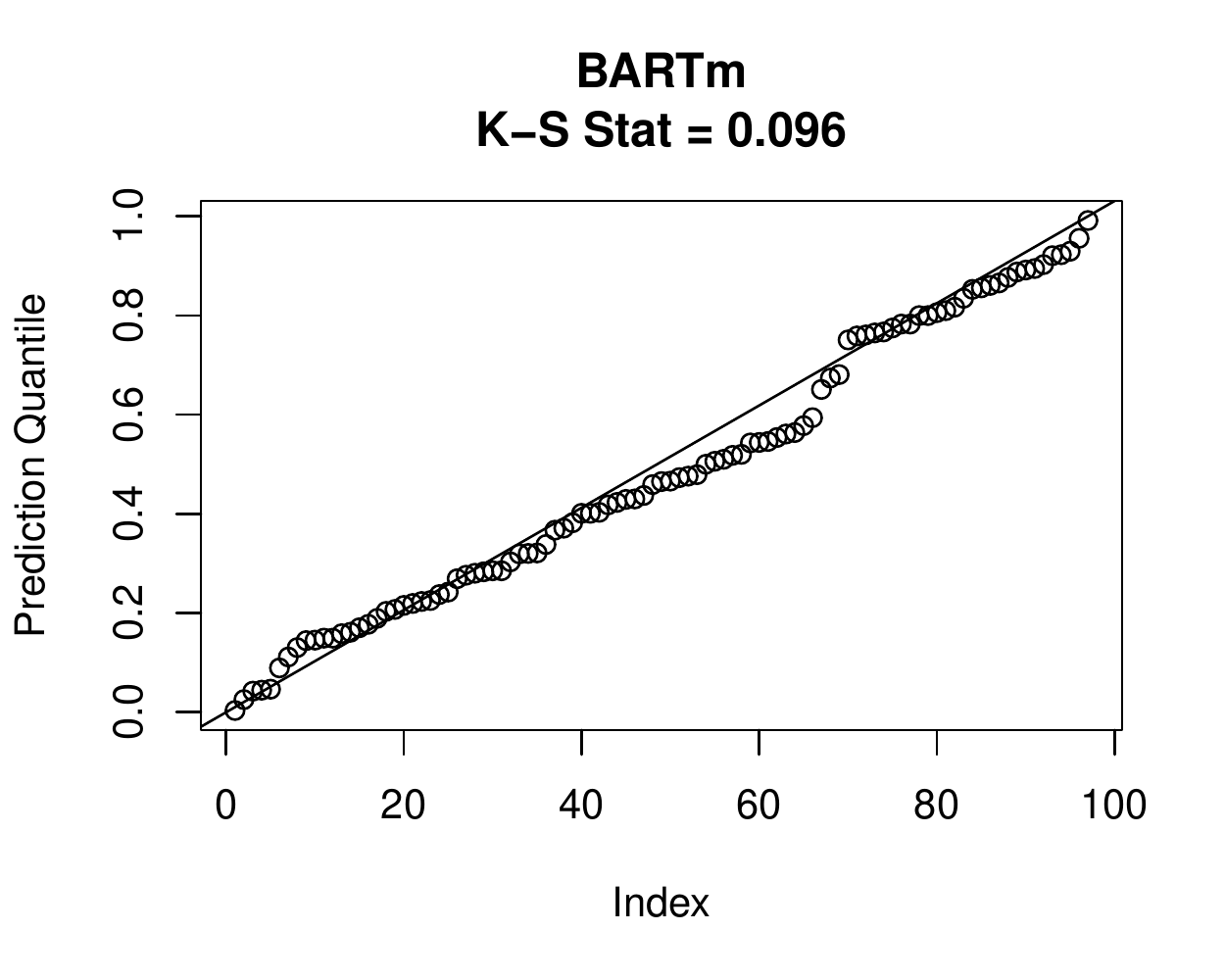}
    \includegraphics[scale=0.7, page=1, trim={0 0 10 0}, clip]{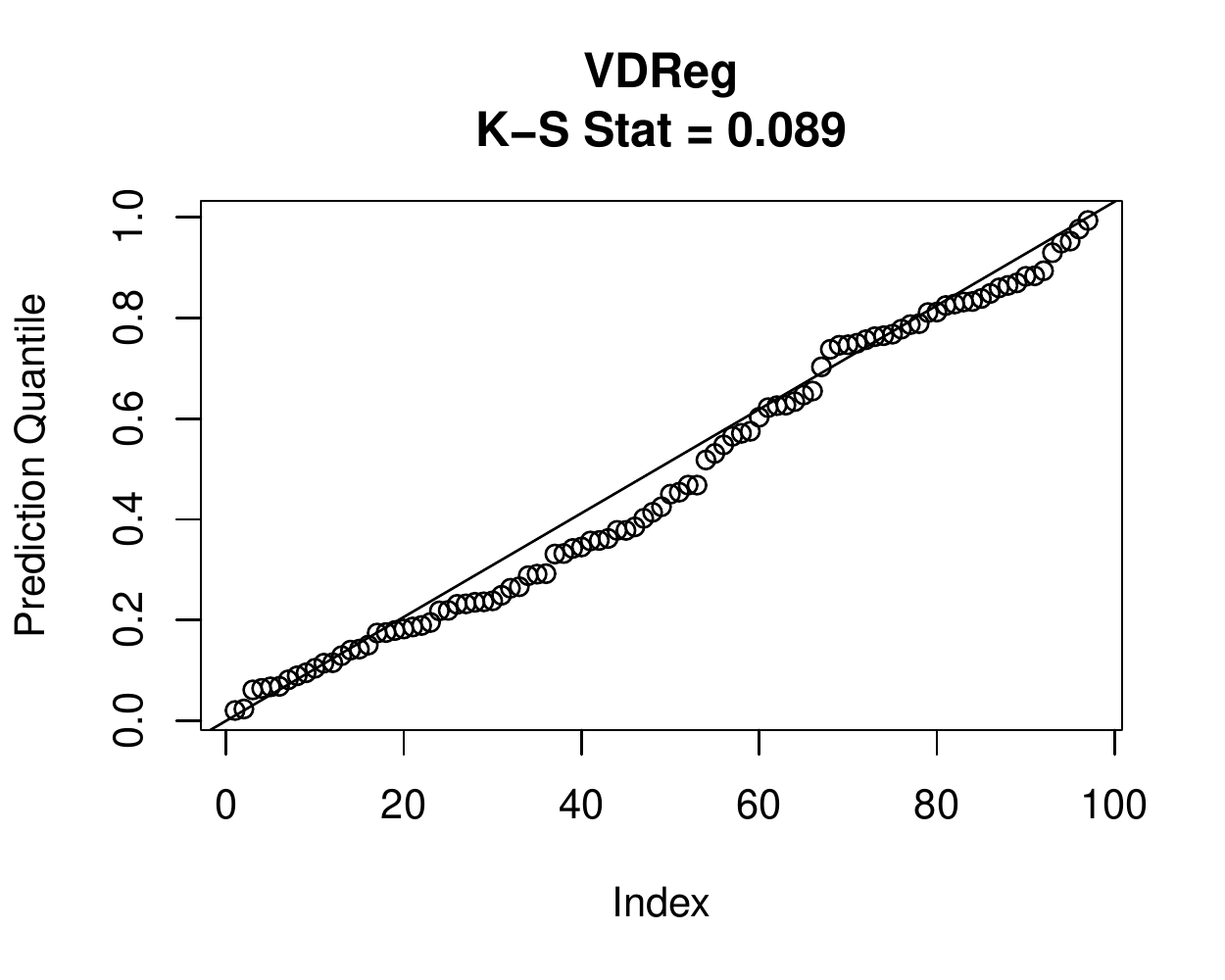}\includegraphics[scale=0.7, page=1, trim={30 0 10 0}, clip]{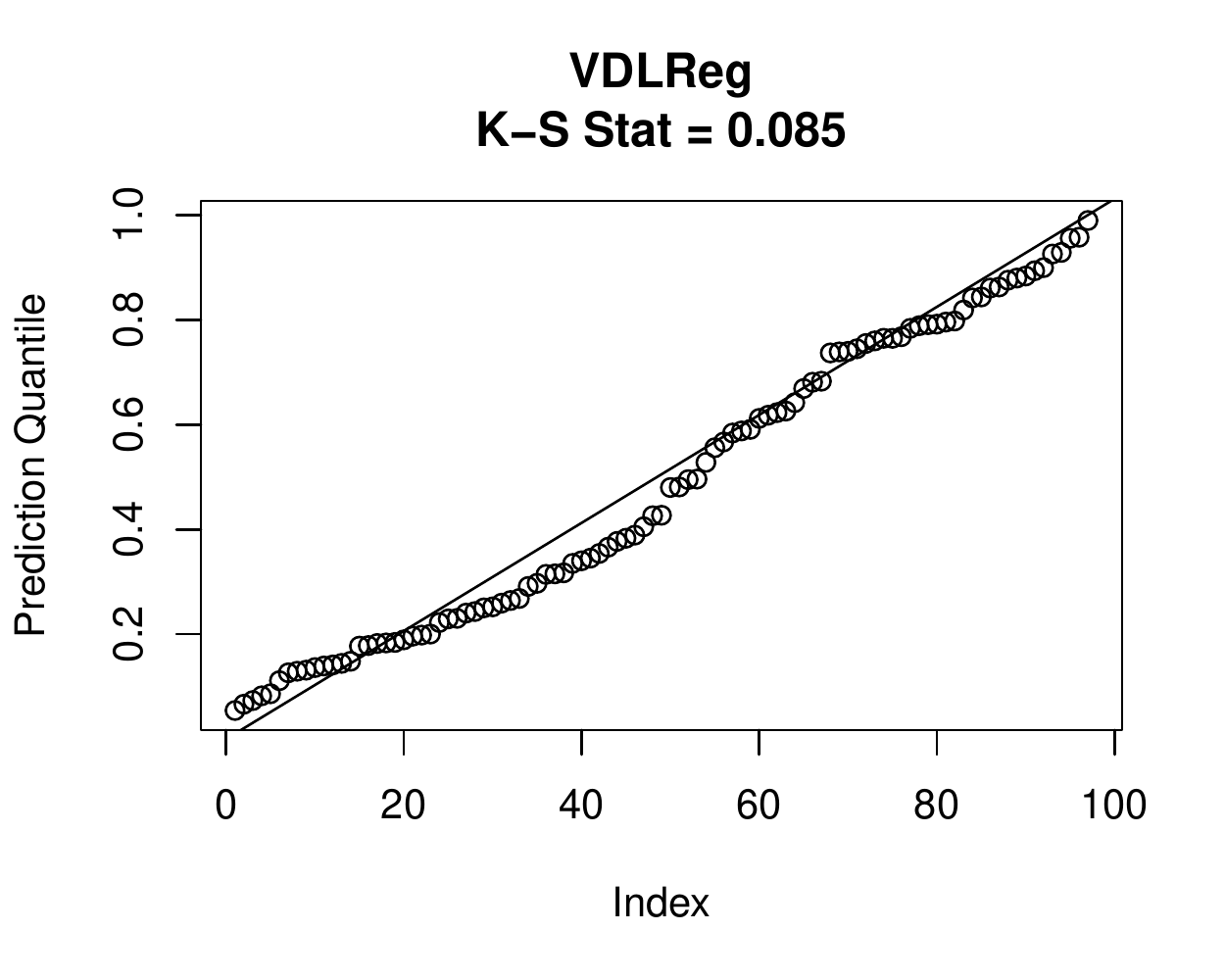}
    
    \caption{Ordered predictive quantile residuals for four fits to one replicate test set of the Old Faithful data using \texttt{w1} and \texttt{d1} as covariates.}
    \label{fig:Qresid-OldFaithful}
\end{figure}

Figures \ref{fig:KS-OldFaithful} and \ref{fig:Qresid-OldFaithful} summarize quantile-residual analysis with the Old Faithful data. All methods appear to generalize to test data acceptably, with exception of MI when \texttt{d1} is excluded from the model. Nevertheless, MI performs surprisingly well in this metric. This does not indicate a successful fit, but rather failure to detect other model inadequacies such as nonlinearity and bimodality. Note that \texttt{d1} is the only covariate with missing values, so MI in this case is a linear model. This underscores the importance of using multiple techniques when assessing goodness of fit. Unlike MI, VDReg and VDLReg successfully capture bimodality in the response distribution in  the \texttt{w1}, \texttt{w2} fit, but curiously exhibit systematic bias in the predictive distribution.

\end{document}